\documentclass[%
 reprint,
superscriptaddress,
 amsmath,amssymb,
 aps,
prb,
floatfix,
]{revtex4-1}

\usepackage[utf8]{inputenc} %Supporting package for xcolor
\usepackage[english]{babel} %Supporting package for xcolor
 
\usepackage[dvipsnames]{xcolor} %Note: xcolor is more powerful than color

\colorlet{color_random_v1}{green!10!orange!90!} % To define custom color (this is the color of every correlation curve produced with random state v1 of DMRG code)

\usepackage{lipsum} %To put one figure on top of another Source: https://tex.stackexchange.com/questions/62071/how-to-put-a-pdf-figure-on-top-of-another-one
\usepackage{graphicx} % Include figure files
\usepackage{subcaption} %To use subfigure

\usepackage[section]{placeins} %This prevents placing floats before a section. 
\usepackage{dcolumn}% Align table columns on decimal point
\usepackage{bm}% bold math
\usepackage{hyperref} % add hypertext capabilities
\hypersetup{
    colorlinks=true,
    linkcolor=blue,
    filecolor=blue,      
    urlcolor=blue,
}
\usepackage{color}
\usepackage{braket} % To write quantum mechanical bra and ket

\newcommand{\hide}[1]{}

\newcommand{\fig}[1]{Fig.\,\ref{#1}}

\begin{document}

\preprint{APS/123-QED}

%\title{Interesting phases of plane-polarized dipoles in a quasi-one-dimensional zigzag chain}
\title{Frustrated plane-polarized dipoles in one dimension}
\author{Niraj R. Ghimire}
\affiliation{Department of Physics, University of Connecticut, Storrs, CT 06269, USA}
\author{Susanne F. Yelin}
\affiliation{Department of Physics, University of Connecticut, Storrs, CT 06269, USA}
\affiliation{Department of Physics, Harvard University, Cambridge, MA 02138, USA}

\date{\today}% It is always \today, today,

\begin{abstract}

We investigate the zero-temperature quantum phases of a quasi-one-dimensional zigzag chain of dipoles that are polarized in a plane by an external electric field. Since the Hamiltonian contains nearest-neighbor (NN) and next-nearest-neighbor (NNN) hopping and interaction terms, this model allows frustration which induces phases that can be interesting and unusual. By using the density matrix renormalization group (DMRG) algorithm, we produce a complex phase diagram. This is an extension of an earlier work by Wang et. al. [\href{https://journals.aps.org/pra/pdf/10.1103/PhysRevA.96.043615}{Phys. Rev. A \textbf{96},\ 043615 (2017)}]. 
\end{abstract}

\maketitle

%\tableofcontents

\section{Introduction}

Ultracold atoms in optical lattices serve as ideal platform for quantum simulation, which is known to be a difficult problem even for the most advanced supercomputers of today, especially when the system size is large \cite{georgescu2014}. Because the geometry, dimension, and depth of an optical lattice can be controlled to a high degree, ultracold atom-based simulators have already been used to investigate quantum many-body problems applicable to fields ranging from condensed matter physics to high energy physics \cite{georgescu2014, gross2017}. Although atoms interact via short-range contact interactions in most cold atom experiments, many-body systems with longer-range interactions are predicted to exhibit intriguing quantum phases ~\cite{lahaye2009,baranov2012,hazzard2014,gross2017}. 

%\textcolor{blue}{In the presence of geometrical frustration, a situation where not all the interactions are satisfied, the system is expected to exhibit even more interesting features, for instance, Refs. ~\cite{bramwell2001,grohol2005,okamoto2007,murg2009,balents2010,struck2011,buessen2018}. One of the major challenges of quantum magnetism has been the search for spin liquid behavior ever since Anderson proposed it \cite{anderson1973,anderson1987}}. Spin liquid phases, which are phases with no magnetic long-range Neel order, are expected to be stable in systems where quantum fluctuations can strongly suppress magnetism, and these situations are found in low dimensions and in frustrated systems \cite{diep2005}. Our model is comprised of both. Moreover, a couple of recent papers on models that are similar to ours have depicted the existence of Haldane phase: a recent work by Furukawa et. al. \cite{furukawa2012} on a spin$-1/2$ frustrated ferromagnetic XXZ chain and a more recent one by Xu et. al. \cite{xu2018} on an experimentally realizable spin$-$1 model of bosons on a zigzag optical lattice. One of the questions that therefore arises is whether frustration in a zigzag lattice of plane-polarized dipoles leads to phases with non-trivial correlations between lattice points.

In the presence of geometrical frustration, a situation where not all the interactions are satisfied, the system is expected to exhibit even more interesting features. For instance, quantum spin liquid phases have been found in frustrated spin$-1$ diamond antiferromagnets \cite{buessen2018} and in frustrated spin$-1/2$ Heisenberg antiferromagnet on the kagome lattice \cite{yan2011}. Similarly, Haldane phases have been shown in a spin$-1/2$ frustrated ferromagnetic XXZ chain \cite{furukawa2012} and in a frustrated zigzag optical lattice of ultracold bosons \cite{greschner2013}. One of the questions that therefore arises is whether frustration in a zigzag lattice of plane-polarized dipoles leads to phases with non-trivial correlations between lattice points.

Wang et. al. \cite{wang2017} have shown a rich phase diagram for this system with the chain opening angle $\gamma \geq 2\pi/3$ (see \fig{fig:zigzag_chain}), a parameter regime with nearest-neighbor (NN) and next-nearest-neighbor (NNN) interactions, but only NN hopping. We produce a phase diagram for the same system, but setting NNN hopping to non-zero values, thus also allowing for much smaller chain opening angles $\gamma$. With the introduction of the NNN hopping, it becomes impossible to do exact calculations for a system size large enough to exhibit many-body effects, we therefore need a numerical approximation method. We use the Density Matrix Renormalization Group (DMRG) method \cite{white1992,white1993} because it is the most powerful numerical method to simulate one-dimensional systems ~\cite{schollwock2005, hallberg2006, schollwock2011}.   

\section{The model}

%To crop a picture: trim={<left> <lower> <right> <upper>}
\begin{figure}[htb!]
	\begin{subfigure}[b]{.5\textwidth}
    	\includegraphics[trim={2cm 7cm 2cm 3cm},clip,width=\textwidth,scale=1]{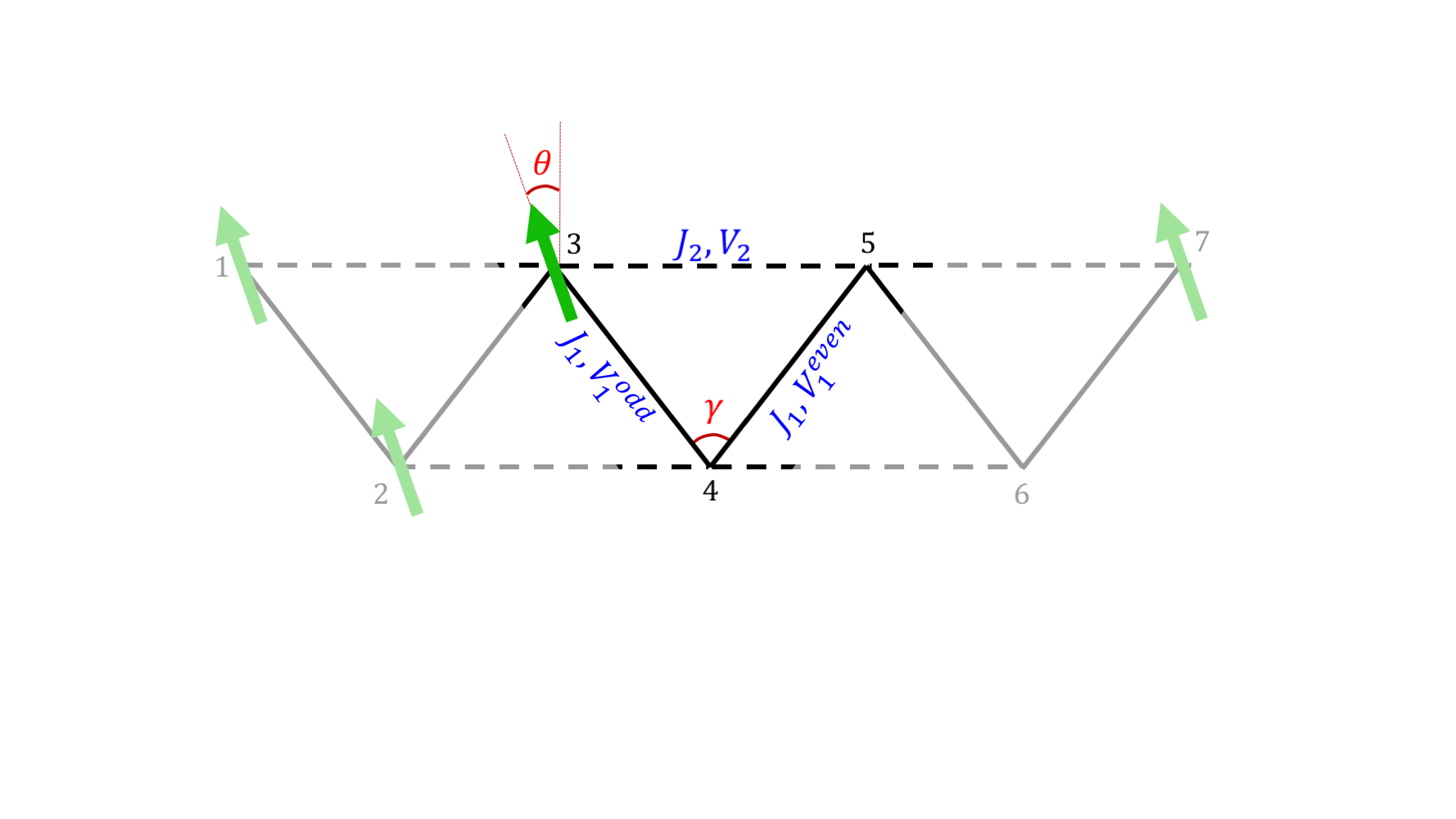}
    	\caption{Dipoles polarized at an angle $\theta$ in the plane of the zigzag chain}
		\label{label-zigzag_chain_dipoles}
	\end{subfigure}\hfill
	\begin{subfigure}[b]{.5\textwidth}
    	\includegraphics[trim={2cm 7cm 2cm 4cm},clip,width=\textwidth,scale=1]{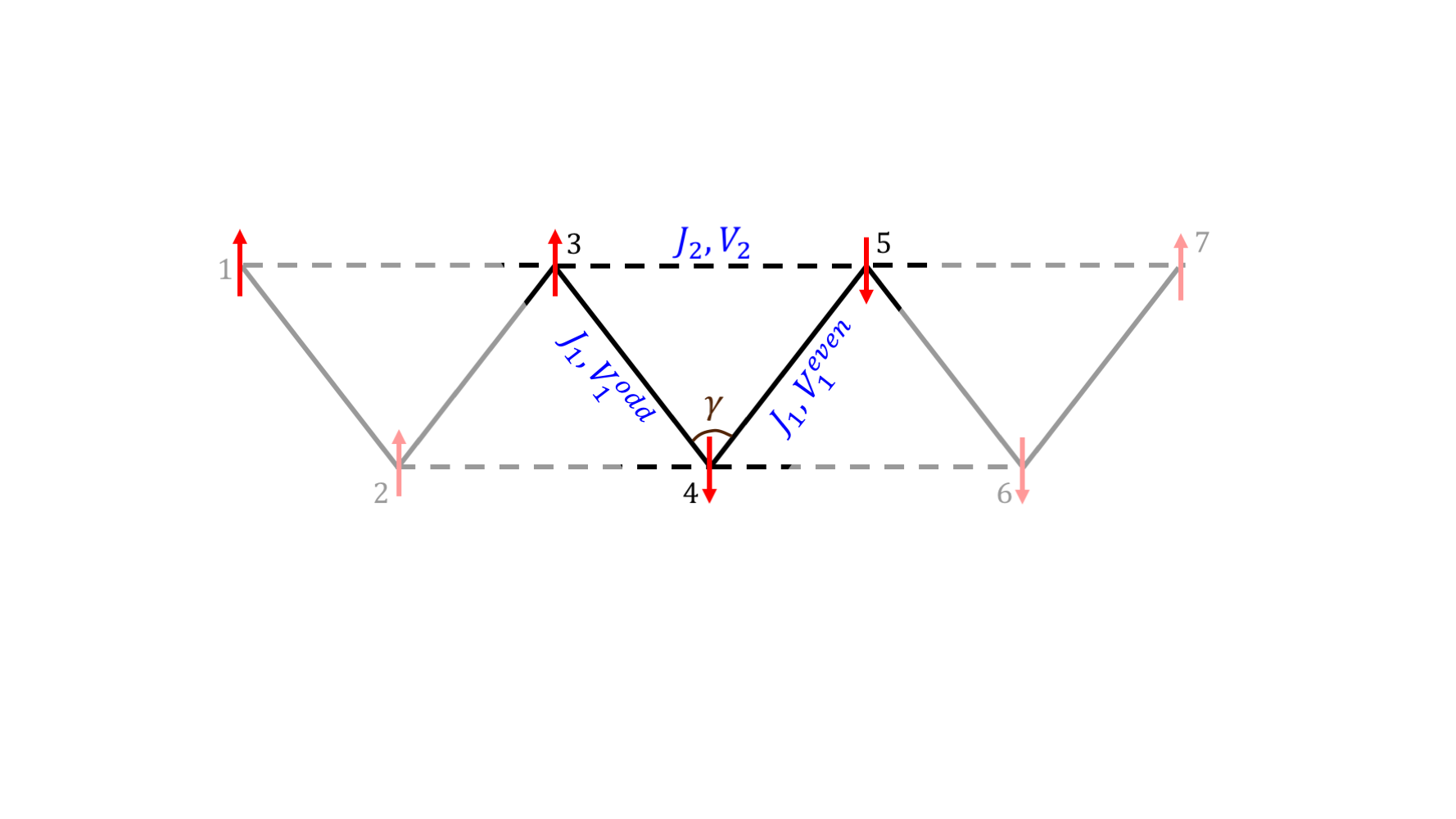}
        \caption{Spin$-1/2$ particles replacing the dipoles}
    	\label{label-zigzag_chain_spin_half}
	%\caption{3D plot for $a_e$.}%\label{label-b}
	\end{subfigure}
%\caption{(Color online) Mapping of $\alpha_o, \alpha_e$ and $\alpha_2$ to the physical parameter regime of $\gamma$ and $\theta$.}
	\caption{(Color online) A zigzag chain of dipoles mapped to one of spin$-1/2$ particles.  For our DMRG simulations, we have considered $N = 100$ sites but the figure shows only seven sites labeled 1 through 7. The hopping is allowed in a leg/direction (odd, even or NNN) of the chain only if the ends of the leg contain opposite spins.}
     \label{fig:zigzag_chain}
\end{figure}

\fig{fig:zigzag_chain} shows the spin$-1/2$ representation of the zigzag chain of dipoles. A dipole at a site is represented by a spin up, $\ket{1} \equiv \ket{\uparrow}$, while an empty site is represented by a spin down, $\ket{0} \equiv \ket{\downarrow}$. With the constraint that double occupancy is not allowed on any lattice sites, we map this quasi-one-dimensional model of dipoles to a spin$-1/2$ chain. %Note that this does not imply fermionic statistics; 
We treat these particles as hardcore bosons because two parallel dipoles on the same lattice site would experience an infinite on-site potential \cite{wang2017}.

Over the years, there has been a lot of work to study the phase diagram of frustrated two-leg spin ladders using various models, for instance, Refs.~\cite{azuma1994,lauchli2013,tonegawa2003,giri2017,wessel2017}. As compared to those, our model is simple because it is one-dimensional, has fewer degrees of freedom, and still exhibits frustration.

The Hamiltonian of the system is written as
\begin{align}
H = & -\ J_1 \sum_j (S^+_j S^-_{j+1} + h.c.)\ -\ J_2 \sum_j (S^+_j S^-_{j+2} + h.c.) \nonumber\\
& +\ V_1^{odd} \sum_{j = odd} S^z_j S^z_{j+1}\ +\ V_1^{even} \sum_{j = even} S^z_j S^z_{j+1} \nonumber \\ 
& +\ V_2 \sum_j S^z_j S^z_{j+2} +\ h \sum_j S^z_j
\label{eqn:Hamiltonian_in_terms_of_J1_and_J2}
\end{align}

\noindent
where $J_1 > 0$ and $J_2 > 0$ are NN and NNN hopping amplitudes and $h$ is the magnetic field. The system is half-filled, therefore the field term can be neglected. The spin operator $S^z$ is defined such that $S^z \ket{\uparrow} = \ket{\uparrow}$ and $S^z \ket{\downarrow} = -\ket{\downarrow}$. %Since the kinetic energy of a particle decreases when it hops to a neighboring site, it is natural to put a negative sign in front of the hopping amplitudes for ultracold systems. 
$V_1^{even}$ and $V_1^{odd}$ are NN dipolar interactions along even and odd legs of the chain respectively and $V_2$ is the NNN dipolar interaction. The interactions are related to the dipole coupling strength $\epsilon_{\rm dd} = \mu_e^2/{(4\pi \epsilon_o |\vec{r}_1 - \vec{r_2}|^3)}$, chain opening angle $\gamma$ and polarization angle $\theta$ as \cite{wang2017}:
\begin{align}
V_1^{even} & = \epsilon_{dd} \bigg [ 1 - 3 \cos^2 \bigg (\pi - \frac{\gamma}{2} - \theta \bigg ) \bigg ] \\
V_1^{odd} & = \epsilon_{dd} \bigg [ 1 - 3 \cos^2 \bigg (\frac{\gamma}{2} - \theta \bigg ) \bigg ] \\
V_2 & = \frac{\epsilon_{dd}}{[2(1-\cos(\gamma))]^{3/2}} \bigg [ 1 - 3 \cos^2 \bigg (\frac{\pi}{2} - \theta \bigg ) \bigg ]
\end{align}

\noindent
where $\epsilon_o$ and $\mu_e$ are the vacuum permittivity and electric dipole moment, and $\vec{r}_1$ and $\vec{r}_2$ are the position of the two interacting molecules.

Before running any numerical simulations, we want to get an intuitive understanding of the model. We start with some fundamental questions: Is there any regime where we can predict the ground state of the system and then use numerics to validate our prediction? Can we identify the frustrated and non-frustrated regimes and map them to the physical parameter regime of $\gamma$ and $\theta$? How are the NN and NNN hopping amplitudes related to one another and to $\gamma$ and lattice depth? How different do the ground state phase diagrams look like for different lattice depths? As shown in \fig{fig:zigzag_chain}, there are pairwise interactions in odd, even and NNN directions, each of which can be attractive or repulsive. We will study the effect of each interaction separately and put them together afterwards to analyze their collective effect on the system.

We write the Hamiltonian for any two interacting sites $i$ and $j$, where $j = i+1$ or $i+2$, as
\begin{align}
H_{\text{two-site-term}} & = \beta \Bigg (-\frac{1}{2} (S^+_i S^-_j + h.c.) + \alpha S^z_i S^z_j \Bigg)
\end{align}
where $\beta = 2J$ and $\alpha = V/2J$, and we refer to them as ``relative'' hopping and interaction strengths respectively. If we exactly solve this ``two site term'' in the basis \{$\ket{\uparrow\uparrow}, \ket{\uparrow\downarrow}, \ket{\downarrow\uparrow}, \ket{\downarrow\downarrow}$\}, we will obtain the following result: Regardless of the value of $\beta$, the two sites prefer parallel alignment, $\uparrow \uparrow$ or $\downarrow \downarrow$, represented by the letter ``F'' (for ``ferromagnetic'') if the pairwise interaction $\alpha < -1/4$, and antiparallel alignment, $\uparrow \downarrow$ or $\downarrow \uparrow$, represented by the letter ``A'' (for ``antiferromagnetic'') if $\alpha > -1/4$. It is worth noting that the critical value $\alpha_c = -1/4$ lies at the boundary between the two different configurations.

We can rewrite the full Hamiltonian as

\begin{align} \label{eq:H_in_terms_of_B_and_a}
H = & \sum_{j = odd} \beta_1 \Bigg (-\frac{1}{2} (S^+_j S^-_{j+1} + h.c.) + \alpha_{o} S^z_j S^z_{j+1} \Bigg) \nonumber\\
& + \sum_{j = even} \beta_1 \Bigg (-\frac{1}{2} (S^+_j S^-_{j+1} + h.c.) + \alpha_{e} S^z_j S^z_{j+1} \Bigg)\nonumber\\
 & + \sum_j \beta_2 \Bigg (-\frac{1}{2} (S^+_j S^-_{j+2} + h.c.) + \alpha_2 S^z_j S^z_{j+2} \Bigg)
\end{align}

\noindent
which is the sum of all the two-site terms in the three directions, where
\begin{eqnarray}
\beta_1 &=& 2 J_1,\ \beta_2 \;=\; 2 J_2,\nonumber\\
\alpha_{o} &=& \frac{V_1^{odd}}{2J_1},\ \alpha_{e} \;=\; \frac{V_1^{even}}{2J_1},\ \alpha_2 \;=\; \frac{V_2}{2J_2}.
\end{eqnarray}

The Hamiltonian written in this form helps us identify the frustrated and non-frustrated regimes and predict the ground state of the system prior to any simulations as we will discuss in the next section. 

%To crop a picture: trim={<left> <lower> <right> <upper>}
\begin{figure*}[htb!]
\begin{minipage}[b]{.3\textwidth}
    \includegraphics[trim={0cm 0cm 0cm 0cm},clip,width=\textwidth,scale=0.25]{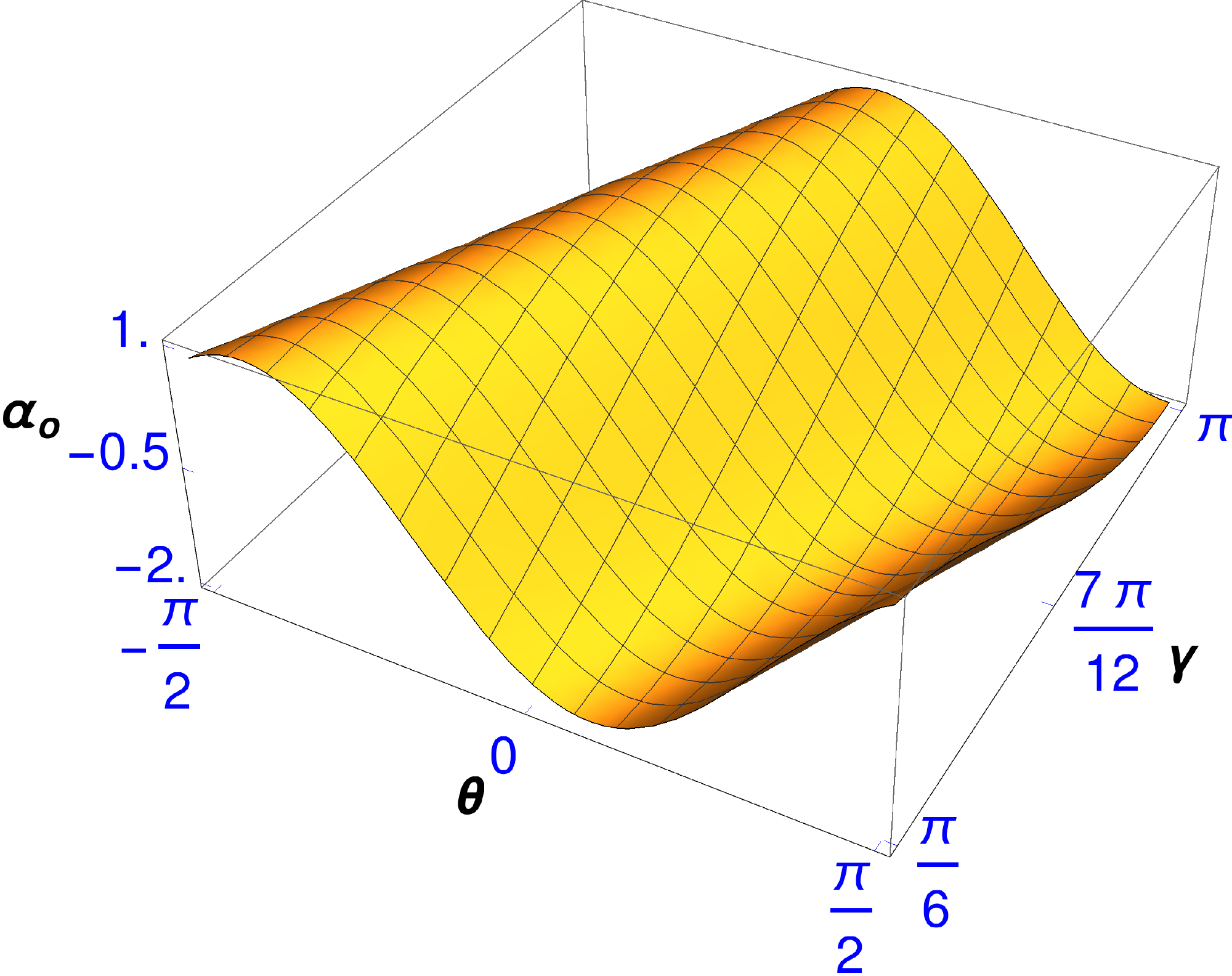}
	%\caption{3D plot for $a_o$. }%\label{label-a}
\end{minipage}\hfill
\begin{minipage}[b]{.3\textwidth}
    \includegraphics[trim={0cm 0cm 0cm 0cm},clip,width=\textwidth,scale=0.25]{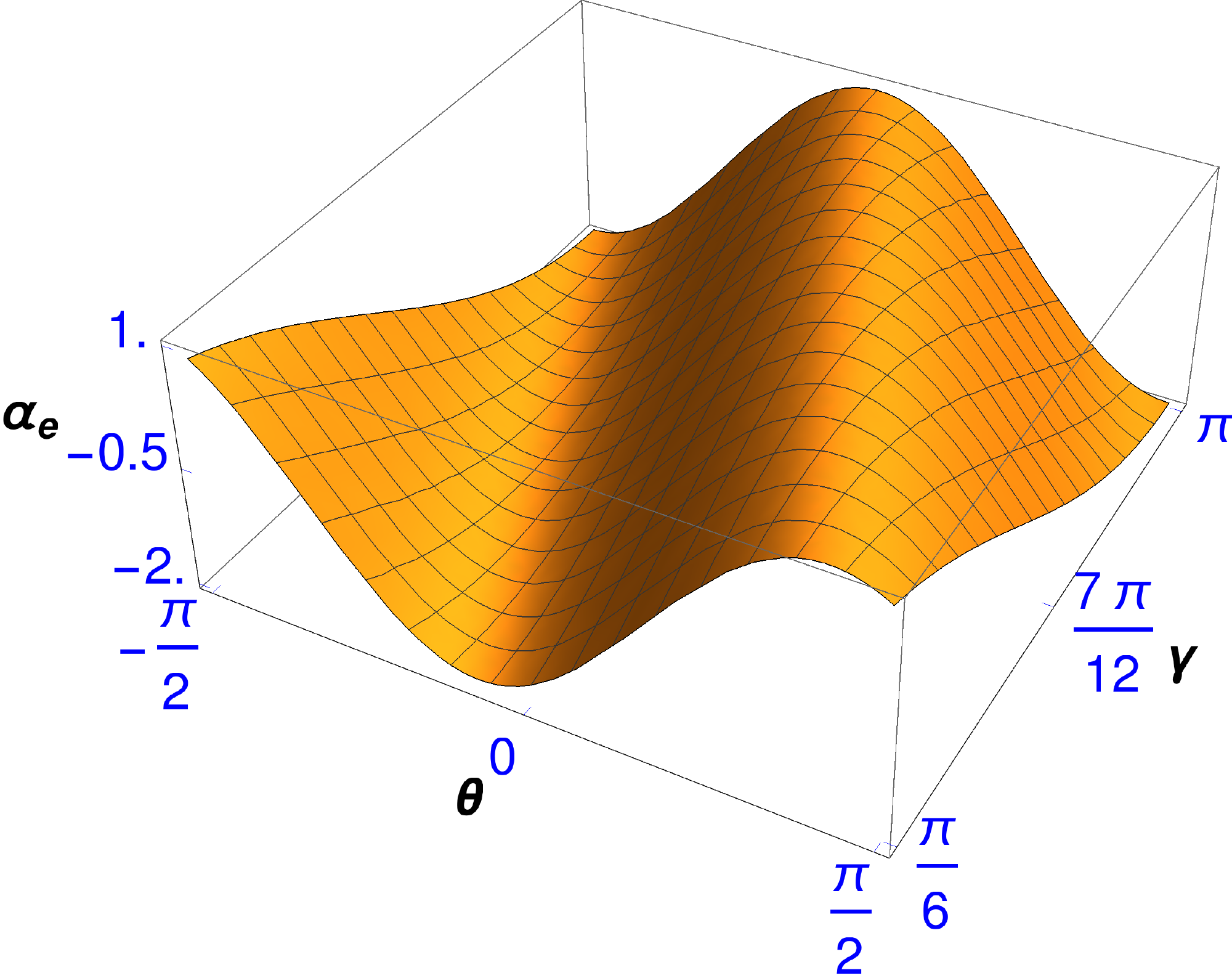}
	%\caption{3D plot for $a_e$.}%\label{label-b}
\end{minipage}\hfill
\begin{minipage}[b]{.3\textwidth}
    \includegraphics[trim={0cm 0cm 0cm 0cm},clip,width=\textwidth,scale=0.25]{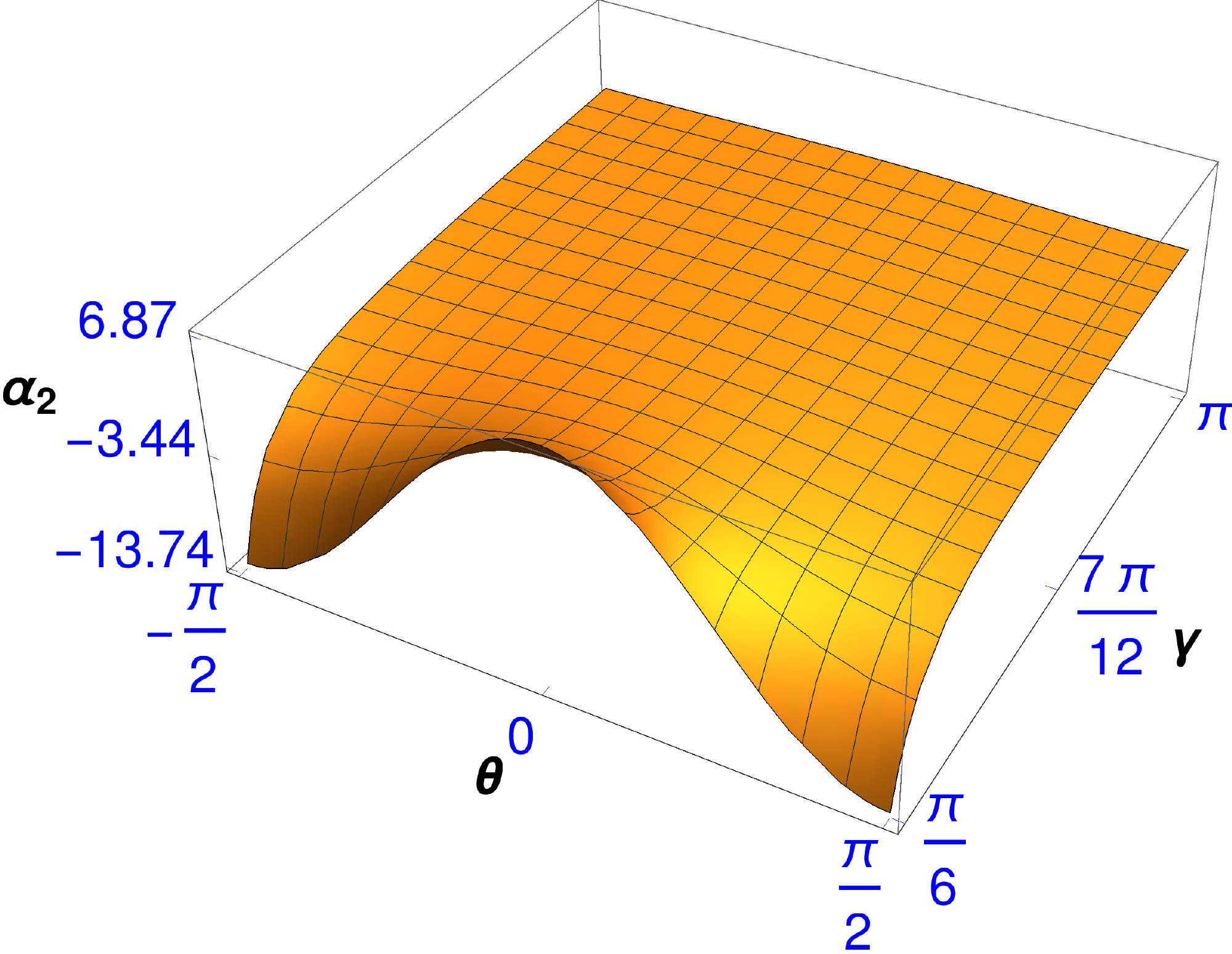}
	%\caption{3D plot for $a_2$.}%\label{label-b}
\end{minipage}\hfill
\caption{(Color online) Mapping of the relative interaction strengths $\alpha_o, \alpha_e$ and $\alpha_2$ to the physical parameter regime of the lattice, chain opening angle $\gamma$ and polarization angle $\theta$: Since $\beta_2$ and $\alpha_2$ diverge as $\gamma \rightarrow 0$, we take $\pi/6$ as an appropriate lower bound for $\gamma$. With $\pi/6 \leq \gamma \leq \pi$ and $-\pi/2 \leq \theta \leq \pi/2$, we observe that both $\alpha_o$ and $\alpha_e$ vary between $-2.00$ and $1.00$, while $\alpha_2$ varies between $-13.74$ and $6.87$.}
\label{fig:3D_plots_of_ao_ae_and_a2}
\end{figure*}

%To crop a picture: trim={<left> <lower> <right> <upper>}
\begin{figure}
    \centering
    \includegraphics[trim={0cm 0cm 0cm 0cm},clip,width=0.45\textwidth,scale=1.0]{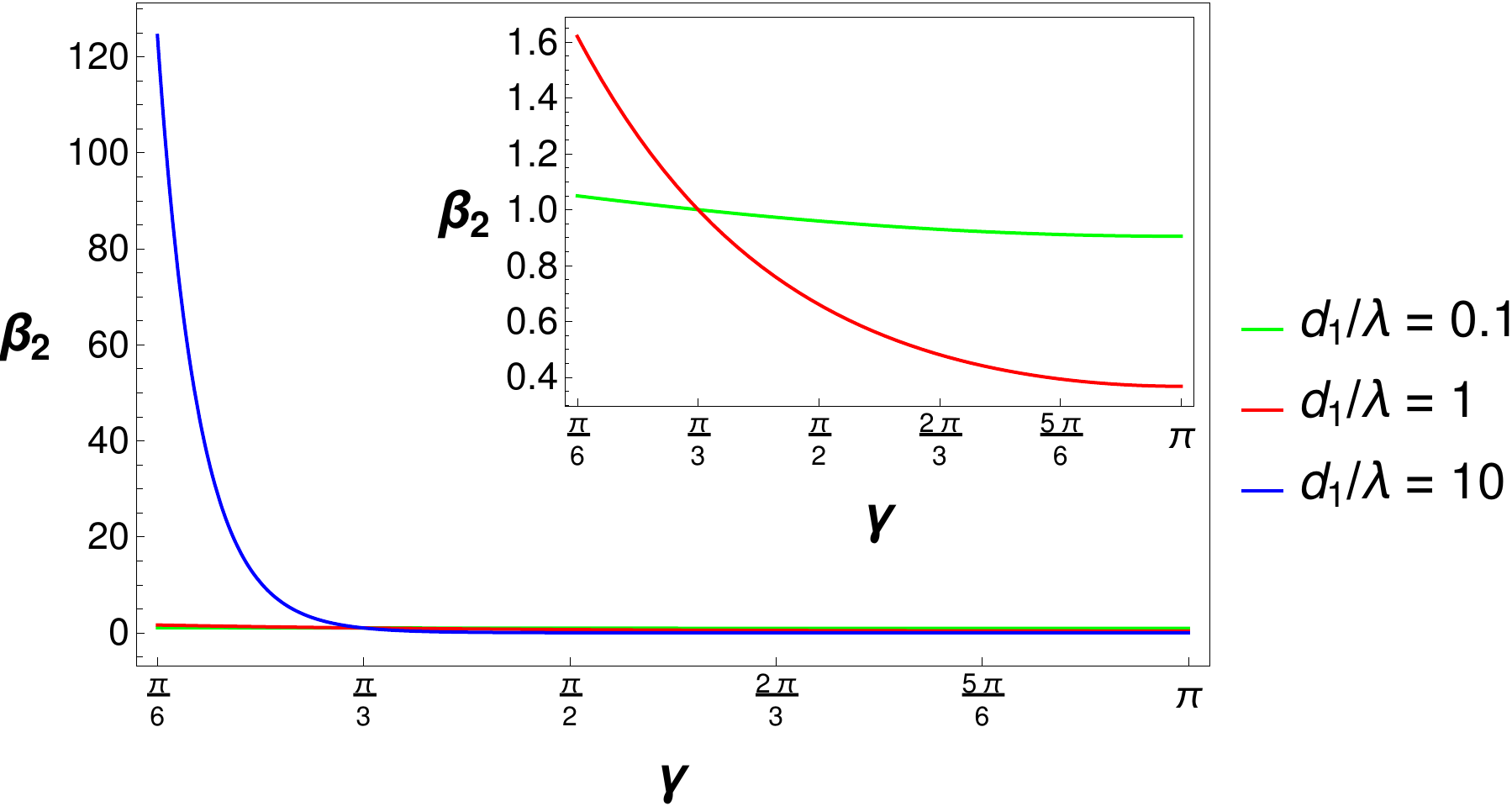}
    \caption{\protect\label{fig:plot_of_B2_versus_gamma} (Color online) Plot of $\beta_2$ against $\gamma$ for three different values of $d_1/\lambda$. Since $d_1/\lambda = 10$ corresponds to a deep lattice, $\beta_2$ increases much more exponentially with decreasing $\gamma$ as compared to the other two values of $d_1/\lambda$. The inset shows a zoomed in plot for $d_1/\lambda = 0.1$ which corresponds to a shallow lattice and for $d_1/\lambda = 1$ which corresponds to a lattice of intermediate depth as compared to the other two ratios.}
\end{figure}

The relative hopping amplitudes $\beta_1$ and $\beta_2$ depend on the distance between interacting sites, chain opening angle, and lattice depth. If $d_1$ and $d_2$ are the lengths of the odd (or even) and NNN legs respectively, then $d_2 = 2\ d_1 \text{sin}(\gamma/2)$. Using this relation and the fact that $\beta_1$ and $\beta_2$ decrease exponentially with distance, we can show that
\begin{align}
\frac{\beta_2}{\beta_1} & = \text{exp} \bigg [ -\frac{d_1}{\lambda} \Big (2 \sin(\gamma/2)-1 \Big ) \bigg ]
\end{align}
where $\lambda$ is a function of the lattice depth, and has the units of length. Although $d_1$ and $\lambda$ can change when $\gamma$ is varied, we can always set the ratio $d_1/\lambda$ to a desired value by tuning the lattice depth and thereby fixing $\lambda$ independent of $d_1$ or $\gamma$. The larger the value of the ratio $d_1/\lambda$, the deeper the lattice. Since $\gamma, \theta$ and $d_1/\lambda$ can be varied independently in real experiments, our model and all the results associated with it depend on these three parameters.

Throughout this paper, we use zero temperature, open boundary conditions,  and $\epsilon_{dd} = 1$, and unless otherwise stated, $d_1/\lambda = 0.1$. In addition, we set $\beta_1 = 1$, and with this choice of $\beta_1$ we allow the interactions to be much stronger than the hopping. 

\fig{fig:3D_plots_of_ao_ae_and_a2} shows how $\alpha_{o}, \alpha_{e}$ and $\alpha_2$ depend on $\gamma$ and $\theta$ while \fig{fig:plot_of_B2_versus_gamma} illustrates how $\beta_2$ varies with $\gamma$ for different lattice depths.

Before we proceed to the next section, we want to clarify that by setting the temperature to absolute zero we nullify thermal fluctuations. However, the experimental realization of this model would be a system at nanokelvin temperature with small but negligible thermal fluctuations. An example of such a system would be an ultracold bosonic gas of $^{23}$Na$^{87}$Rb molecules that are stable against chemical reaction in their absolute ground state \cite{zuchowski2010}, have a large permanent electric dipole moment (for instance, as large as 3.3 Debye \cite{aymar2005}) which can lead to strong dipolar interactions, and can be easily polarized by a moderate electric field. For instance, a $5 {\rm \ kV \ cm^{-1}}$ electric field can induce a dipole moment larger than 2 Debye \cite{wang2015}. As for the zigzag optical lattice, which can be produced by using three laser beams as explained in Ref. \cite{becker2010}, it would be natural to set $d_1 \sim 1  \ {\rm micrometer}$ because lattice constant is typically of that order. With a dipole moment of 5 Debye (since experimentally realizable systems consist of molecules with dipole moment $1-5$ Debye \cite{lemeshko2012}), the dipolar coupling strength $\epsilon_{\rm dd} \approx \mu_e^2/{4\pi \epsilon_o d_1^3} \approx 2.5 \times 10^{-30}$ Joules. A natural energy scale for molecules in optical lattice potentials is the molecular recoil energy $E_r = {\hbar^2 k^2}/{2m}$ where $m$ is the molecular mass. Since recoil energies (divided by the Plank constant $h$) are of the order of several kilohertz \cite{eckardt2017}, we estimate that $E_r/h \sim 10$ kilohertz for molecular dipoles which means $E_r \approx 6.63 \times 10^{-30}$ Joules. With this estimate, we obtain $\epsilon_{dd} \approx 2.65 E_r$. By setting $\beta_1 = 1$ and $\epsilon_{dd} = 1$, we are using $\epsilon_{dd}$ as our energy scale so that $J_1 = 0.5 \epsilon_{dd}$, a value that might be too small to probe experimentally but could be increased by using smaller lattice constant (i.e., $< 1 \ \rm micrometer$) or larger dipole moment (i.e., $> 5 \ \rm Debye$). With this value of $J_1$, we can readily see how the interaction strength in each of the three directions scales with the corresponding hopping strength. For instance, when $(\gamma, \theta) = (\pi/3,\pi/3)$, we obtain $|J_1/{V^{\rm even}_1}| = 0.5, |J_1/{V^{\rm odd}_1}| = 0.4$ and $|J_2/{V_2}| = 0.4$.

\section{Frustrated and non-frustrated regimes}

%To crop a picture: trim={<left> <lower> <right> <upper>}

\begin{figure}[htb!]
\centering
     \includegraphics[trim={0cm 0cm 0cm 0cm},clip,width=0.4\textwidth, scale=1.0]{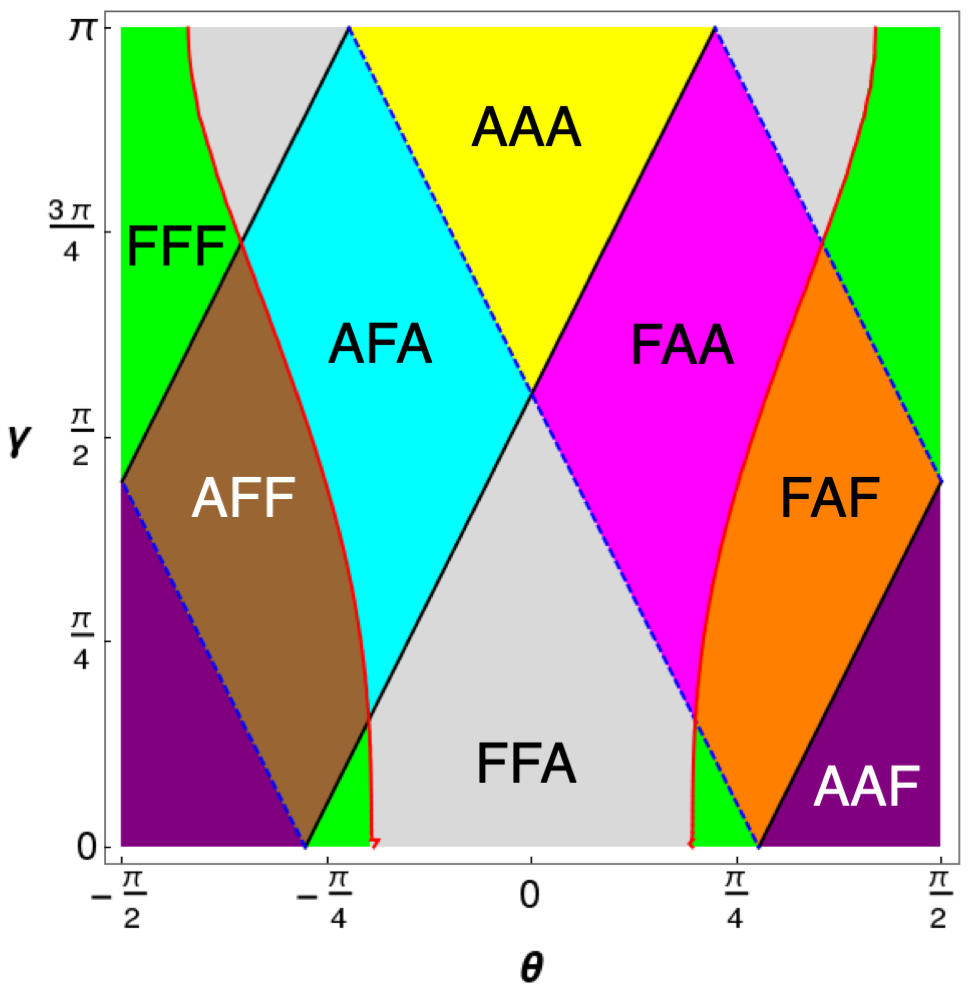}
	\caption{(Color online) Mapping of the frustrated and non-frustrated regimes to the physical parameter regime of chain opening angle $\gamma$ and polarization angle $\theta$. There are eight regions each with a unique color and labeled with three letters which correspond, from left to right, to the odd, even and NNN directions respectively (Frustrated: AAA, AFF, FAF and FFA; non-frustrated: FFF, AAF, AFA and FAA). The black solid, blue dashed and red solid lines represent the contours for $\alpha_o, \alpha_e$ and $\alpha_2$ respectively, each of which is equal to $-1/4$.}
	\label{fig:contourplot}
\end{figure}

As mentioned in the previous section, the pairwise interaction $\alpha$ in any direction is ferromagnetic or attractive if $\alpha <-1/4$, and antiferromagnetic or repulsive if $\alpha > -1/4$. If we arrange the interactions in all the directions based on whether they are attractive or repulsive, we find eight different combinations/regions as shown in \fig{fig:contourplot}. Although this figure corresponds to the value of $d_1/\lambda$ equal to $0.1$, we get qualitatively similar plots for any other value of $d_1/\lambda$ (see the Appendix); this implies that the phase diagrams should also be similar regardless of the value of $d_1/\lambda$. Of the eight regions, four (AAA, AFF, FAF and FFA) are in the frustrated regime while the other four (FFF, AAF, AFA and FAA) are in the non-frustrated regime. 

We will first explain and analyze non-frustrated regions in the absence of hopping and then discuss the potential scenario when the hopping is allowed. The simplest case of a non-frustrated regime is the region FFF where the pairwise interactions in all the directions are ferromagnetic (FM). In the absence of hopping, the spins would be classical and since the system is half-filled, the two equal energy states \{$\ket{\ldots\uparrow\uparrow\uparrow\downarrow\downarrow\downarrow \ldots}, \ket{\ldots \downarrow\downarrow\downarrow\uparrow\uparrow\uparrow \ldots}$\} would be the exact ground states (from now on, the curly braces \{\} will represent states with the same energy). %In the presence of hopping, however, we expect the ground state to be similar but not exactly the same.
Another non-frustrated region is AAF where the pairwise interactions in the odd and even directions prefer antiferromagnetic (AFM) alignment while that in the NNN direction prefers FM alignment. In the absence of hopping, the two Neel states \{$\ket{\uparrow\downarrow\uparrow\downarrow\uparrow\downarrow\ldots},\ket{\downarrow\uparrow\downarrow\uparrow\downarrow\uparrow\ldots}$\} are equally likely configurations to have the lowest energy and therefore, we expect the ground state to be AFM. %In the presence of hopping, we expect the ground state to be approximately AFM.
Similarly, the ground state is expected to be a dimer of the type \{$\ket{\uparrow\uparrow\downarrow\downarrow\uparrow\uparrow\downarrow\downarrow\ldots},\ket{\downarrow\downarrow\uparrow\uparrow\downarrow\downarrow\uparrow\uparrow\ldots}$\} in the non-frustrated region FAA, and of the type \{$\ket{\uparrow\downarrow\downarrow\uparrow\uparrow\downarrow\downarrow\ldots},\ket{\downarrow\uparrow\uparrow\downarrow\downarrow\uparrow\uparrow\ldots}$\} in the non-frustrated region AFA. In the presence of hopping, however, the four non-frustrated regions could feature phases that become superfluid instead of solid, particularly when the hopping dominates over the interactions.

The four regions in the frustrated regime are potentially more interesting. The first such region is AFF where the pairwise interaction in the odd leg prefers AFM alignment while those in the even and NNN legs prefer FM alignment. It is impossible for the spins to satisfy the interactions in all directions simultaneously, and hence the system is frustrated. We can make similar arguments to conclude that the other three regions FAF, FFA and AAA are also frustrated. As we will see later, there are regions in the frustrated regime where the pairwise interactions in the three directions are of similar strength and thus compete against one another. These regions require particular attention.

\section{Phase Diagram}

\begin{figure}[htb!]
	\includegraphics[trim={0cm 0cm 0cm 0cm},clip,width=0.45\textwidth, scale=1.0]{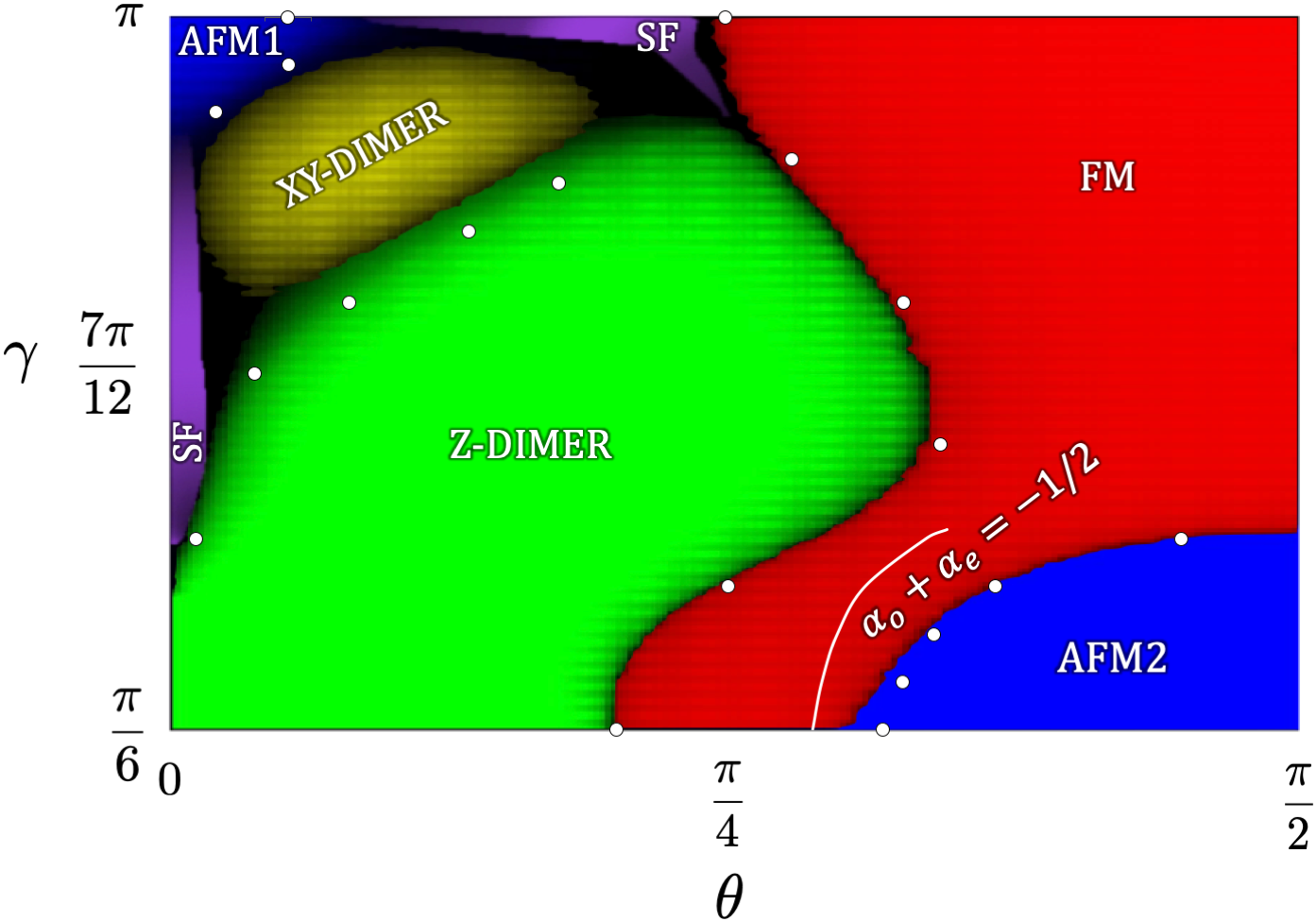}
	\caption{(Color online) Ground state phase diagram. These results depend on three independent parameters: chain opening angle $\gamma$, polarization angle $\theta$, and the ratio $d_1/\lambda$ which we have set equal to $0.1$. Each color is associated with a different phase; the brighter a color, the deeper the system in that phase. The black color corresponds to the region where the order parameters vanish for all phases. AFM1 and AFM2 are both antiferromagnetic phases labeled differently because of the nature of the ground state returned by DMRG. The white curve labeled as ``$\alpha_o + \alpha_e = -1/2$'' represents the physical parameter regime where one of the pairwise interactions in NN directions is attractive while the other repulsive, and they both are the same distance away from their critical values $\alpha_{o,c} = \alpha_{e,c} = -1/4$. The superfluid phase has been drawn using the values of the correlation function for the finite system size $N = 100$. All the other phases and their boundaries have been drawn using the values of order parameters for the aforementioned system size. The white dots with very small error bars, obtained using finite-size scaling analysis, represent the phase boundaries in the thermodynamic limit $N \rightarrow \infty$.}
    \label{fig:phase_diagram}        
\end{figure}

\fig{fig:phase_diagram} shows the zero-temperature ground state phase diagram of the system for different values of $\gamma$ and $\theta$. This diagram has been produced with several DMRG trials each with a different initial state/condition, and the most appropriate ground state (the one with the lowest energy possible) has been considered. The different phases, the order parameters and correlation functions used to identify them, and the crossover between those phases will be discussed in the subsequent paragraphs (see the Appendix for additional correlations). We label the initial state as $\ket{\text{init}}$. We name the initial state with spins randomly distributed in the lattice as ``random initial state'' and label it as $\ket{\text{random}}$. The letter ``$E$'' with a value attached to it will represent the energy of the ground state returned by a simulation. We will often show ground states for two different initial states to demonstrate how the initial conditions affect the final results obtained from DMRG simulations. When we show the results for only one initial state, it means that the state has led to the most appropriate ground state. The color brightness for each phase represents the value of its order parameter while the black color represents the region where all the order parameters vanish. We produce this phase diagram for the finite system size $N = 100$ and we extrapolate the boundary between phases in the thermodynamic limit $N \rightarrow \infty$ using finite-size scaling analysis which we will discuss later. We find a sharp transition between FM and AFM phases, and hence DRMG pinpoints the boundary between these two phases, while we find a smooth transition everywhere else as we will discuss later.

It should be noted that the Hamiltonian Eq.\eqref{eq:H_in_terms_of_B_and_a} remains unchanged under the transformation $\theta \to -\theta$ (where $\alpha_o$ and $\alpha_e$ swap their values while $\alpha_2$ stays the same). This implies that the phase diagram gives similar results in the range $\theta \in [-\pi/2, 0]$ as in the range $\theta \in [0,\pi/2]$, and therefore, we can restrict ourselves to the latter.

%To crop a picture: trim={<left> <lower> <right> <upper>}

\subsection{Dimerized phases}

\begin{figure}[htb!]
\begin{subfigure}[b]{0.24\textwidth}
\centering
  \includegraphics[trim={0cm 0cm 0.35cm 0cm},clip,width=\linewidth]{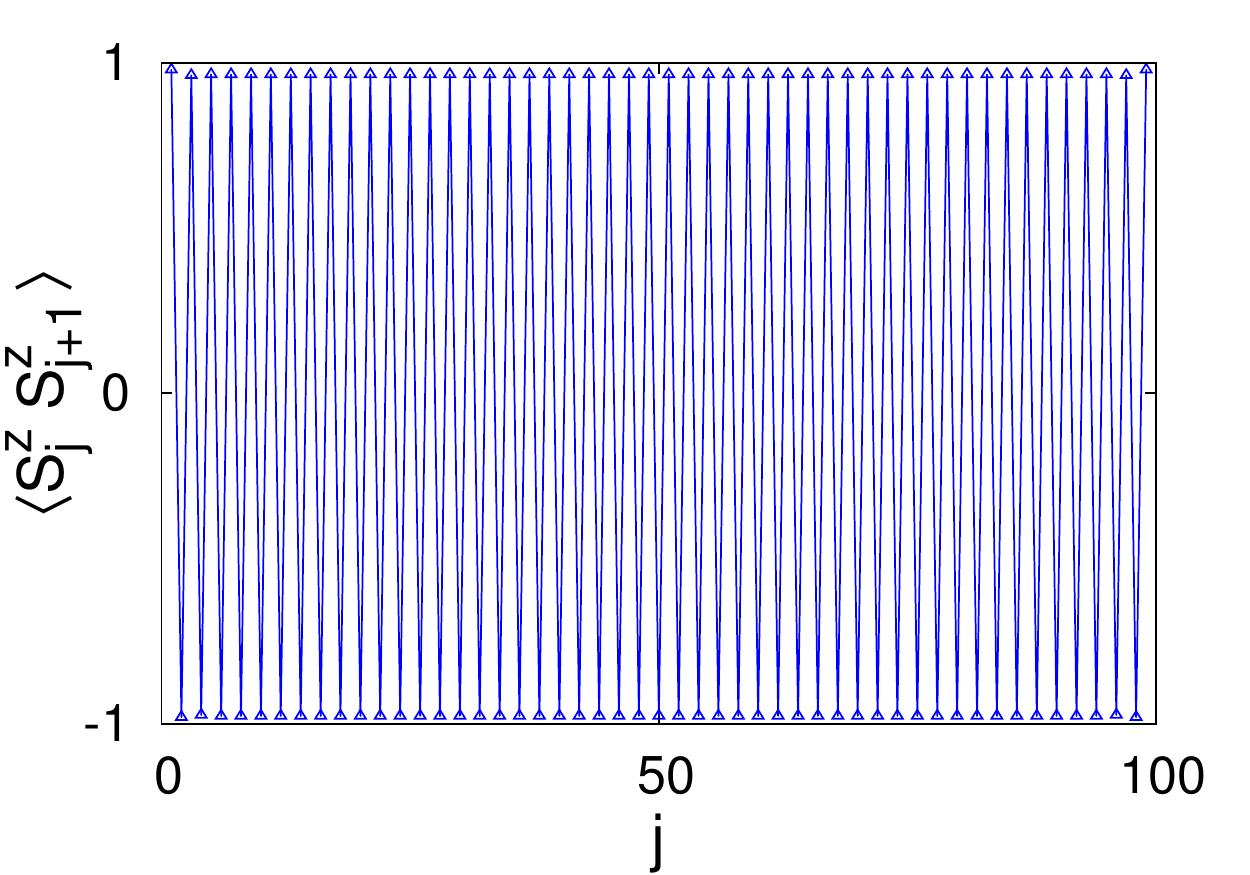}
\caption{$(\gamma, \theta) = (\pi/3, \pi/6)$. Region: FAA.}
\label{label-random_v1_gamma_60_theta_30}
\end{subfigure}\hfill
\begin{subfigure}[b]{0.24\textwidth}
\centering
  \includegraphics[trim={0cm 0cm 0.35cm 0cm},clip, width=\linewidth]{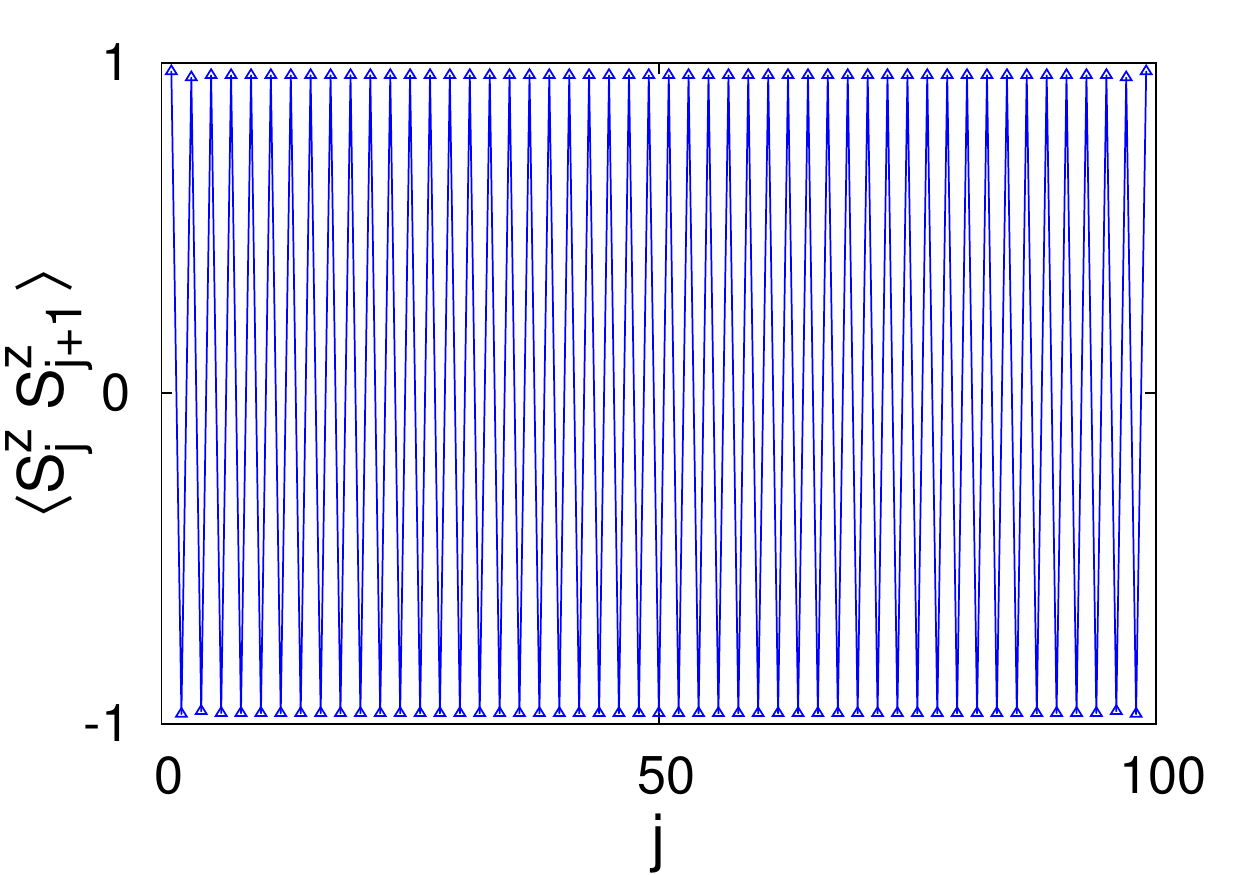}
\caption{$(\gamma, \theta) = (\pi/3, \pi/12)$. Region: FFA.} 
\label{label-random_v1_gamma_60_theta_15}
\end{subfigure}
\caption{(Color online) Z-dimer phase. $\ket{\text{init}} = \ket{\downarrow\uparrow\downarrow\uparrow\downarrow\uparrow \ldots}$. The left plot shows a z-dimer in the non-frustrated region FAA as expected. The right plot shows a similar phase in the frustrated region FFA which clearly indicates that the attractive interaction in the odd (or even) direction and the repulsive interaction in the NNN direction dominate over the attractive interaction in the third direction.}
\label{fig:z_dimer_gamma_60}
\end{figure}

In the earlier section, we mentioned two distinct sets of expected ground states: \{$\ket{\uparrow\uparrow\downarrow\downarrow\uparrow\uparrow\downarrow\downarrow...},\ket{\downarrow\downarrow\uparrow\uparrow\downarrow\downarrow\uparrow\uparrow...}$\} and \{$\ket{\uparrow\downarrow\downarrow\uparrow\uparrow\downarrow\downarrow...},\ket{\downarrow\uparrow\uparrow\downarrow\downarrow\uparrow\uparrow...}$\}. We call this type of dimer a ``z-dimer'' and although the non-frustrated regions FAA and AFA are the natural candidates for this phase, a frustrated region can also exhibit this type of phase as shown in \fig{fig:z_dimer_gamma_60}.

%To crop a picture: trim={<left> <lower> <right> <upper>}

%To crop a picture: trim={<left> <lower> <right> <upper>}

\begin{figure}[htb!]
\begin{subfigure}[b]{0.24\textwidth}
\centering
\includegraphics[trim={0cm 0cm 0.35cm 0cm},clip, width=\linewidth]{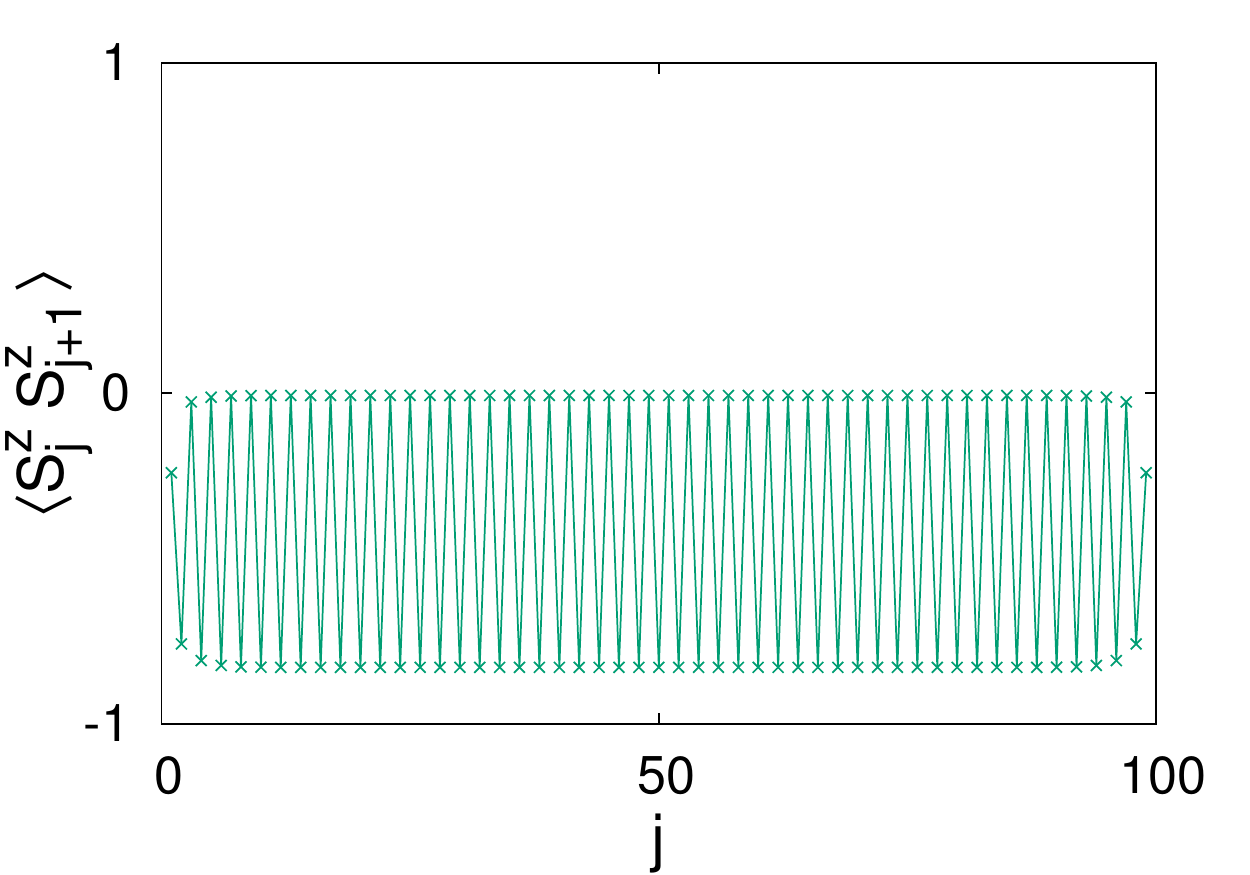}
\caption{} 
\label{label-random_v1_and_v2_gamma_150_theta_16_Szj_Szjp1}
\end{subfigure}\hfill
\begin{subfigure}[b]{0.24\textwidth}
\centering
\includegraphics[trim={0cm 0cm 0.35cm 0cm},clip,width=\linewidth]{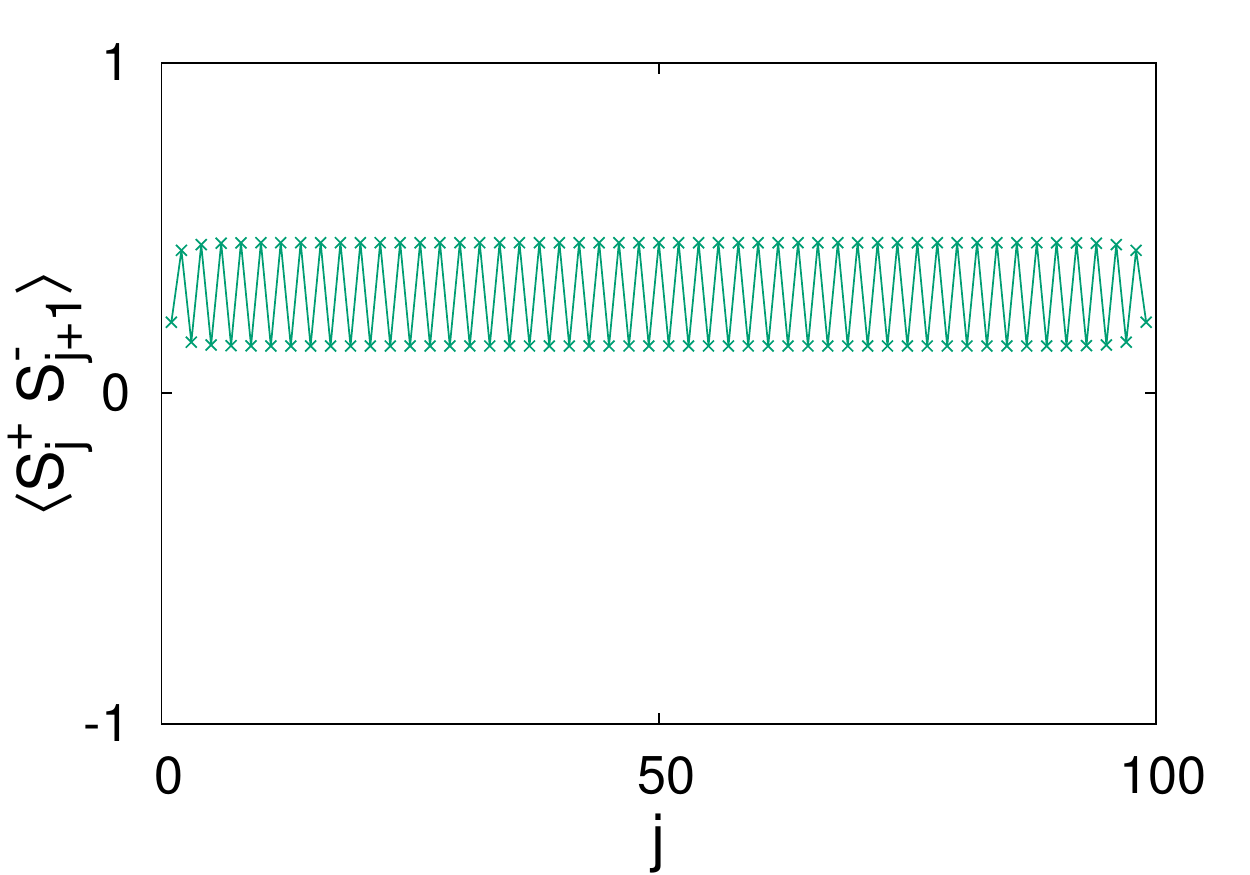}
\caption{}
\label{label-random_v1_and_v2_gamma_150_theta_16_Spj_Smjp1}
\end{subfigure}
\caption{(Color online) XY-dimer phase. $(\gamma, \theta) = (5\pi/6, 0.0889\pi)$. Region: AAA. $\ket{\text{init}} = \ket{\text{random}}$. These two plots have been produced with exactly the same initial condition. What we see is an example of a xy-dimer with dangling spins, which means the repulsive interaction in the odd direction has a dominating effect over that in the even and NNN directions.}
\label{fig:gamma_150_theta_16_xydimer_phase}
\end{figure}

Before discussing the other type of dimer that appears in the phase diagram, let us define $\ket{+} \equiv (1/\sqrt{2})(\ket{\uparrow\downarrow} + \ket{\downarrow\uparrow})$. Then a ``xy-dimer'' is simply the triplet bound state $\ket{+} \otimes ... \otimes \ket{+}$  or the one with free spins at the edges (often referred to as ``dangling spins'') \{$\ket{\uparrow} \otimes \ket{+} \otimes ... \otimes \ket{+} \otimes \ket{\downarrow}, \ket{\downarrow} \otimes \ket{+} \otimes ... \otimes \ket{+} \otimes \ket{\uparrow}$\}. The xy-dimer with dangling spins (or bound spins at the edges) is plausible when the interaction in the even (or odd) direction is highly repulsive while that in the other two directions is weak as shown in \fig{fig:gamma_150_theta_16_xydimer_phase}. If the hopping amplitudes were positive (i.e., $J_1 < 0$ and $J_2 < 0$), as is the case for fermionic statistics, the xy-triplets would be replaced with xy-singlets \footnote{It is worth noting that the xy-dimer with dangling spins will look similar to the valence bond solid state of the AKLT model if we replace the xy-triplets with xy-singlets; however, the absence of non-local correlation in the former makes it strikingly different from the latter.}.

Spin liquid phases, which are phases with no magnetic long-range Neel order, are expected to be stable in systems where quantum fluctuations can strongly suppress magnetism, and these situations are found in low dimensions and in frustrated systems \cite{diep2005}. Our model is comprised of both. In the following paragraph, we explore the possibility of such a phase.

For a finite lattice, a xy-dimer phase with bound spins at the edges is lower in energy than the one with dangling spins at the edges, and the system chooses as its ground state the former or the latter depending on the values of the pairwise interactions. In the thermodynamic limit, however, the two phases would have the same energy. Therefore, one would expect the frustrated region that results in the xy-dimer phase to be an ideal candidate for a spin liquid phase when the interactions in the odd and even directions are equally repulsive; this would allow the ground state to be in the superposition of the two xy-dimer phases, a state similar to a \textit{resonating valence bond} (see Ref. \cite{zhou2017} for a nice review of this state) but with the xy-singlets replaced with xy-triplets. In other words, a spin liquid phase may occur if the triplet bond connecting two adjacent sites can freely switch between odd and even directions. The fact that the pairwise interactions in the two NN directions are always unequal in the xy-dimer regime of our model eliminates the possibility of a spin liquid phase.

Similarly, because of the existence of triplet bonds, the region in the phase diagram where a xy-dimer is observed is the only one where there could potentially be a Haldane phase. The existence of such a phase can be numerically investigated using a string correlation function \cite{nijs1989,tasaki1991,watanabe1993,nishiyama1995,white1996,kim2000}. We consider the one employed by Furukawa et. al. \cite{furukawa2012}:

\begin{align}
    O^z_{\text{str}} (l,l+2r) = & - \Bigg \langle (S^z_l + S^z_{l+1}) \ \text{exp} \Bigg( i\pi \sum_{m=l+2}^{l+2r-1} S^z_m \Bigg) \nonumber\\ 
    & \;\;\;\;\;\; \times \ (S^z_{l+2r} + S^z_{l+2r+1}) \Bigg \rangle
\end{align}

To explain how this correlation function is associated with a Haldane phase, we consider a pair of spins at adjacent sites $l+2j$ and $l+2j+1$. If there were such a phase, the sum of the spins $S^z_{l+2j} + S^z_{l+2j+1}$ measured along the zigzag chain would alternate between $+1$ and $-1$ with one or more $0$'s in between, thus showing a hidden antiferromagnetic order. The correlation function $O^z_{\text{str}} (l,l+2r)$ would detect this hidden order and take non-zero values as $r$ becomes large. We calculate this correlation function for all $j$ and $r$ but we do not see a pattern as explained before, and therefore we claim that we do not find a Haldane phase. And although we are unable to find one, we note that Xu et. al. \cite{xu2018} have shown the existence of such a phase in an experimentally realizable spin-1 model of bosons in a zigzag optical lattice.

\subsection{Superfluid phase}

The reason that there are only small regions of superfluid (SF) phase in our phase diagram is that we choose our parameters such that the interactions are much stronger than the hopping. Depending on the values of $\beta_1$ and $\beta_2$, there can be various regions of SF phase. The existence of this phase is confirmed by the polynomially decaying long-range correlation $\langle S^+_1 S^-_j\rangle$ ~\cite{rossini2012,pandey2015}, known as the ``superfluid correlation'', as shown in \fig{fig:gamma_120_theta_0_and_gamma_180_theta_42_SF} (see the Appendix for additional correlations). 

These two plots also show that that the two different frustrated regions AAA and FFA can feature the same phase (SF in this case). It is worthwhile to look at the values of the pairwise interactions for the left plot: $(\alpha_o, \alpha_e, \alpha_2) = (0.250, 0.250, 0.207)$. While the interactions are equally repulsive in the NN directions, the one in the NNN direction is slightly less repulsive. This means the SF phase we observe is the result of the competition between the interactions in the three directions.

%To crop a picture: trim={<left> <lower> <right> <upper>}

\begin{figure}[htb!]
\begin{subfigure}[b]{0.24\textwidth}
\centering
    \includegraphics[trim={0cm 0cm 0.35cm 0cm},clip,width=\linewidth]{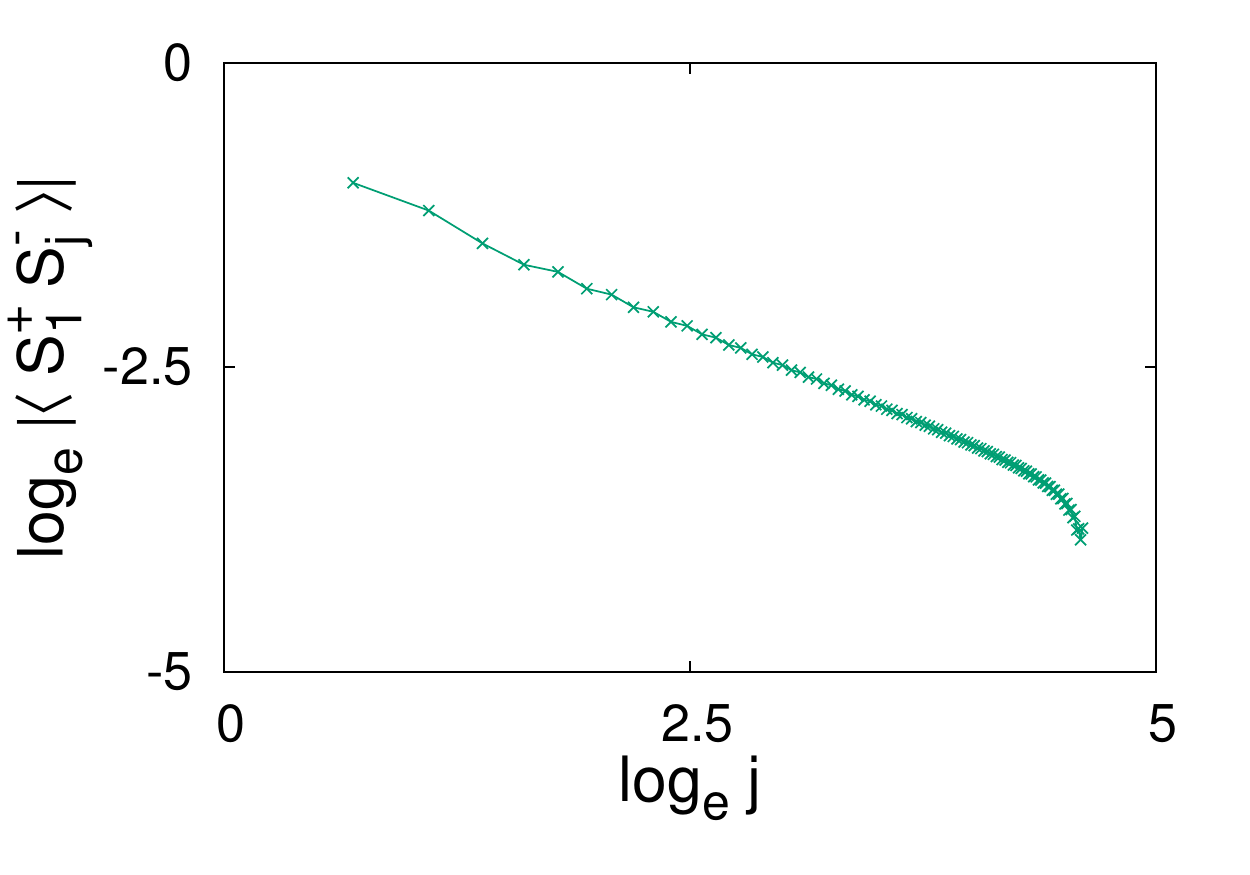}
\caption{$(\gamma, \theta) = (2\pi/3, 0)$. \\Region: AAA.}
\label{label-random_v1_gamma_120_theta_0_loge_Sp_1_Sm_j_vs_loge_j}
\end{subfigure}\hfill
\begin{subfigure}[b]{0.24\textwidth}
\centering
    \includegraphics[trim={0cm 0cm 0.35cm 0cm},clip,width=\linewidth]{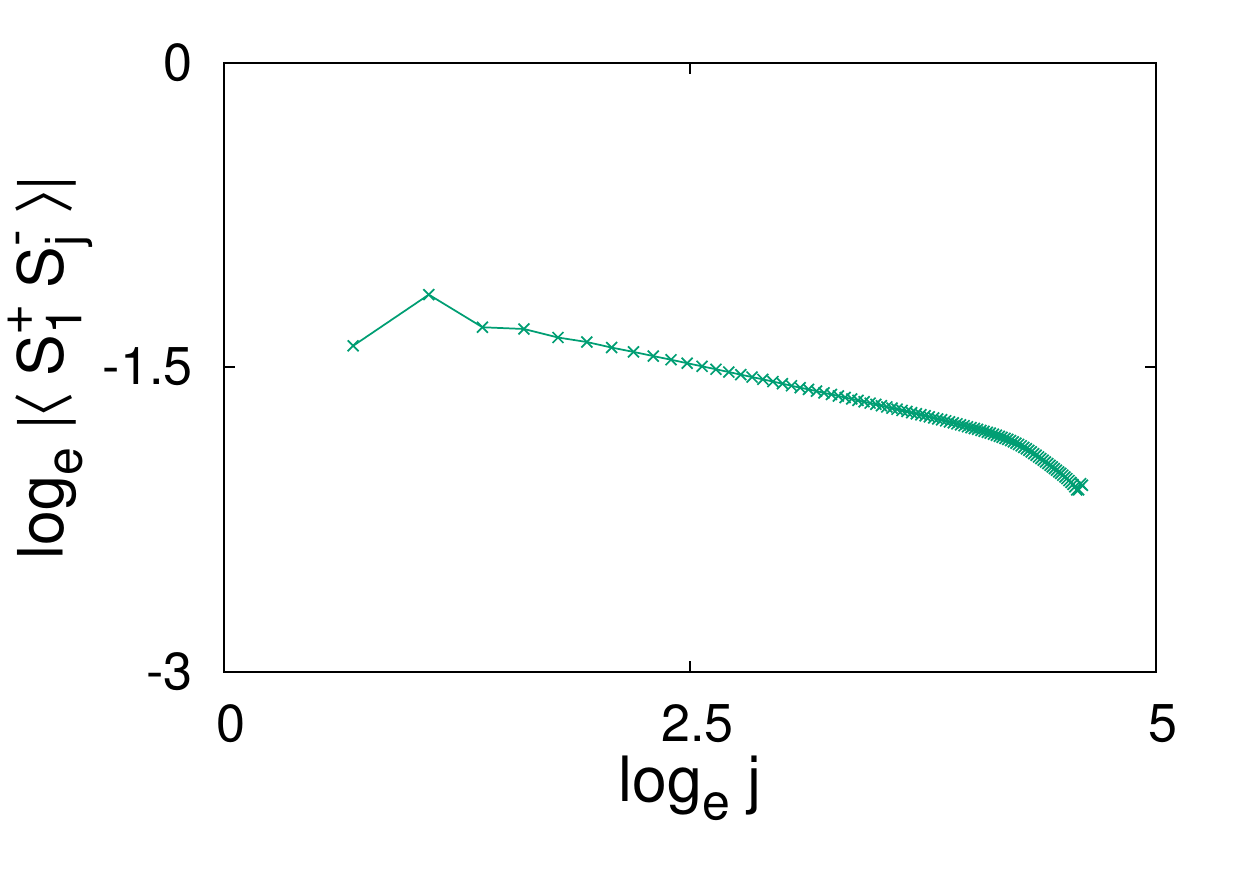}
\caption{$(\gamma, \theta) = (\pi, 0.2333 \pi)$. \\Region: FFA.}
\label{label-random_v1_gamma_180_theta_42_loge_Sp_1_Sm_j_vs_loge_j}
\end{subfigure}
\caption{(Color online) SF phase. $\ket{\text{init}} = \ket{\text{random}}$. The two plots show the polynomially decaying superfluid correlation; the non-polynomial decay near the open ends of the chain is due to the edge effect. }
\label{fig:gamma_120_theta_0_and_gamma_180_theta_42_SF}
\end{figure}

\subsection{Ferromagnetic phase}

\begin{figure}[htb!]
\begin{subfigure}[b]{0.24\textwidth}
\centering
    \includegraphics[trim={0cm 0cm 0.35cm 0cm},clip,width=\linewidth]{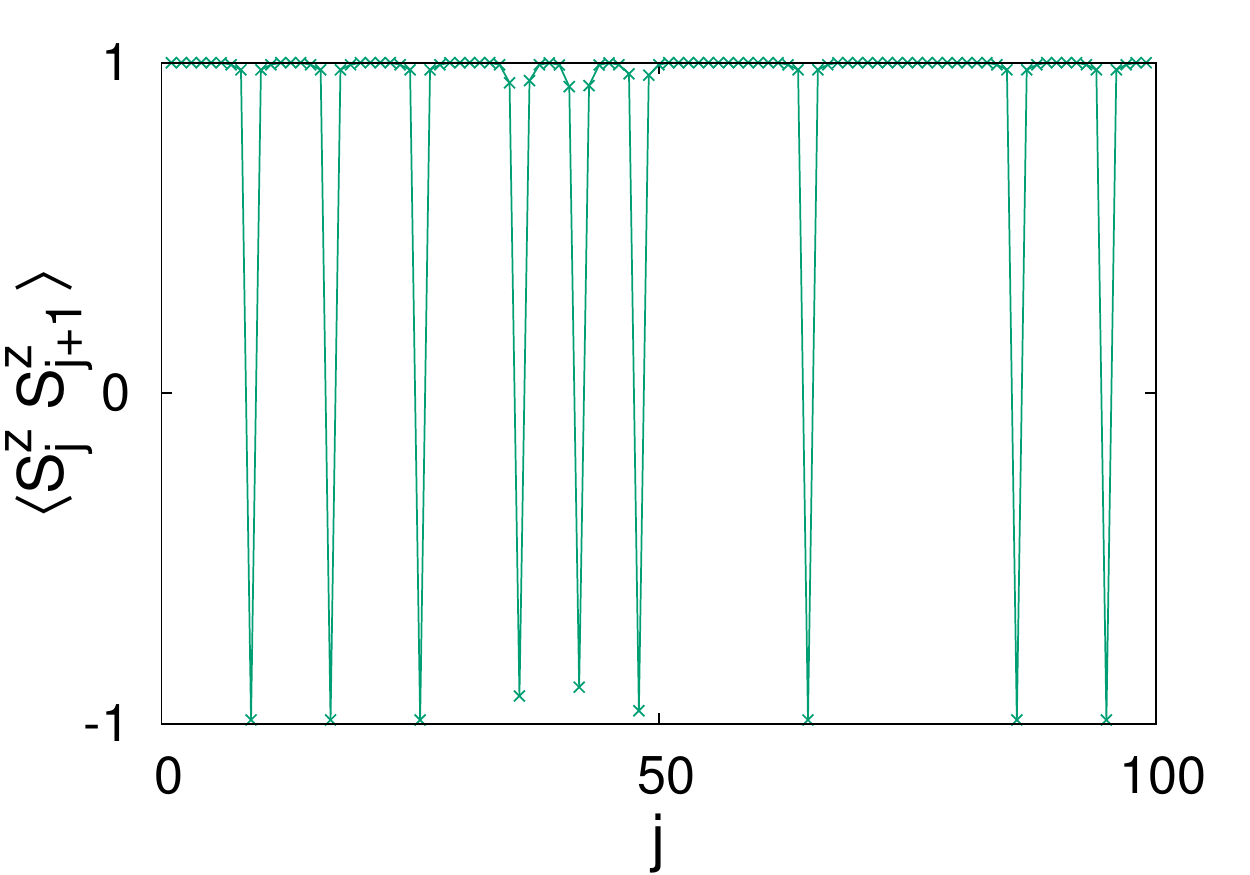}
\caption{$\ket{\text{init}} = \ket{\text{random}}$. $E = -178.26$.} 
\label{label-random_v1_gamma_180_theta_90}
\end{subfigure}\hfill
\begin{subfigure}[b]{0.24\textwidth}
\centering
    \includegraphics[trim={0cm 0cm 0.35cm 0cm},clip,width=\linewidth]{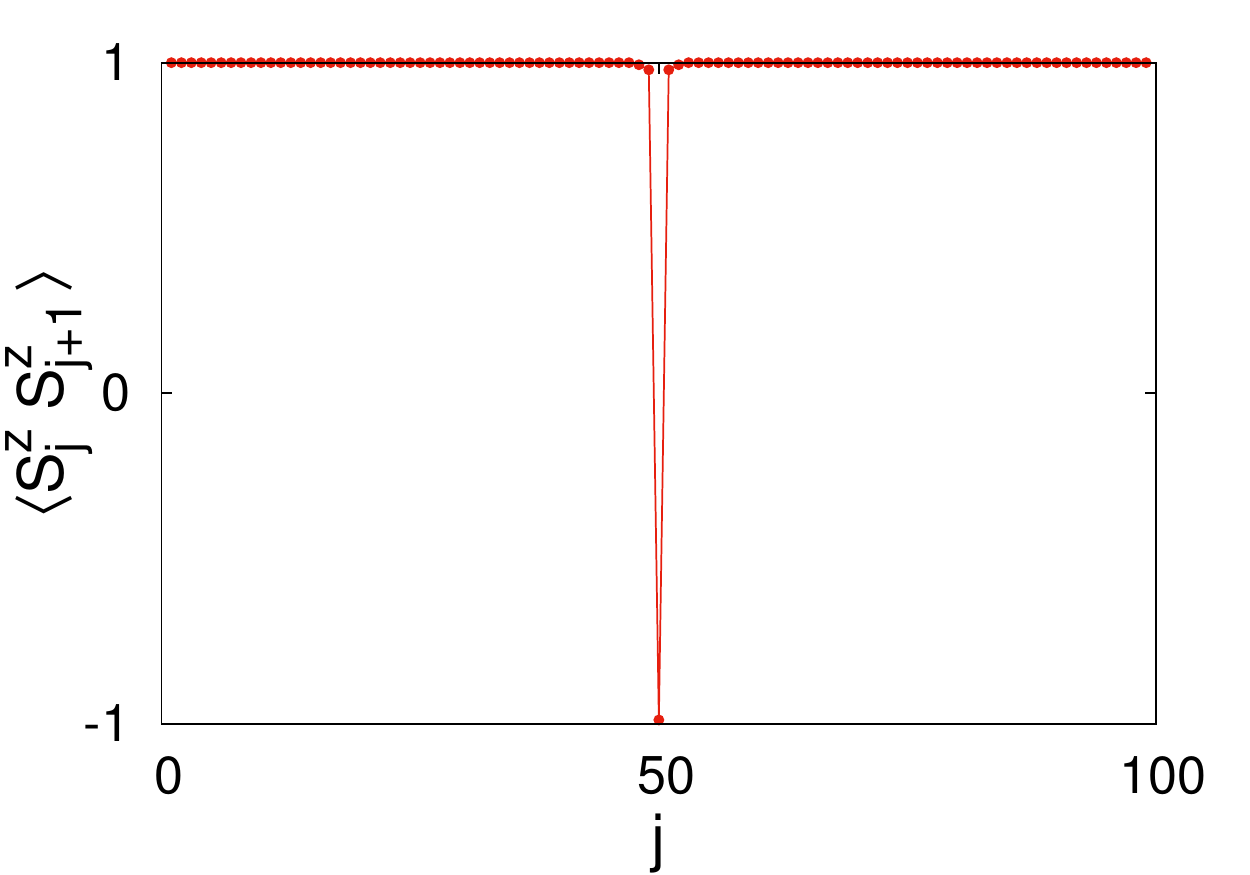}
\caption{$\ket{\text{init}} = \ket{\ldots \downarrow\downarrow\downarrow\uparrow\uparrow\uparrow \ldots}$. $E = -217.58$.}
\label{label-FM_with_single_domain_wall_gamma_180_theta_90}
\end{subfigure}
\caption{(Color online) FM phase. $(\gamma, \theta) = (\pi, \pi/2)$. Region: FFF. Since all the interactions are attractive at this point, the FM phase is expected unless the hopping dominates over the interactions. A single domain wall FM phase is the lowest energy state in this regime and the only way we can obtain this phase is by choosing itself as the initial condition. A simulation with any other initial state, although only the one with random initial state is shown here, results in a FM phase with several domain walls.}
\label{fig:gamma_180_theta_90_FM}
\end{figure}

\fig{fig:gamma_180_theta_90_FM} shows the ferromagnetic (FM) phase in this system. We show results subject to two different initial conditions in order to highlight the nature of the phase returned by DMRG. When the system is in the FM regime, the FM state with a single domain wall is the true ground state because it has the lowest energy as compared to the states produced with any other initial conditions.

%To crop a picture: trim={<left> <lower> <right> <upper>}

The dashed line on the phase diagram which is labeled as ``$\alpha_o + \alpha_e = -1/2$'' represents the points where $\alpha_o$ and $\alpha_e$ are equally far away from their critical values $\alpha_{o,c} = \alpha_{e,c} = -1/4$, one being attractive while the other repulsive. So one would expect a FM phase on one side of this line and an AFM phase on the other. Our results, however, show that the attractive interaction in the odd (or even) direction of the spin chain dominates over the repulsive interaction in the even (or odd) direction to a certain threshold, thus resulting in a FM phase on both sides of this line. It should be noted that this line disappears when $\gamma \rightarrow 0.4467\pi$ because above this value of $\gamma$, the system would be deep in the FM regime and therefore, we do not obtain an AFM phase regardless of the value of $\theta$.

\subsection{Antiferromagnetic phase}

\begin{figure}[htb!]
\begin{subfigure}[b]{0.24\textwidth}
\centering
    \includegraphics[trim={0cm 0cm 0.35cm 0cm},clip,width=\linewidth]{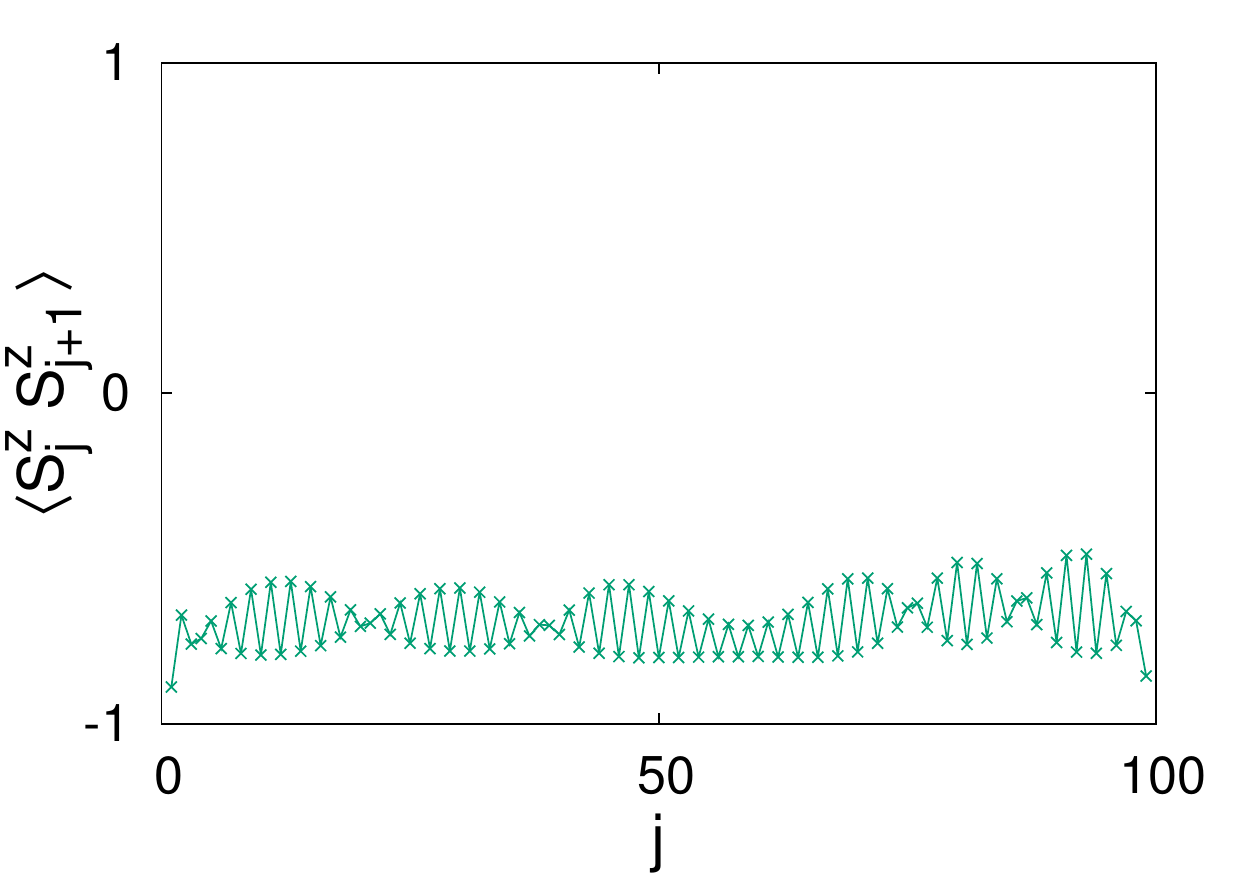}
\caption{$\ket{\text{init}} = \ket{\text{random}}$. $E = -96.13$.} 
\label{label-random_v1_and_v2_gamma_180_theta_0}
\end{subfigure}\hfill
\begin{subfigure}[b]{0.24\textwidth}
\centering
    \includegraphics[trim={0cm 0cm 0.35cm 0cm},clip,width=\linewidth]{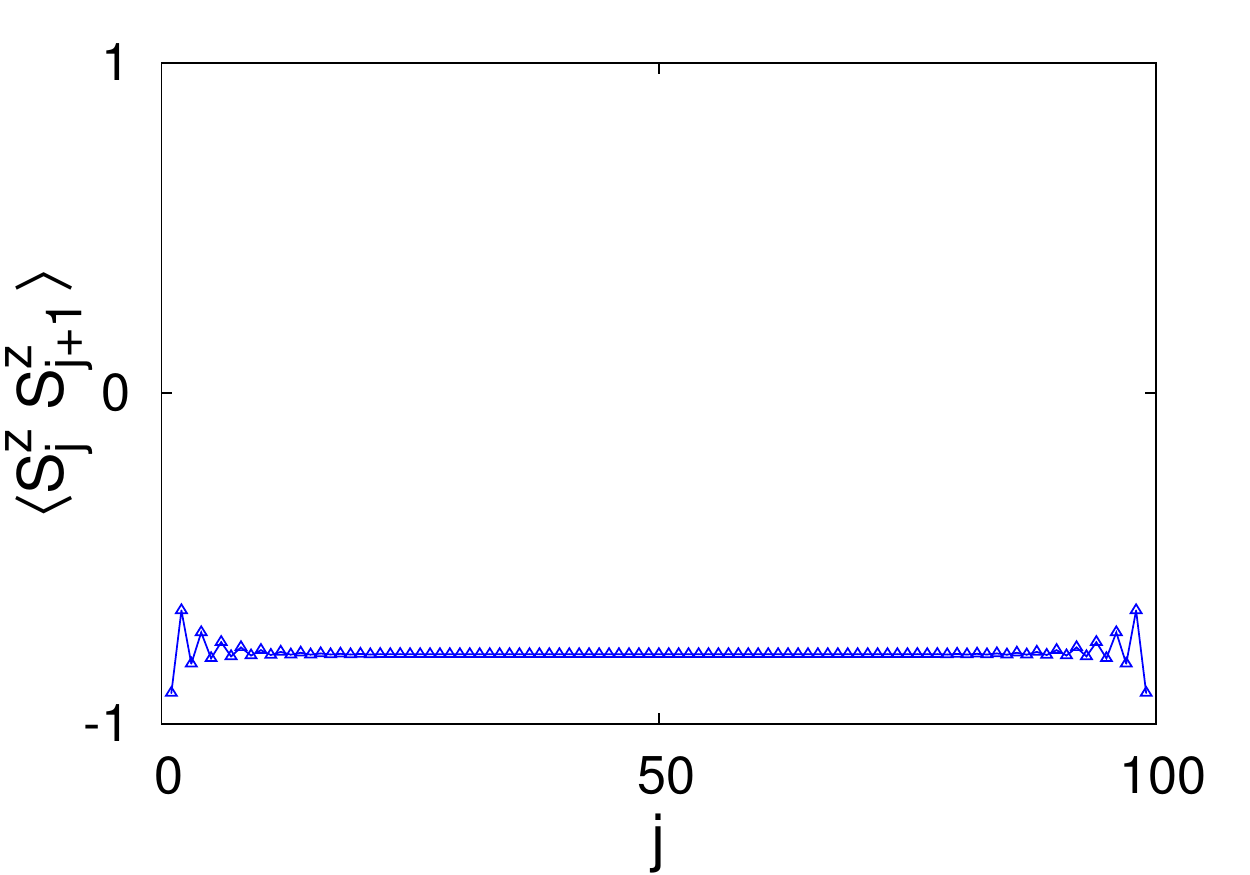}
\caption{$\ket{\text{init}} = \ket{\downarrow\uparrow\downarrow\uparrow\downarrow\uparrow \ldots}$. $E = -98.60$.}
\label{label-neel_gamma_180_theta_0}
\end{subfigure}
\caption{(Color online) AFM1 phase. $(\gamma, \theta) = (\pi, 0)$. Region: AAA. Although both the plots show an AFM phase, the one on the right is a better approximation to the true phase because it has a lower energy.}
\label{fig:gamma_180_theta_0_AFM}
\end{figure}

\begin{figure}[htb!]
\begin{subfigure}[b]{0.24\textwidth}
\centering
    \includegraphics[trim={0cm 0cm 0.35cm 0cm},clip,width=\linewidth]{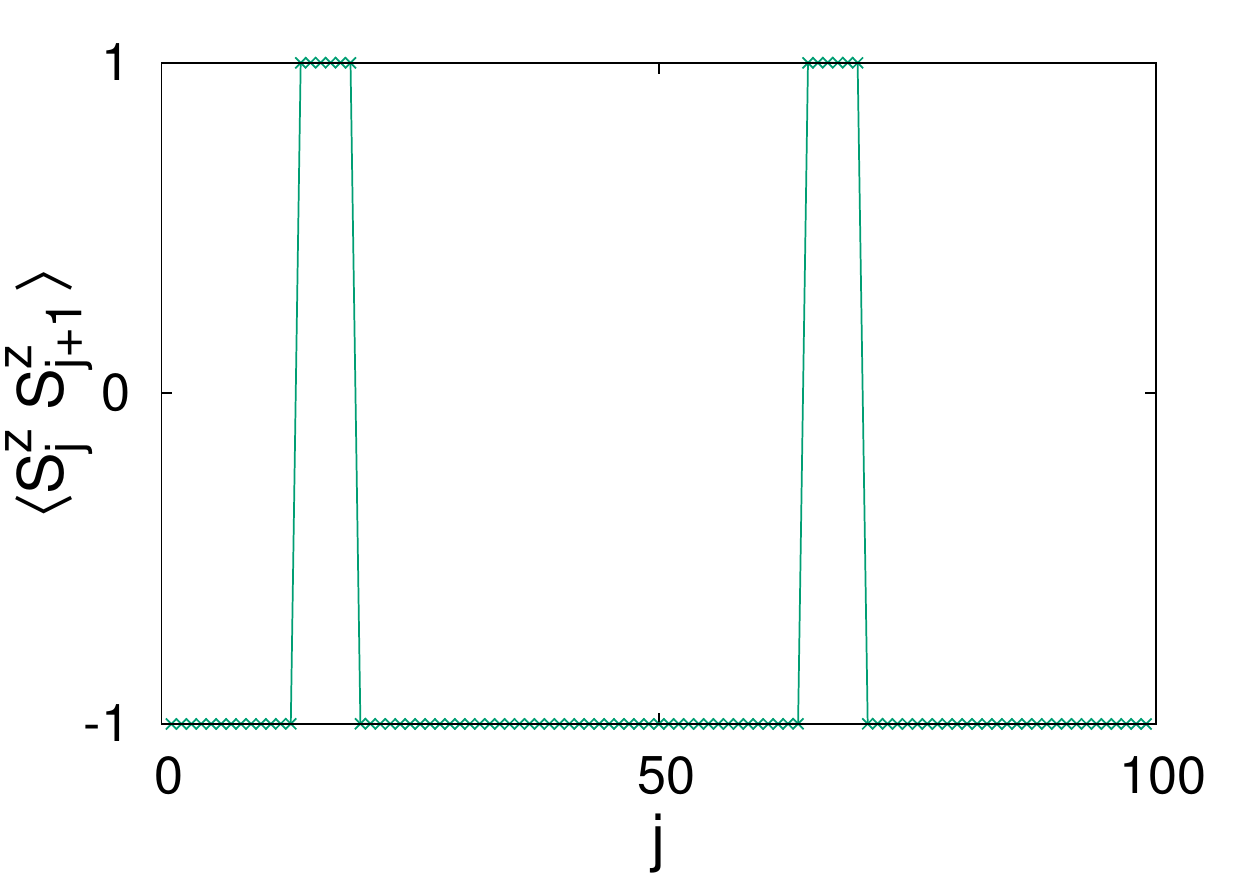}
\caption{$\ket{\text{init}} = \ket{\text{random}}$. $E = -1357.91$.} 
\label{label-random_gamma_30_theta_90}
\end{subfigure}\hfill
\begin{subfigure}[b]{0.24\textwidth}
\centering
    \includegraphics[trim={0cm 0cm 0.35cm 0cm},clip,width=\linewidth]{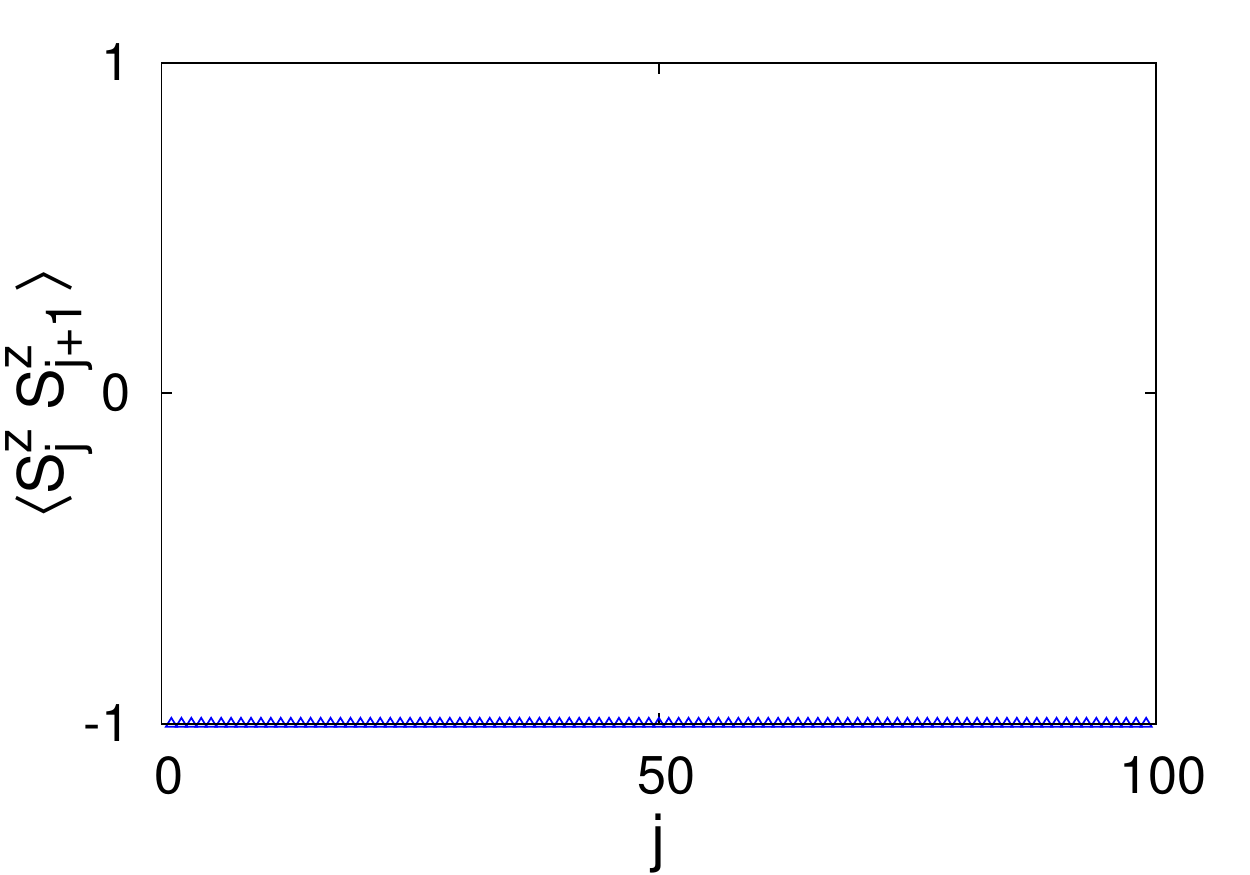}
\caption{$\ket{\text{init}} = \ket{\downarrow\uparrow\downarrow\uparrow\downarrow\uparrow \ldots}$. $E = -1492.43$.} 
\label{label-neel_gamma_30_theta_90}
\end{subfigure}
\caption{(Color online) AFM2 phase. $(\gamma, \theta) = (\pi/6, \pi/2)$. Region: AAF. The left plot shows a phase with mostly AFM correlations except for a couple of trapped regions while the right plot shows a pure AFM phase which is the true phase because it has a much lower energy.}
\label{fig:gamma_30_theta_90_AFM}
\end{figure}

In \fig{fig:gamma_180_theta_0_AFM} and \fig{fig:gamma_30_theta_90_AFM}, it can be seen that the accuracy of DMRG depends on the choice of initial state. There are obviously two different AFM regimes. We label the phase as ``AFM1'' when the NN correlations $\langle S^z_j S^z_{j+1}\rangle$ are negative but greater than -1 for each site index $j$ as shown in \fig{fig:gamma_180_theta_0_AFM}. A look at the values of the long-range correlation $\langle S^z_1 S^z_j \rangle$ (see the Appendix) confirms that this is an AFM phase. Simlarly, we label the phase as ``AFM2'' when the system is deep in the AFM regime so that $\langle S^z_j S^z_{j+1}\rangle \approx -1$. It is worth noting that although a pure AFM phase is expected in the non-frustrated region AAF, a simulation with a random initial state results in a phase that has mostly AFM correlations but with one or more clusters of identical spins, which we call ``trapped regions''. It is clearly not a true phase but still makes sense from an experimental point of view, which we will explain later.

%To crop a picture: trim={<left> <lower> <right> <upper>}

%To crop a picture: trim={<left> <lower> <right> <upper>}

\subsection{Phase transitions and DMRG}

%To crop a picture: trim={<left> <lower> <right> <upper>}

\begin{figure}[htb!]
	\includegraphics[trim={0cm 0cm 0cm 0cm},clip,width=0.45\textwidth, scale=1.0]{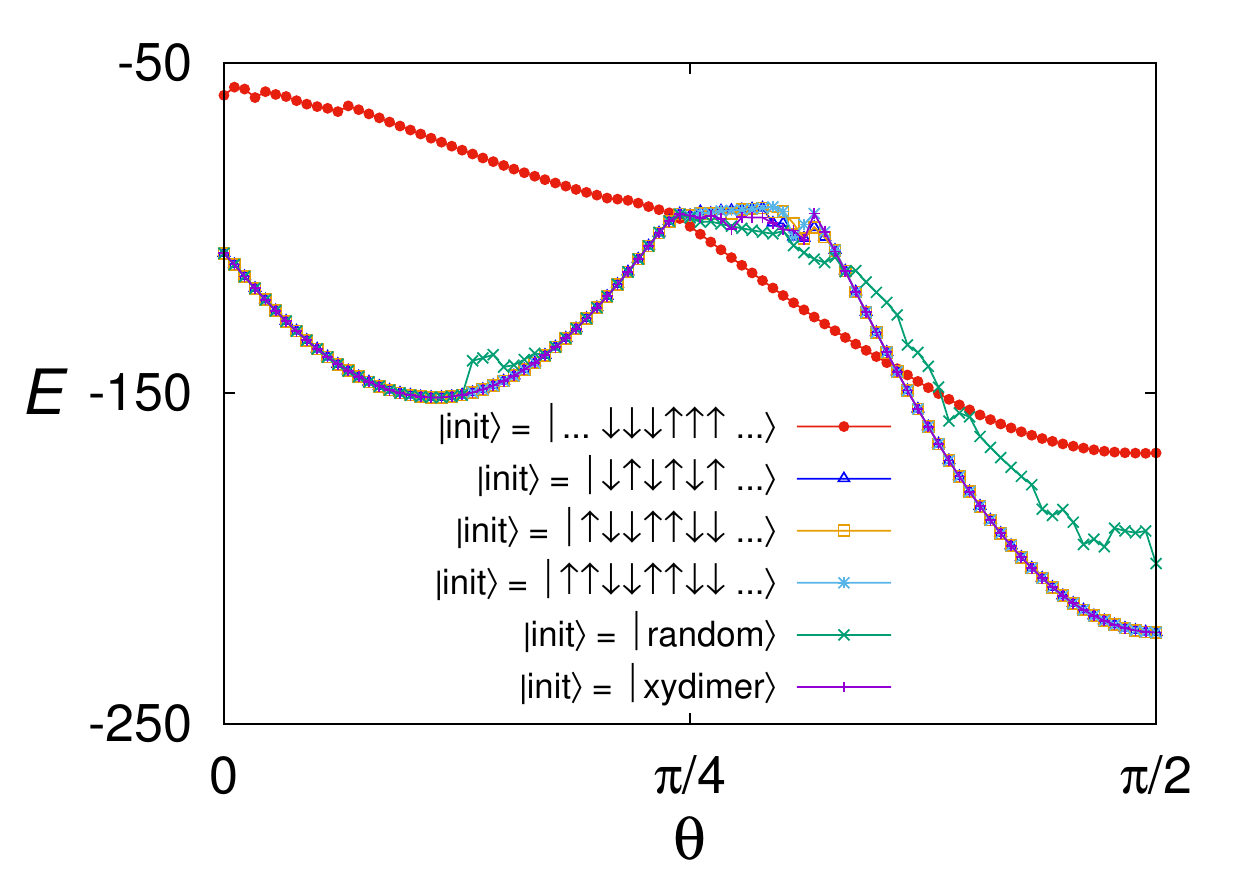}
	\caption{(Color online) Ground state energy of the system, $E$, plotted as a function of polarization angle, $\theta$, for $\gamma = \pi/3$. The state $\ket{\text{init}}$ has been used to denote the ``initial state'' for a DMRG simulation, $\ket{\text{random}}$ denotes the ``random initial state'' and $\ket{\text{xydimer}}$ denotes the triplet bound state $\ket{+} \otimes \ldots \otimes \ket{+}$. This figure clearly shows that in a regime where a FM phase is expected, only a simulation with a FM initial state results in a true ground state. It also shows that several curves meet at two points: $\theta = 0.2424 \pi$, which belongs to a smooth crossover between z-dimer and FM phases (see \fig{fig:Szj_Szjp1_gamma_60_theta_64point77}), and $\theta = 0.3598 \pi$, which lies at a sharp crossover between FM and AFM phases (see \fig{fig:Szj_Szjp1_gamma_60_theta_43point64}).}
	\label{fig:ground_state_energy_vs_theta_gamma_60}
\end{figure}

\fig{fig:ground_state_energy_vs_theta_gamma_60} shows how initial states affect the ground state energy in DMRG simulations and why it is important to perform multiple trials with various initial conditions. If we look at these results with reference to the phase diagram (\fig{fig:phase_diagram}), we can see that in the regime where the ground state is expected to be dimerized or AFM, the best choice for the initial state would be a z-dimer, a Neel state or a xy-dimer because these three states result in exactly the same ground state. Similarly, in the regime where the ground state is expected to be FM, a simulation must start with a single domain wall FM state. 

Simulations with various initial conditions clearly show that there is a sharp transition between FM and AFM phases, and a smooth transition between z-dimer and FM phases and between SF and other phases (see the Appendix for detailed explanation of transition between SF and AFM phases). Experiments, however, can be expected to confirm the unclear DMRG results in the following way: Suppose we build a system from a sample of randomly distributed spins and slowly cool it down so that the spins restribute in the lattice to minimize their energy. If the sample consists of one or more trapped regions, the system must overcome an enormous energy hurdle to flip the spins in these regions, therefore the spin configuration would be expected to show signatures of these trapped regions (as we saw earlier in \fig{fig:gamma_30_theta_90_AFM}) although it is not the lowest energy configuration. 

\begin{figure}[htb!]
\begin{subfigure}[b]{0.24\textwidth}
\centering
    \includegraphics[trim={0cm 0cm 0.35cm 0cm},clip,width=\linewidth]{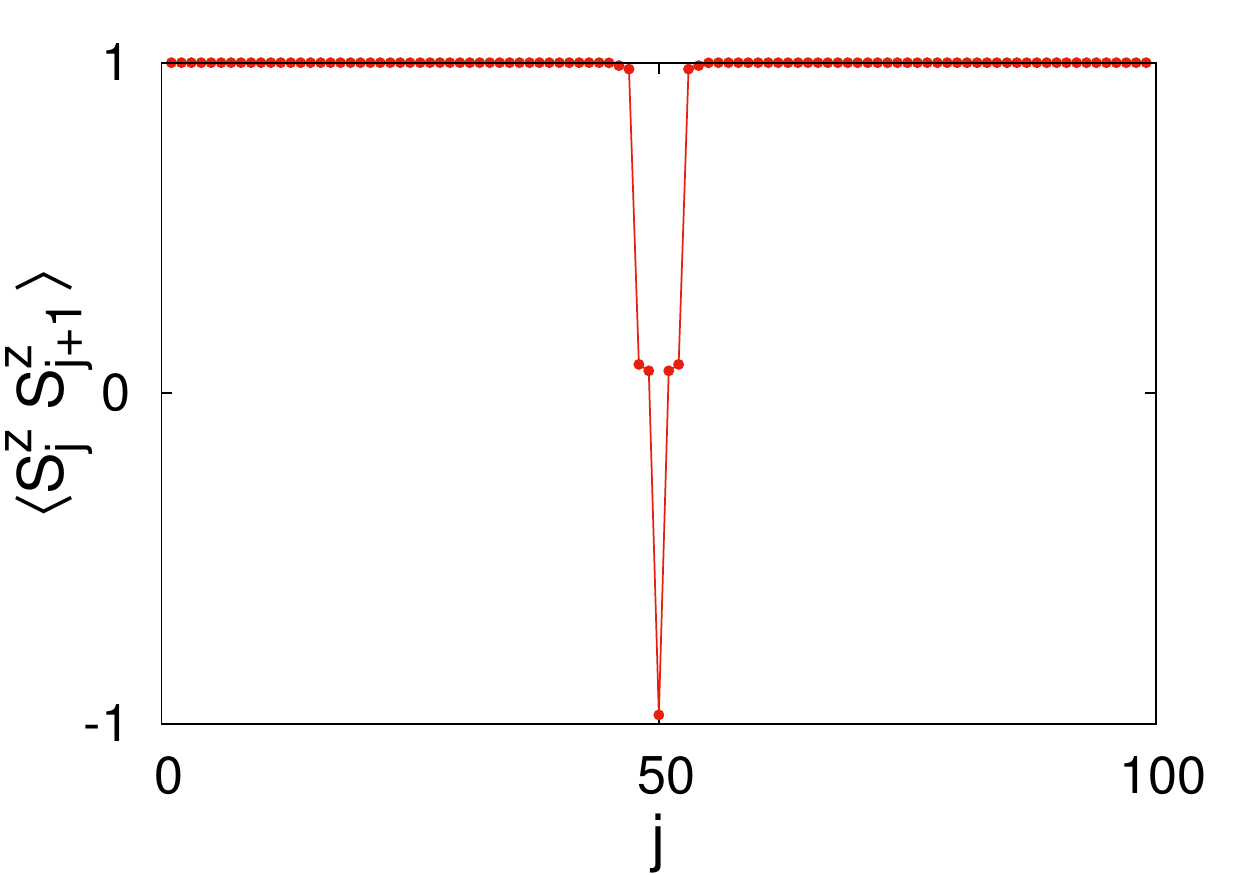}
\caption{$\ket{\text{init}} = \ket{\ldots \downarrow\downarrow\downarrow\uparrow\uparrow\uparrow \ldots}$. $E = -142.14$.}
%\label{label-random_v1_and_v2_gamma_60_theta_30}
\end{subfigure}\hfill
\begin{subfigure}[b]{0.24\textwidth}
\centering
    \includegraphics[trim={0cm 0cm 0.35cm 0cm},clip,width=\linewidth]{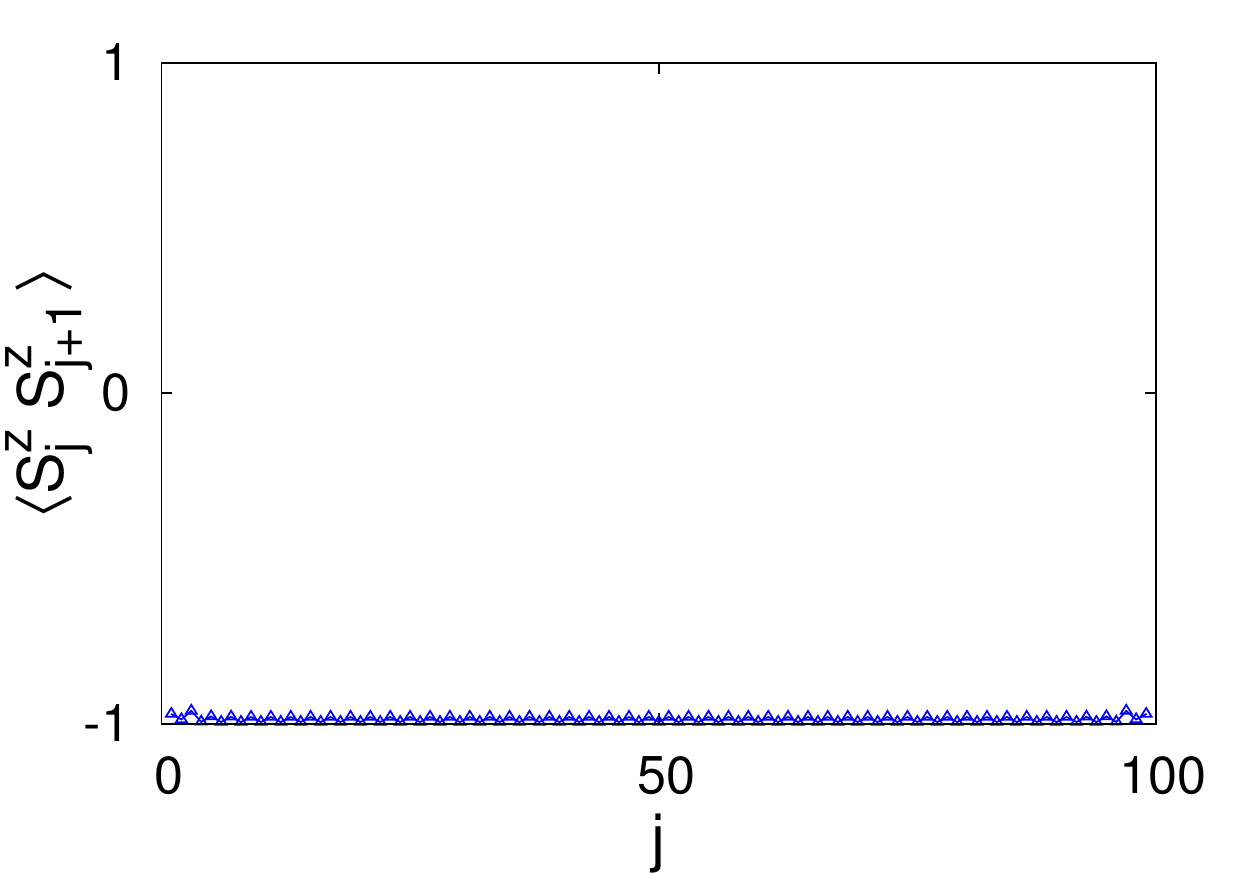}
\caption{$\ket{\text{init}} = \ket{\downarrow\uparrow\downarrow\uparrow\downarrow\uparrow \ldots}$. $E = -142.13$.}
%\label{label-random_v1_and_v2_gamma_60_theta_15}
\end{subfigure}
\caption{(Color online) Ground states and their energies subject to two initial conditions. $(\gamma, \theta) = (\pi/3,0.3598 \pi)$. Region: FAF. }
\label{fig:Szj_Szjp1_gamma_60_theta_64point77}
\end{figure}

\begin{figure}[htb!]
\begin{subfigure}[b]{0.24\textwidth}
\centering
    \includegraphics[trim={0cm 0cm 0.35cm 0cm},clip,width=\linewidth]{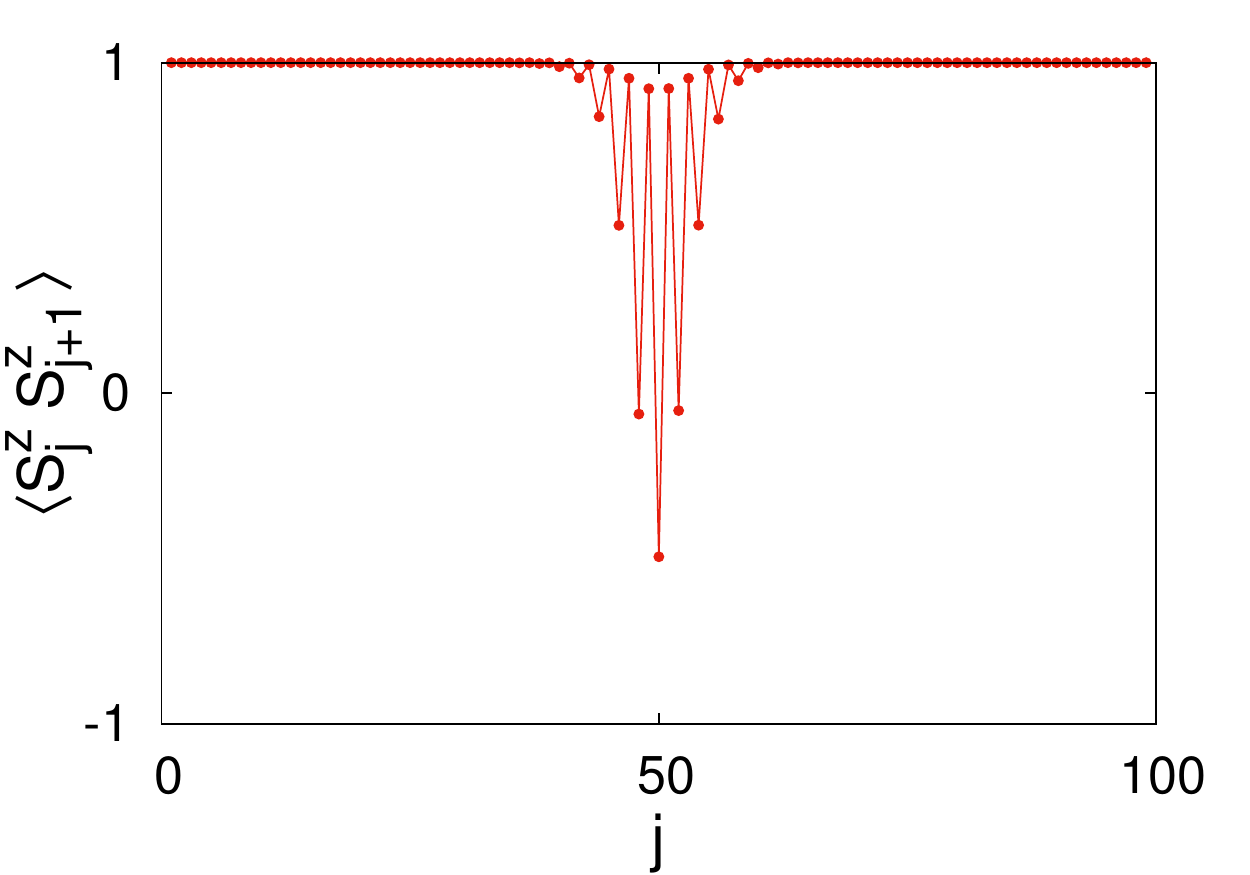}
\caption{$\ket{\text{init}} = \ket{\ldots \downarrow\downarrow\downarrow\uparrow\uparrow\uparrow \ldots}$. $E = -96.31$.}
%\label{label-random_v1_and_v2_gamma_60_theta_30}
\end{subfigure}\hfill
\begin{subfigure}[b]{0.24\textwidth}
\centering
    \includegraphics[trim={0cm 0cm 0.35cm 0cm},clip,width=\linewidth]{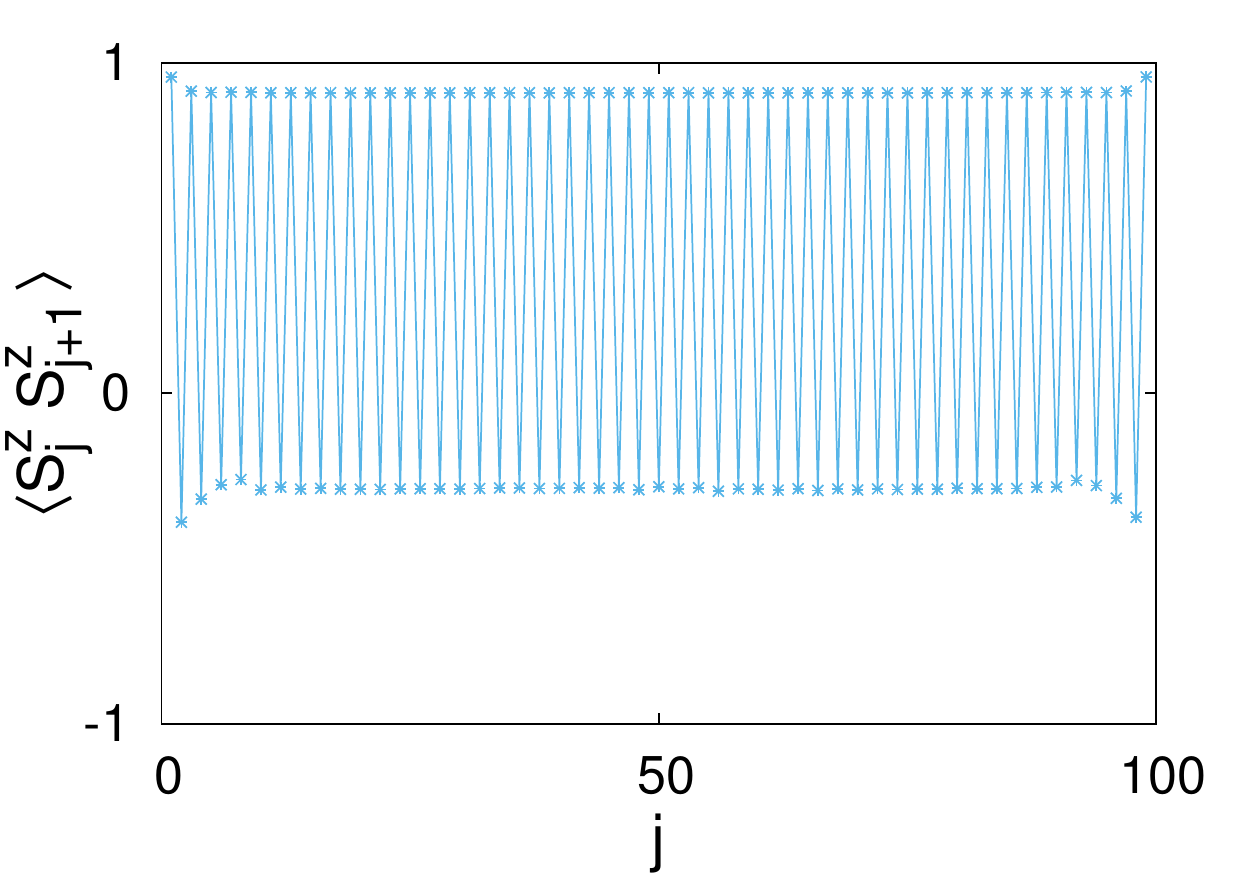}
\caption{$\ket{\text{init}} = \ket{\uparrow\uparrow\downarrow\downarrow\uparrow\uparrow\downarrow\downarrow \ldots}$. $E = -96.30$.}
%\label{label-random_v1_and_v2_gamma_60_theta_15}
\end{subfigure}
\caption{(Color online) Ground states and their energies subject to two initial conditions. $(\gamma, \theta) = (\pi/3,0.2424 \pi)$. Region: FAF.}
\label{fig:Szj_Szjp1_gamma_60_theta_43point64}
\end{figure}

\subsection{Order parameters}

We define the order parameters for ferromagnetic, antiferromagnetic, z-dimer and xy-dimer phases as follows:

\begin{align}
    O_{\rm ferro} & = \frac{4}{N} \sum_{i= \frac{N}{4}+1}^{\frac{N}{2}} \Bigg| \sum_{j = \frac{3N}{4}}^{\frac{3N}{4} + 3} \langle S^z_i S^z_j \rangle \Bigg|
\end{align}

\begin{align}
    O_{\rm neel} & = \frac{4}{N} \sum_{i= \frac{N}{4}+1}^{\frac{N}{2}} \Bigg| \sum_{j = \frac{3N}{4}}^{\frac{3N}{4} + 3} (-1)^j \langle S^z_i S^z_j \rangle \Bigg|
\end{align}

\begin{align}
    O_{\rm zdimer} & = \frac{2}{N} \sum_{i= \frac{N}{4} + 1}^{\frac{3N}{4}} \Big|\langle S^z_i S^z_{i+1} - S^z_{i+1} S^z_{i+2} \rangle \Big|
\end{align}

\begin{align}
    %In my notes, O_xydimer was defined like this initially (I need to double check this): O_{xydimer} & = \Bigg|\sum_{i = N/4}^{N/4 + 3} (-1)^i \Big|\langle S^z_i S^z_{i+1} \rangle \Big| \Bigg|
	O_{\rm xydimer} & = \Bigg|\sum_{i = \frac{N}{4} + 1}^{\frac{N}{4} + 4} (-1)^i \Big|\langle S^z_i S^z_{i+1} \rangle \Big| \Bigg|
\end{align}

\begin{figure}[htb!]
\begin{subfigure}[b]{0.4\textwidth}
	\centering
	\includegraphics[trim={0cm 0cm 0cm 0cm},clip,width=\linewidth]{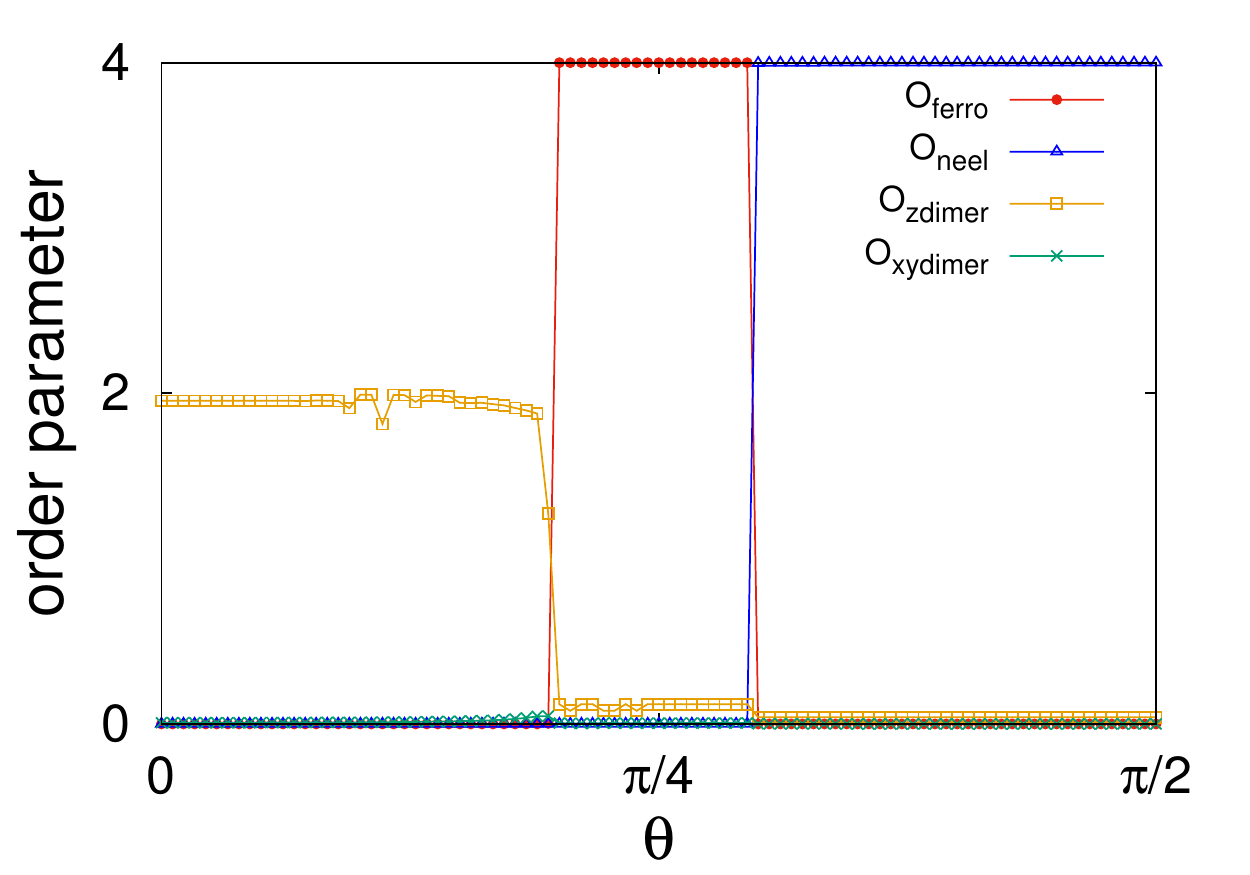}
%\caption{Energy gap as a function of polarization angle for different system sizes}
	\caption{$\gamma = \pi/6$}
	\label{label-FM_NEEL_xydimer_zdimer_order_parameter_vs_theta_gamma_30}
\end{subfigure}\hfill
\begin{subfigure}[b]{0.4\textwidth}
	\centering
	\includegraphics[trim={0cm 0cm 0cm 0cm},clip, width=\linewidth]{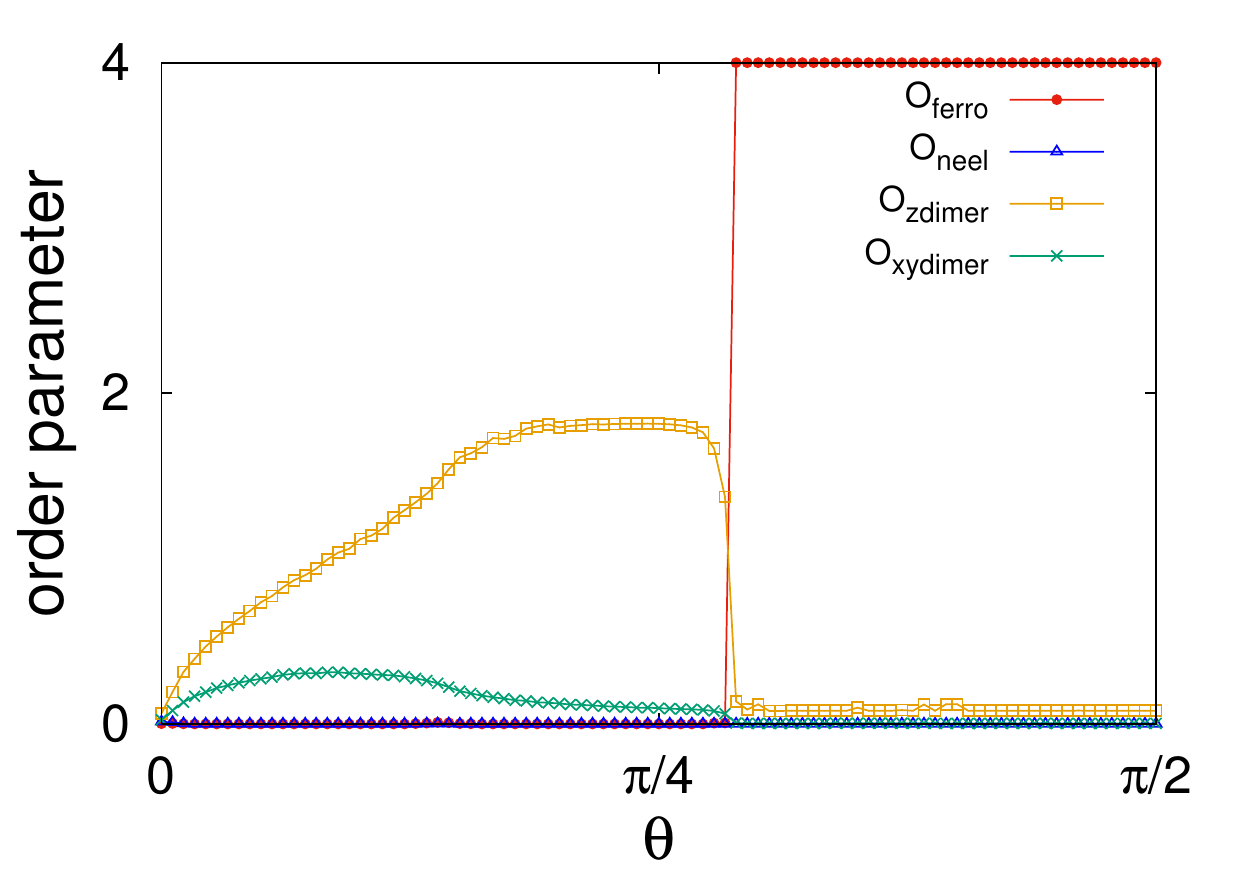}
%\caption{Energy gap as a function of system size near a phase transition point}
	\caption{$\gamma = 7\pi/9$}
	\label{label-FM_NEEL_xydimer_zdimer_order_parameter_vs_theta_gamma_140}
\end{subfigure}
\caption{Order parameter for various phases as a function of polarization angle $\theta$}
\label{fig:FM_NEEL_xydimer_zdimer_order_parameter_vs_theta_gamma_30_and_140}
\end{figure}

\noindent
Although we use correlation functions to explain how we identify each phase, we use order parameters to find how deep the system is in a given phase and also to find the crossover between the phases. For the dimerized phases, we use the definitions given by Furukawa et. al.\cite{furukawa2012} To minimize the edge effects due to open boundaries, we use the method employed by Rossini et. al.\cite{rossini2012} $-$ we define the order parameters for ferromagnetic, antiferromagnetic and z-dimer phases as average expectation values of the correlators between spins in the middle part of the chain. For the xy-dimer phase, however, we only consider the the correlations $N/4$ sites away from the left end of the chain but not their average expectation values. For the superfluid phase, we use the values of the superfluid correlation function $\langle S^+_1 S^-_j\rangle$ which, as mentioned earlier, decays polynomially in this phase. It should be noted that we have defined the order parameters such that they are always non-negative.

\fig{fig:FM_NEEL_xydimer_zdimer_order_parameter_vs_theta_gamma_30_and_140} shows how the order parameters for different phases vary with polarization angle $\theta$ for a given value of $\gamma$. By definition, the order parameter for a given phase should vanish in all other phases and our results for FM and AFM phases are consistent with this. However, the dimerized phases consist of two flavors, xy and z, which pair neighboring spins in different directions. Therefore, their order parameters overlap. The finite size scaling, which we will discuss later, along with the values of correlation functions allows us to find the boundary between these two phases.

\subsection{Finite-size scaling and extrapolation}

\begin{figure}[htb!]
\begin{subfigure}[b]{0.4\textwidth}
	\centering
	\includegraphics[trim={0cm 0cm 0cm 0cm},clip,width=\linewidth]{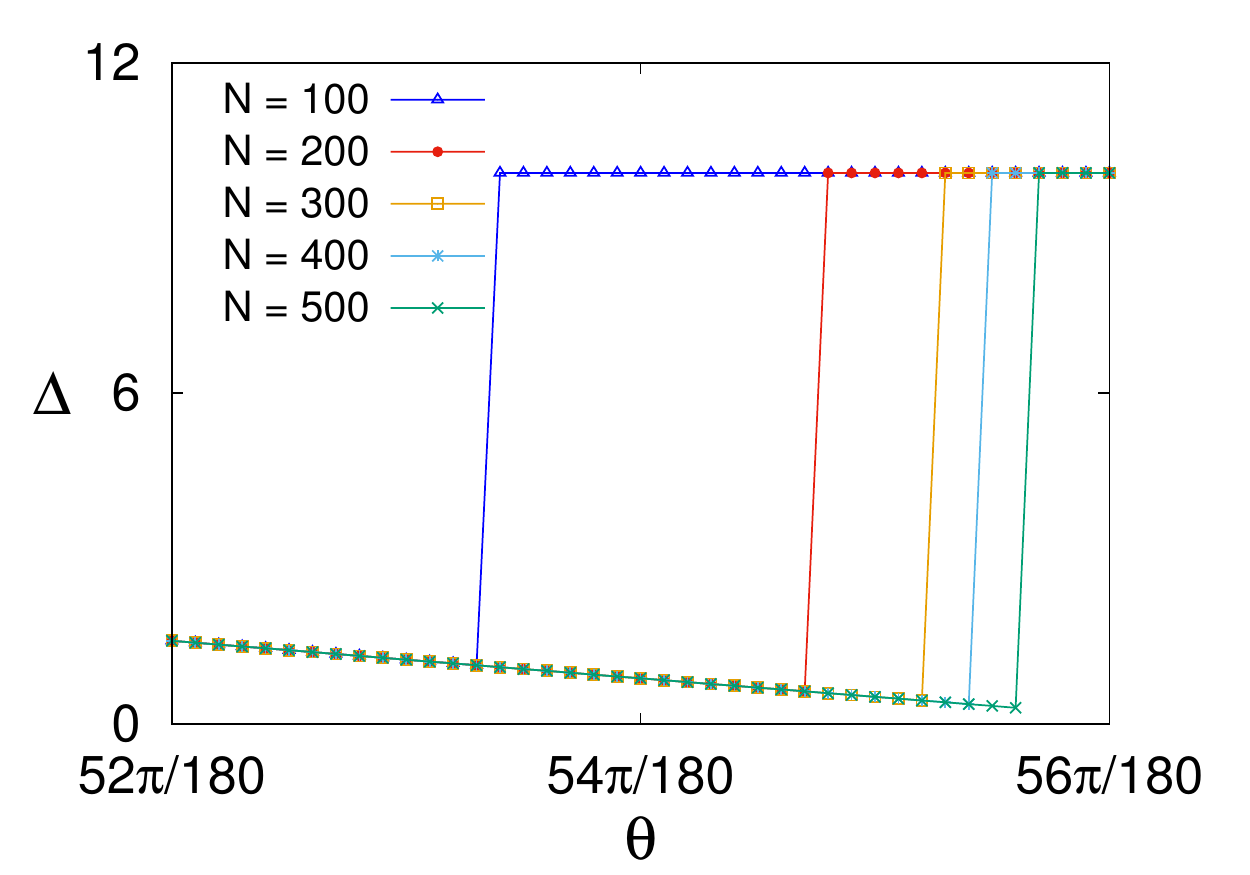}
%\caption{Energy gap as a function of polarization angle for different system sizes}
	\caption{}
	\label{label-energy_gap_vs_theta_gamma_30}
\end{subfigure}\hfill
\begin{subfigure}[b]{0.4\textwidth}
	\centering
	\includegraphics[trim={0cm 0cm 0cm 0cm},clip, width=\linewidth]{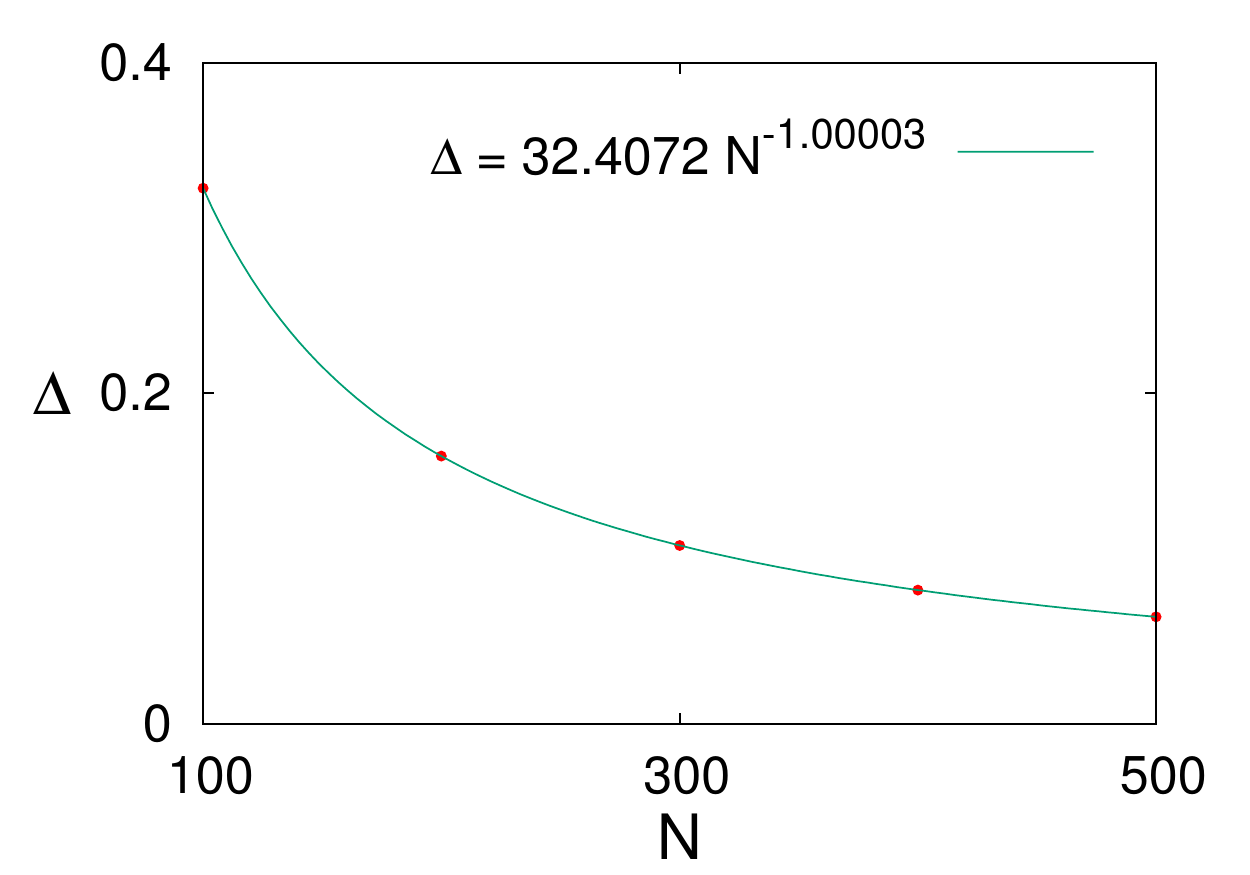}
%\caption{Energy gap as a function of system size near a phase transition point}
	\caption{}
	\label{label-energy_gap_vs_N_gamma_30_theta_57}
\end{subfigure}
\begin{subfigure}[b]{0.4\textwidth}
	\centering
	\includegraphics[trim={0cm 0cm 0cm 0cm},clip,width=\linewidth]{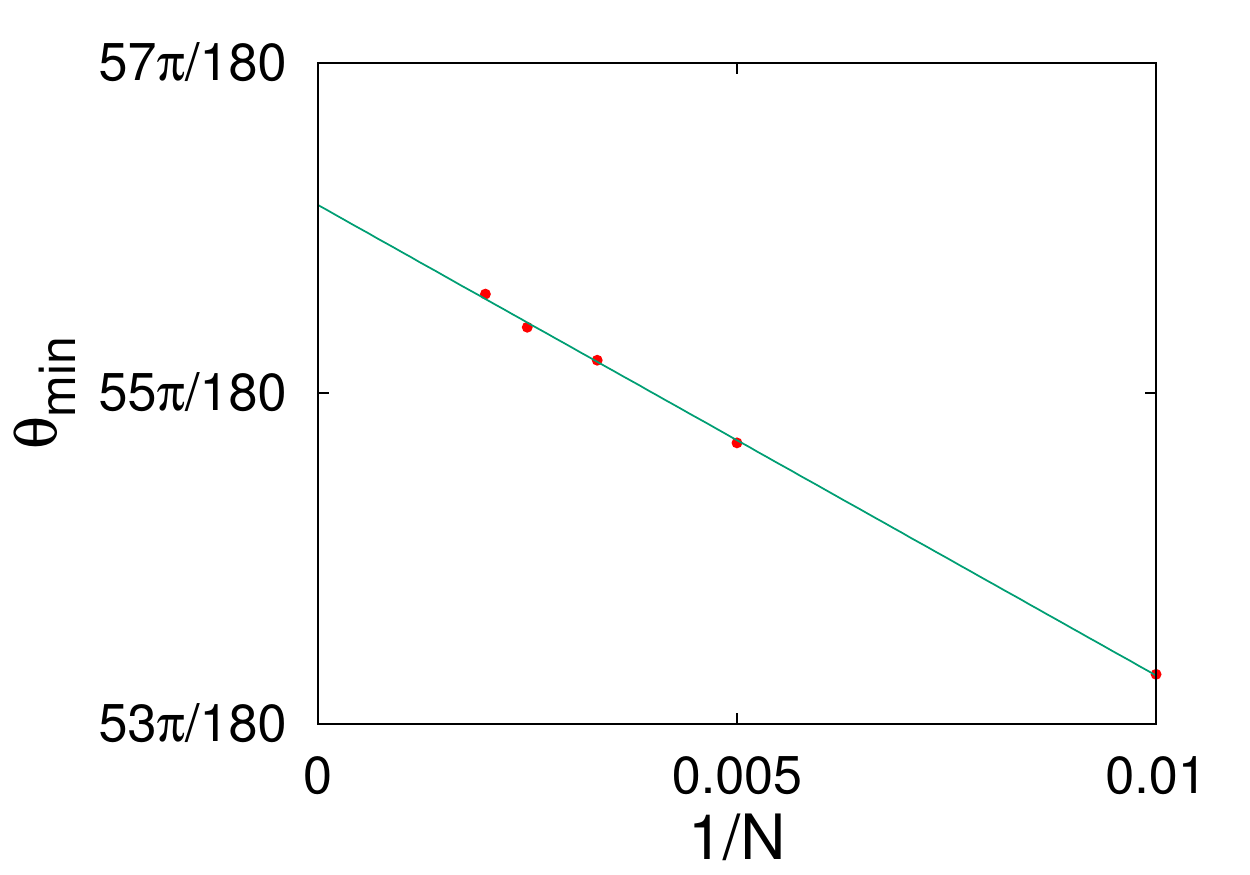}
	\caption{}
	\label{label- minimum_theta_vs_1_by_N_gamma_30}
\end{subfigure}\hfill
\caption{Finite-size scaling of the energy gap: $\gamma = \pi/6$. (a) The energy gap $\Delta$ is plotted as a function of the polarization angle $\theta$ and different system sizes $N$. The gap is minimum at the phase transition point, we denote the corresponding value of $\theta$ by $\theta_{\text{min}}$. (b) The energy gap is plotted as a function of the system size at $\theta = 0.3167 \pi$ which is near the phase transition point. The line of best fit is $\Delta = 32.4072 N^{-1.00003}$, which implies that the energy gap scales polynomially with the system size. (c) By plotting $\theta_{\text{min}}$ against $1/N$, we extrapolate the phase transition point in the thermodynamic limit $N \rightarrow \infty$ as $\theta = (0.3119 \pm 1.2155 \times 10^{-4}) \pi$.}
%NOTE: For gamma = 30 degrees, the phase boundary in the thermodynamic limit is (56.1398 +- 0.02188) degrees. I converted this to radians and put in the caption above.
\label{fig:energy_gap_N_100_to_500_gamma_30}
\end{figure}

As mentioned earlier, the phase diagram (\fig{fig:phase_diagram}) has been drawn using the values of order parameters and correlation functions for the finite system size $N = 100$. We extrapolate the phase boundaries in the thermodynamic limit $N \rightarrow \infty$ using the finite-size scaling method explored by Rossini et. al. \cite{rossini2012} We calculate the energy gap for different system sizes $N$ and find the value of $\theta$ for which the gap is minimum for each $N$, as shown in \fig{label-energy_gap_vs_theta_gamma_30}. We call this value $\theta_{\rm min}$. We then plot these $\theta_{\rm min}$ against $1/N$ and extrapolate the value of $\theta_{\text{min}}$ when $1/N \rightarrow 0$ as shown in \fig{label- minimum_theta_vs_1_by_N_gamma_30}. Although difficult to see, the boundaries denoted by white dots in the phase diagram have small error bars that are due to uncertainty in the fitting of the curves for different values of $N$. Our analysis shows that the energy gap scales polynomially with the system size near the boundary between z-dimer and ferromagnetic phases as shown in \fig{label-energy_gap_vs_N_gamma_30_theta_57}. We are unable to find the boundary between xy-dimer and superfluid phases. 

\section{Conclusion}

%Mapping a zigzag chain of dipoles to one of spin$-1/2$ particles, we have thus explored the many-body properties of our system. 
In conclusion, we have numerically studied the ground-state properties of a quasi-one-dimensional model that contains hopping and interactions up to second neighbors. Even though this is a rather simple model, it comprises of frustrated regimes that lead to a rich phase diagram. We have used a novel approach to write the Hamiltonian that gives an intuitive understanding of the model, makes it convenient to identify frustrated and non-frustrated regimes, and helps predict the ground states beforehand so that the results obtained from numerical simulations can be verified. We have observed all the phases that Wang et. al. \cite{wang2017} investigated. Nevertheless, in contrast to what was shown in their phase diagrams, we have observed a sharp transition between FM and AFM phases. We are, however, unable to find any spin liquid, Haldane or topological phase in this system.

\section*{ACKNOWLEDGMENTS}

We thank Q. Wang, H. Pichler, A. V. Balatsky and J. Javanainen for many helpful discussions. DMRG simulations were performed using the ITensor library \cite{itensor}. We are extremely grateful to E. M. Stoudenmire, the lead developer of ITensor, for helping us write the codes for our model and checking them for errors. This research project is supported by National Science Foundation.

\FloatBarrier

\section*{Appendix}

%Source: https://tex.stackexchange.com/questions/279/how-do-i-ensure-that-figures-appear-in-the-section-theyre-associated-with Just put \FloatBarrier above and below every section/subsection/subsubsection.

\subsection{Frustrated and non-frustrated regimes}

%To crop a picture: trim={<left> <lower> <right> <upper>}

\begin{figure}[htb!]
\begin{subfigure}[b]{0.4\textwidth}
\centering
    \includegraphics[trim={0cm 0.35cm 0cm 0cm},clip,width=\linewidth]{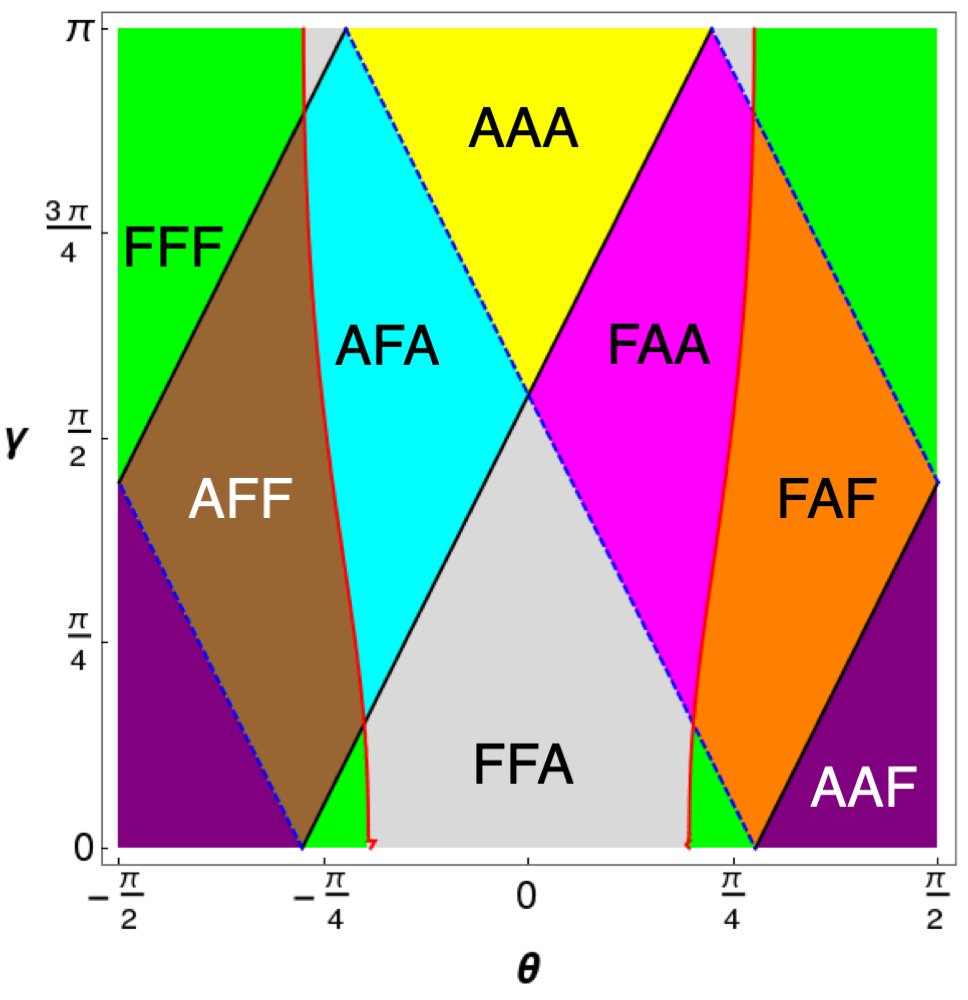}
    \label{label-contourplot_v11_Bo_1_c_1}
    \caption{$d_1/\lambda = 1$}
	\end{subfigure}\hfill
\begin{subfigure}[b]{0.4\textwidth}
\centering
    \includegraphics[trim={0cm 0.35cm 0cm 0cm},clip,width=\linewidth]{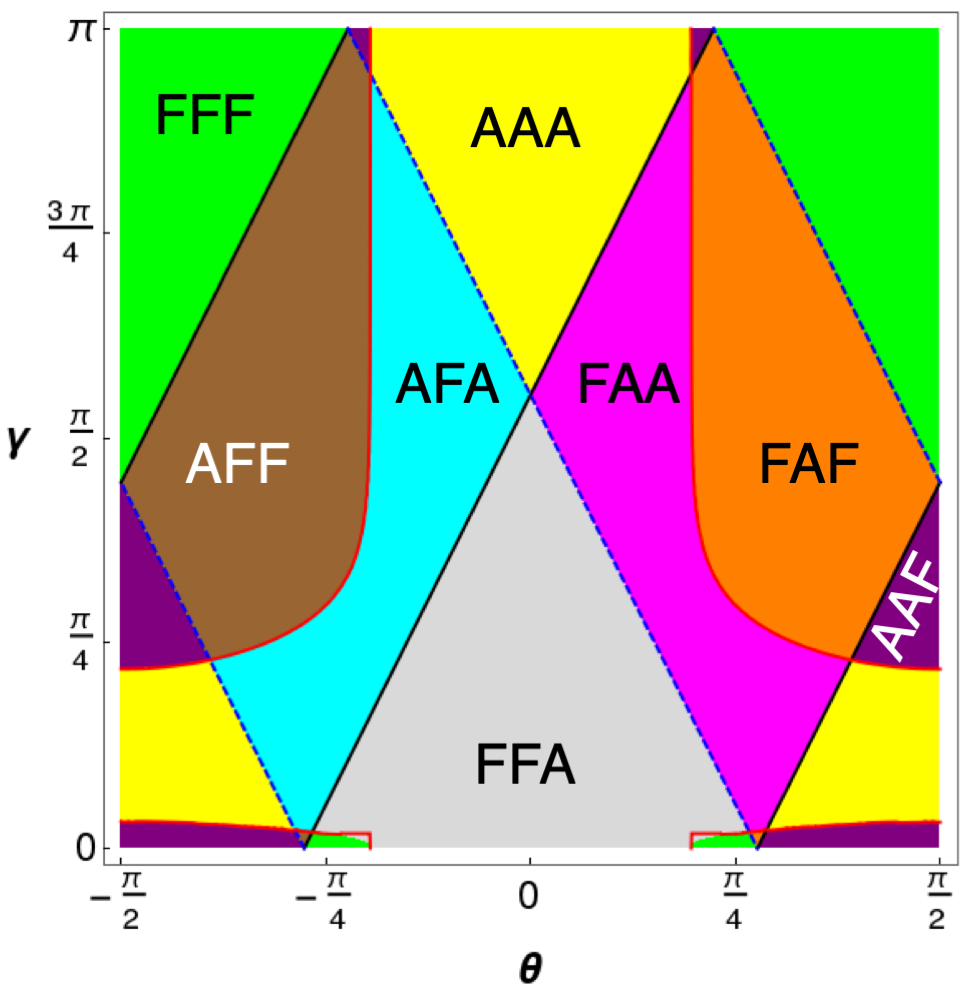}
    \label{label-contourplot_v11_Bo_1_c_10}
    \caption{$d_1/\lambda = 10$}
	\end{subfigure}
	\caption{(Color online) Frustrated and non-frustrated regions for other lattice depths.}
	\label{fig:contourplot_2}
\end{figure}

In \fig{fig:contourplot}, we saw how the eight regions - four frustrated (AFF, FAF, FFA and AAA) and four non-frustrated (FFF, AAF, AFA and FAA) - were related to the chain opening angle $\gamma$ and polarization angle $\theta$ given the ratio $d_1/\lambda = 0.1$. \fig{fig:contourplot_2} illustrates how these regions depend on the angles $\gamma$ and $\theta$ for other lattice depths. We find that all the eigght regions exist in our system, although their shape and size vary, regardless of the value of $d_1/\lambda$.

\FloatBarrier 

\subsection{Correlation functions for various phases}

In the body of this paper, we have shown the values of only one or two correlation functions to confirm a given phase. In this section, we will show additional plots to support our claim. We will also include the values of the interactions to show which frustrated/non-frustrated region the example point under consideration belongs to.

\subsubsection{Z-dimer phase}

\begin{figure}[htb!]
\begin{subfigure}[b]{0.24\textwidth}
\centering
    \includegraphics[trim={0cm 0cm 0.35cm 0cm},clip,width=\linewidth]{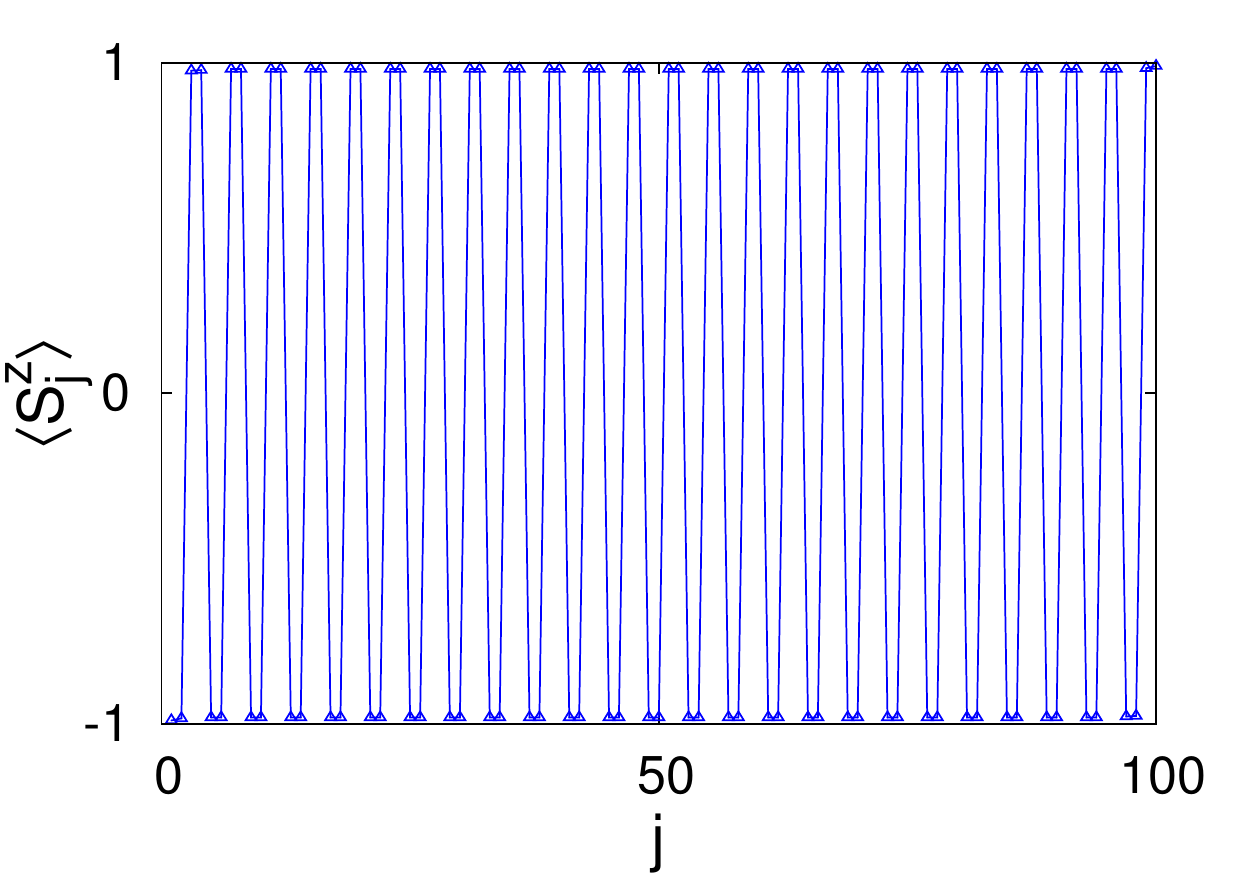}
\caption{}
%\label{label-random_v1_and_v2_gamma_60_theta_30}
\end{subfigure}\hfill
\begin{subfigure}[b]{0.24\textwidth}
\centering
    \includegraphics[trim={0cm 0cm 0.35cm 0cm},clip,width=\linewidth]{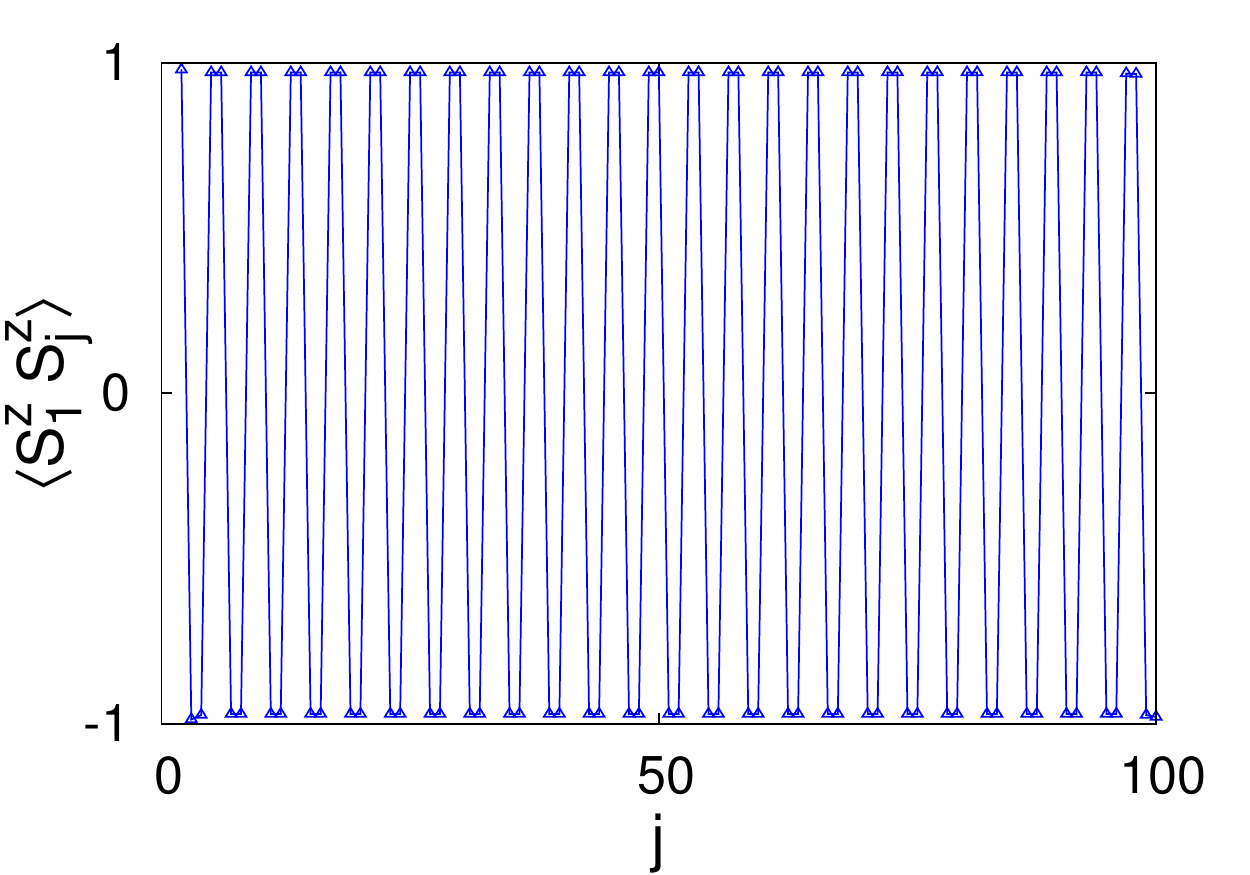}
\caption{}
%\label{label-random_v1_and_v2_gamma_60_theta_15}
\end{subfigure}\hfill
\begin{subfigure}[b]{0.24\textwidth}
\centering
    \includegraphics[trim={0cm 0cm 0.35cm 0cm},clip,width=\linewidth]{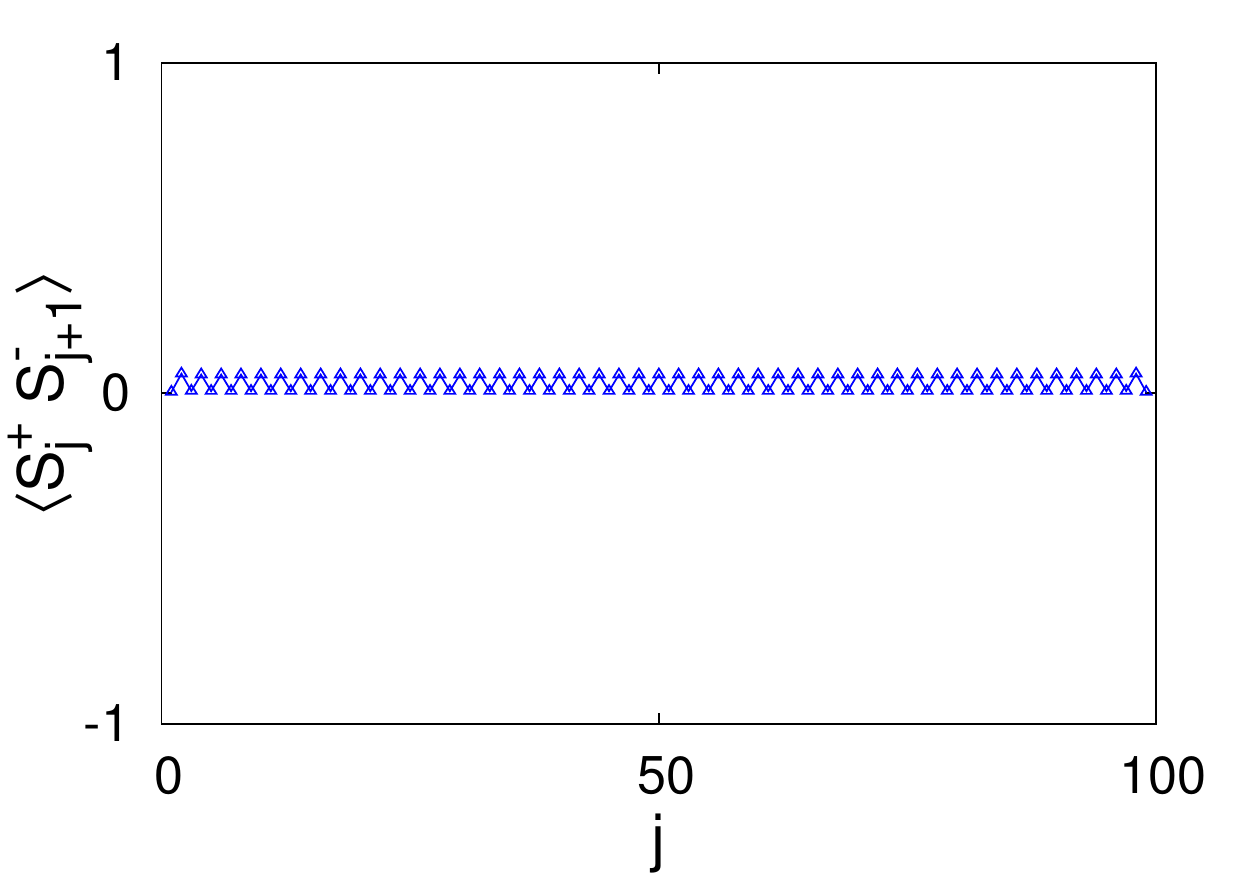}
\caption{} 
%\label{label-random_v1_and_v2_gamma_60_theta_15}
\end{subfigure}\hfill
\begin{subfigure}[b]{0.24\textwidth}
\centering
    \includegraphics[trim={0cm 0cm 0.35cm 0cm},clip,width=\linewidth]{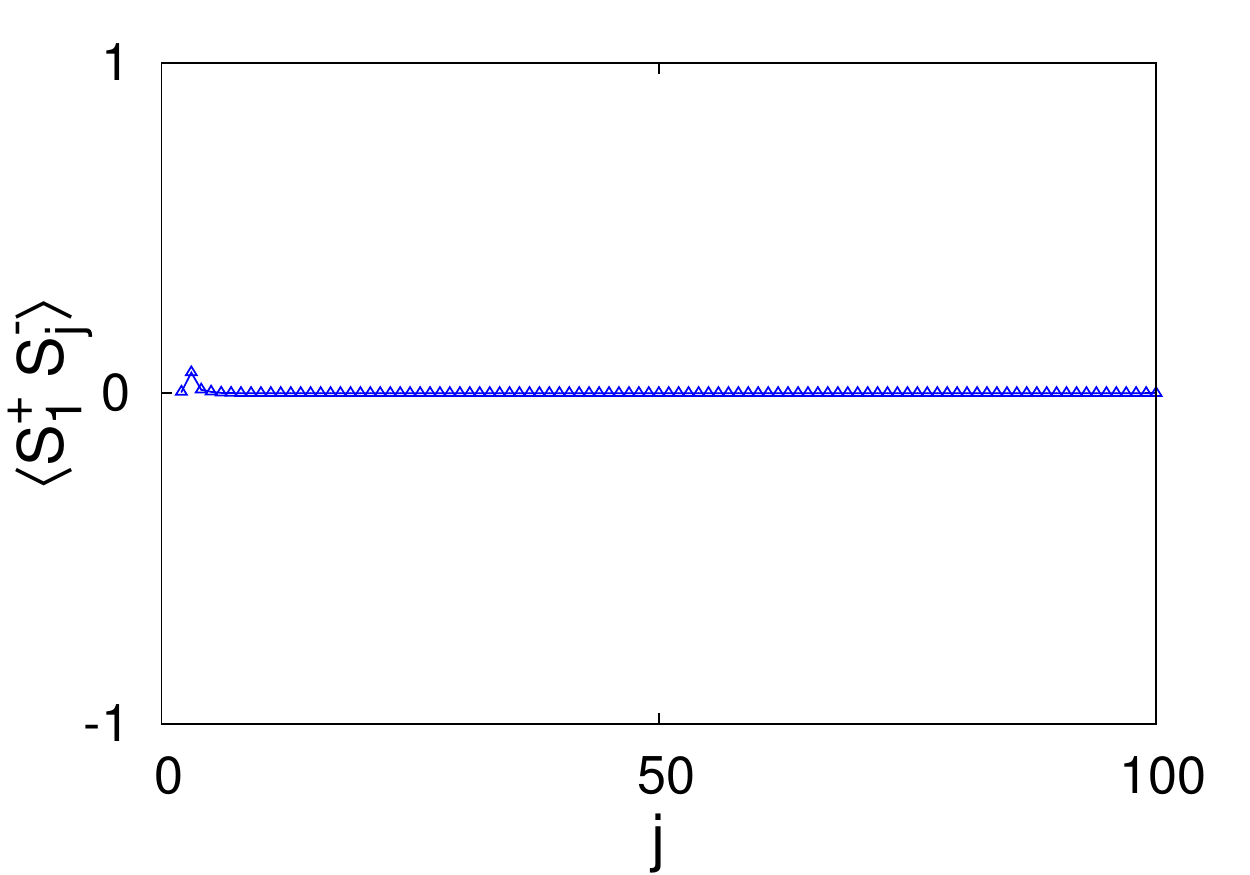}
\caption{} 
%\label{label-random_v1_and_v2_gamma_60_theta_15}
\end{subfigure}
\caption{Additional correlations for the z-dimer phase. $(\gamma, \theta) = (\pi/3,\pi/6)$. $(\alpha_o, \alpha_e, \alpha_2) = (-2.000, 0.250, 0.250)$. Region: FAA. $\ket{\text{init}} = \ket{\downarrow\uparrow\downarrow\uparrow\downarrow\uparrow \ldots}$.}
\label{fig:correlations_gamma_60_theta_30_zdimer_phase_appendix}
\end{figure}

\fig{fig:correlations_gamma_60_theta_30_zdimer_phase_appendix} shows additional plots for the z-dimer phase shown in \fig{label-random_v1_gamma_60_theta_30}, which belongs to the non-frustrated region FAA. In principle, one should obtain $\langle S^z_j \rangle = 0$ for each site index $j$ because the ground state is expected to be a superposition of the two states \{$\ket{\downarrow\downarrow\uparrow\uparrow\downarrow\downarrow\uparrow\uparrow \ldots}, \ket{\uparrow\uparrow\downarrow\downarrow\uparrow\uparrow\downarrow\downarrow \ldots}$\}. However, DMRG returns one of these two states rather than a superposition. A similar argument is valid for all other phases. 

The other three plots are straightforward. We would expect the same results regardless of whether the ground state is a single z-dimer state, as is the result from DMRG, or a superposition of two degenerate z-dimer states, as is the result from ab-intio calculations. A similar argument is valid for all other phases.

\FloatBarrier

\subsubsection{XY-dimer phase}

\begin{figure}[htb!]
\begin{subfigure}[b]{0.24\textwidth}
\centering
    \includegraphics[trim={0cm 0cm 0.35cm 0cm},clip,width=\linewidth]{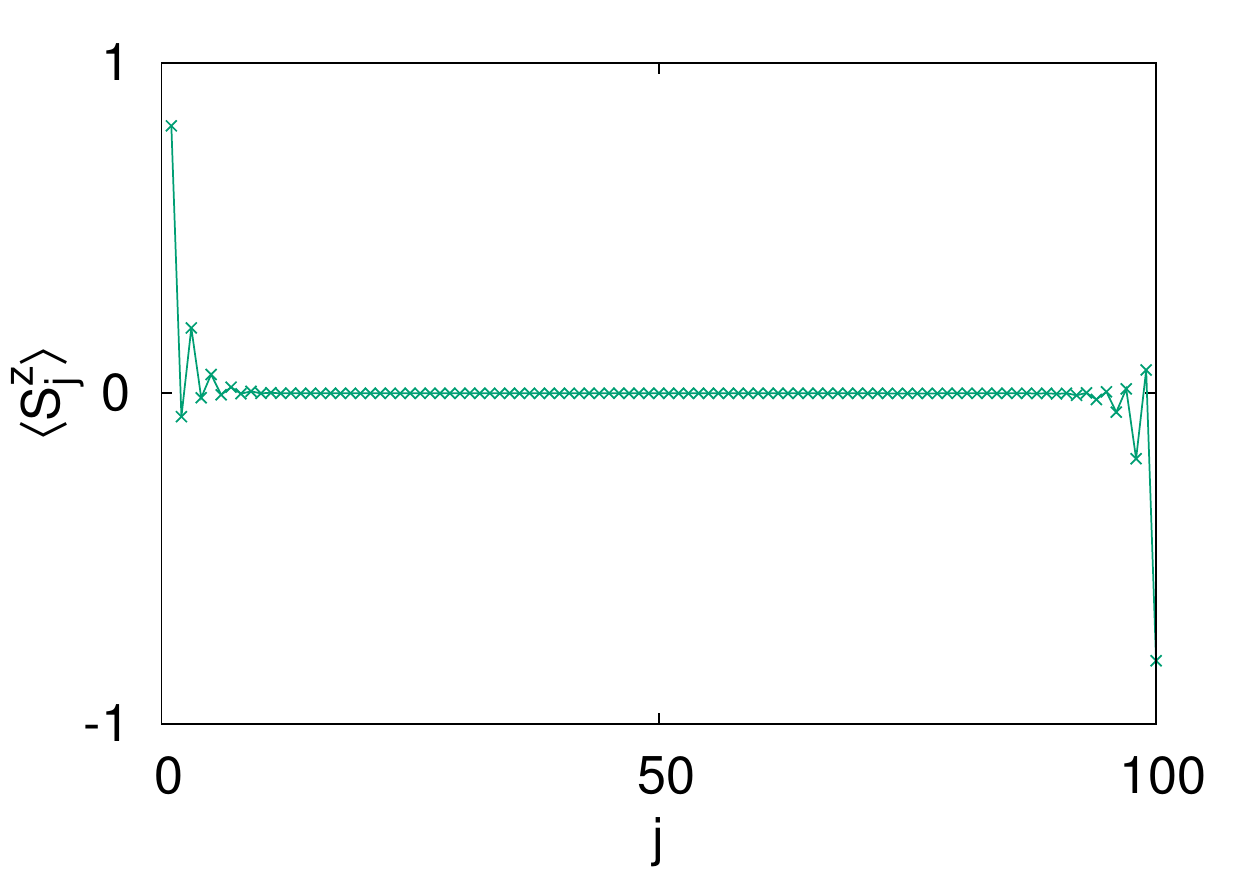}
\caption{}
%\label{label-random_v1_and_v2_gamma_60_theta_30}
\end{subfigure}\hfill
\begin{subfigure}[b]{0.24\textwidth}
\centering
    \includegraphics[trim={0cm 0cm 0.35cm 0cm},clip,width=\linewidth]{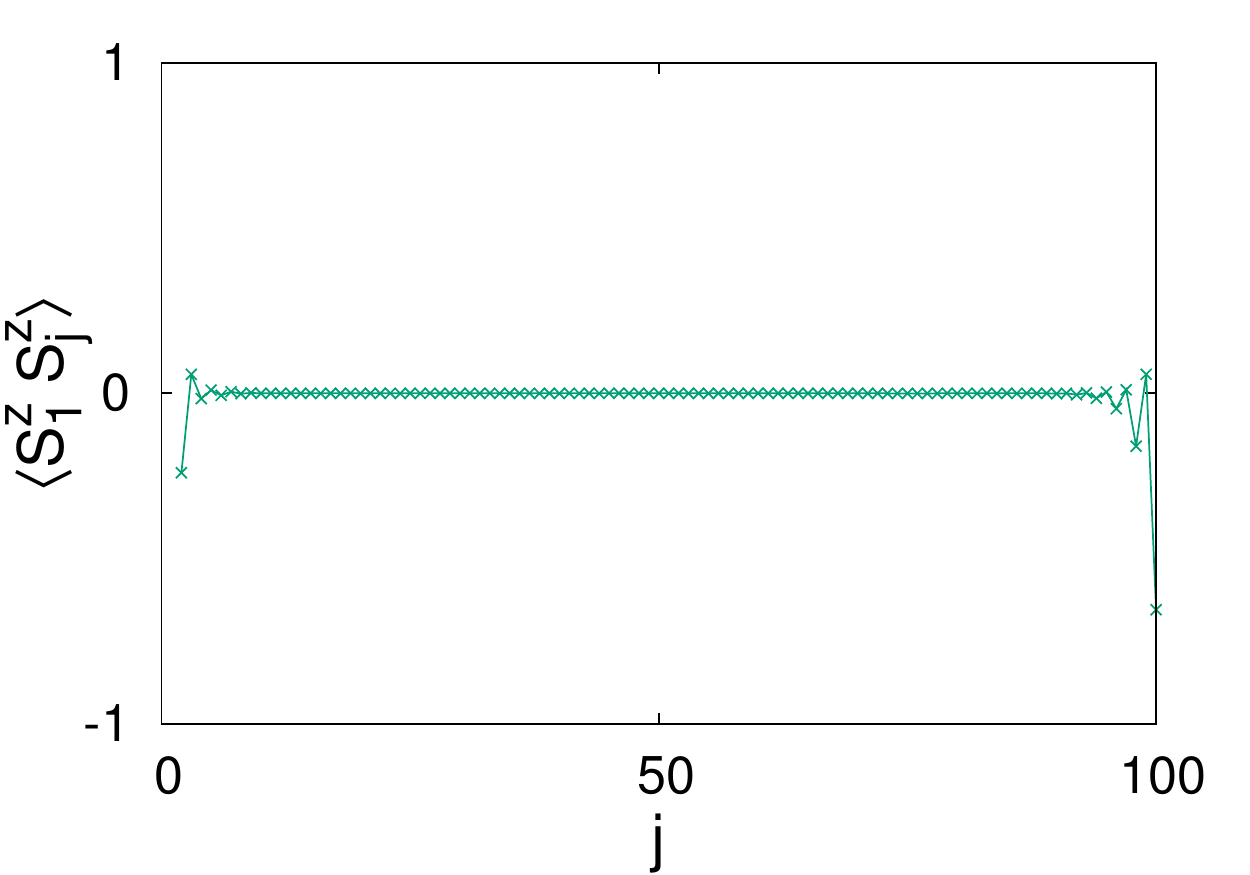}
\caption{} 
%\label{label-random_v1_and_v2_gamma_60_theta_15}
\end{subfigure}\hfill
\begin{subfigure}[b]{0.24\textwidth}
\centering
    \includegraphics[trim={0cm 0cm 0.35cm 0cm},clip,width=\linewidth]{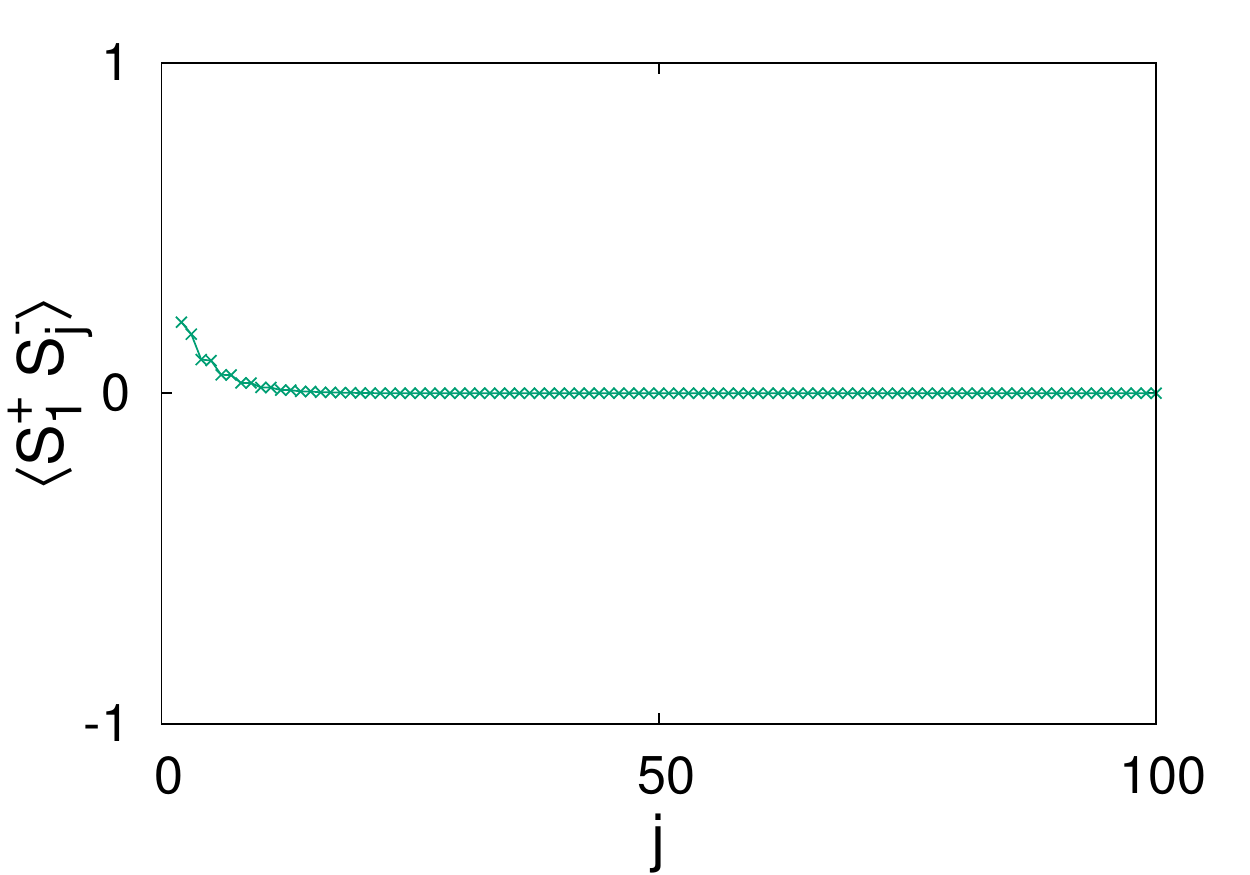}
\caption{} 
%\label{label-random_v1_and_v2_gamma_60_theta_15}
\end{subfigure}
\caption{Additional correlations for the xy-dimer phase. $(\gamma, \theta) = (5 \pi/6, 0.0889 \pi)$. $(\alpha_o, \alpha_e, \alpha_2) = (0.204, 0.999, 0.117)$. Region: AAA. $\ket{\text{init}} = \ket{\text{random}}$.}
\label{fig:correlations_gamma_150_theta_16_xydimer_phase_appendix}
\end{figure}

\fig{fig:correlations_gamma_150_theta_16_xydimer_phase_appendix} shows additional plots for the xy-dimer phase shown in \fig{fig:gamma_150_theta_16_xydimer_phase}, which belongs to the frustrated region AAA. 

\FloatBarrier 

\subsubsection{Superfluid phase}

\begin{figure}[htb!]
\begin{subfigure}[b]{0.24\textwidth}
\centering
    \includegraphics[trim={0cm 0cm 0.35cm 0cm},clip,width=\linewidth]{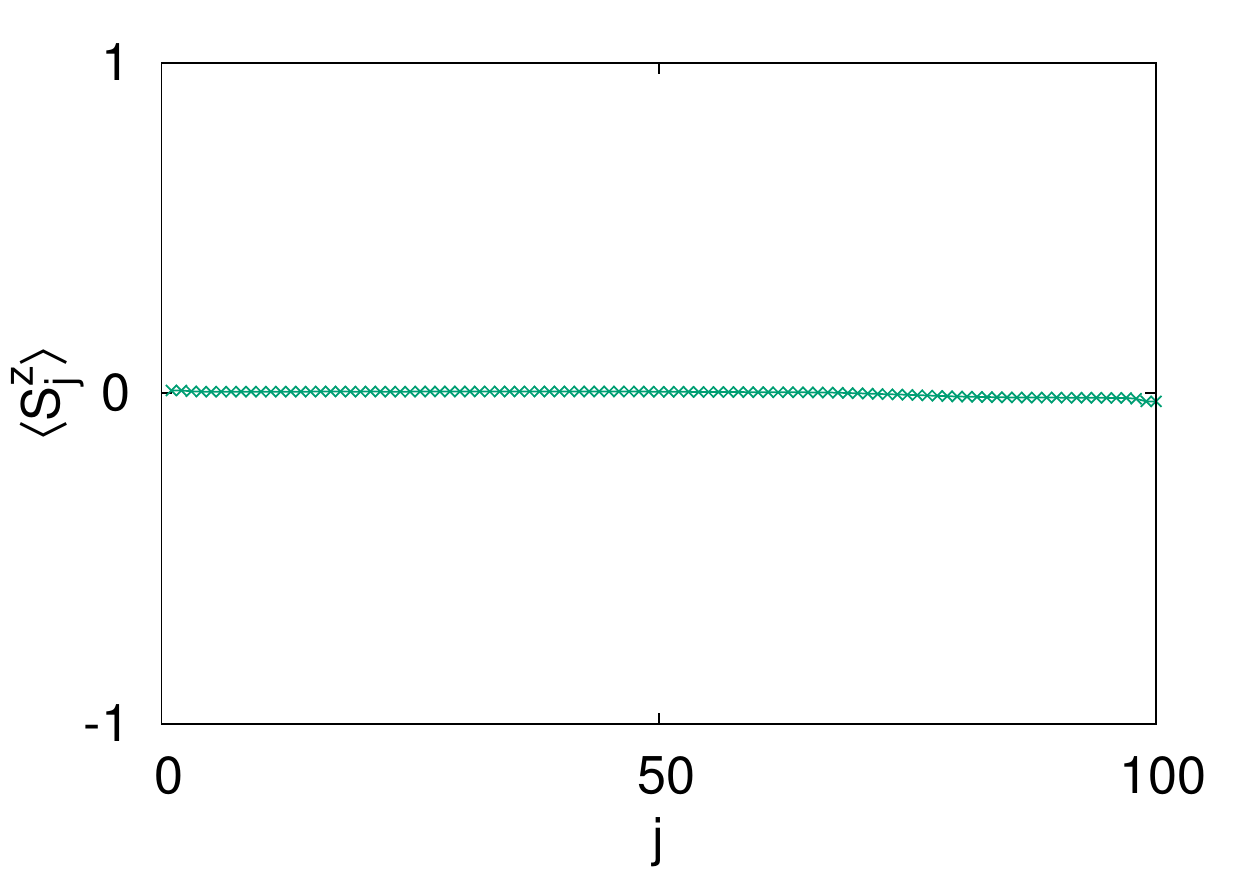}
\caption{}
%\label{label-random_v1_and_v2_gamma_60_theta_30}
\end{subfigure}\hfill
\begin{subfigure}[b]{0.24\textwidth}
\centering
    \includegraphics[trim={0cm 0cm 0.35cm 0cm},clip,width=\linewidth]{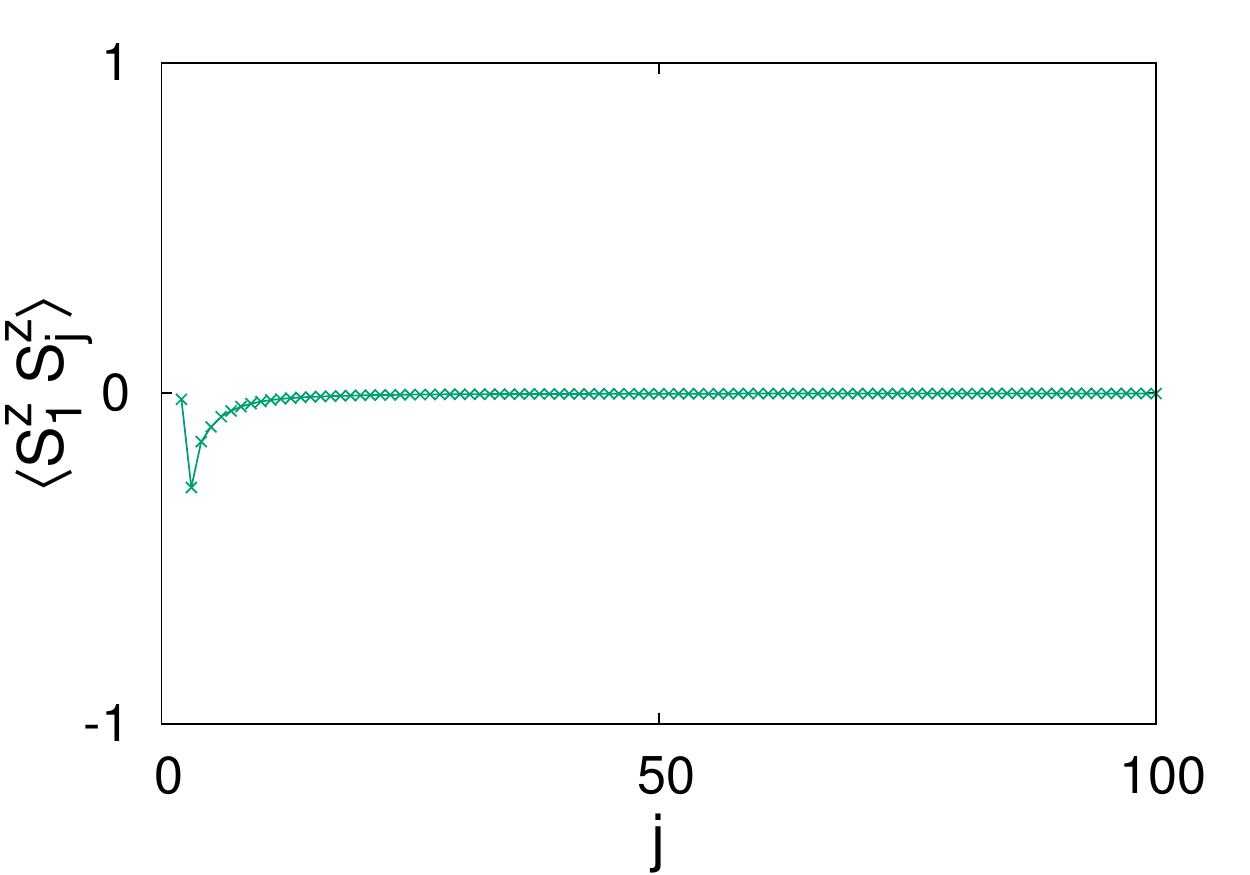}
\caption{} 
%\label{label-random_v1_and_v2_gamma_60_theta_15}
\end{subfigure}\hfill
\begin{subfigure}[b]{0.24\textwidth}
\centering
    \includegraphics[trim={0cm 0cm 0.35cm 0cm},clip,width=\linewidth]{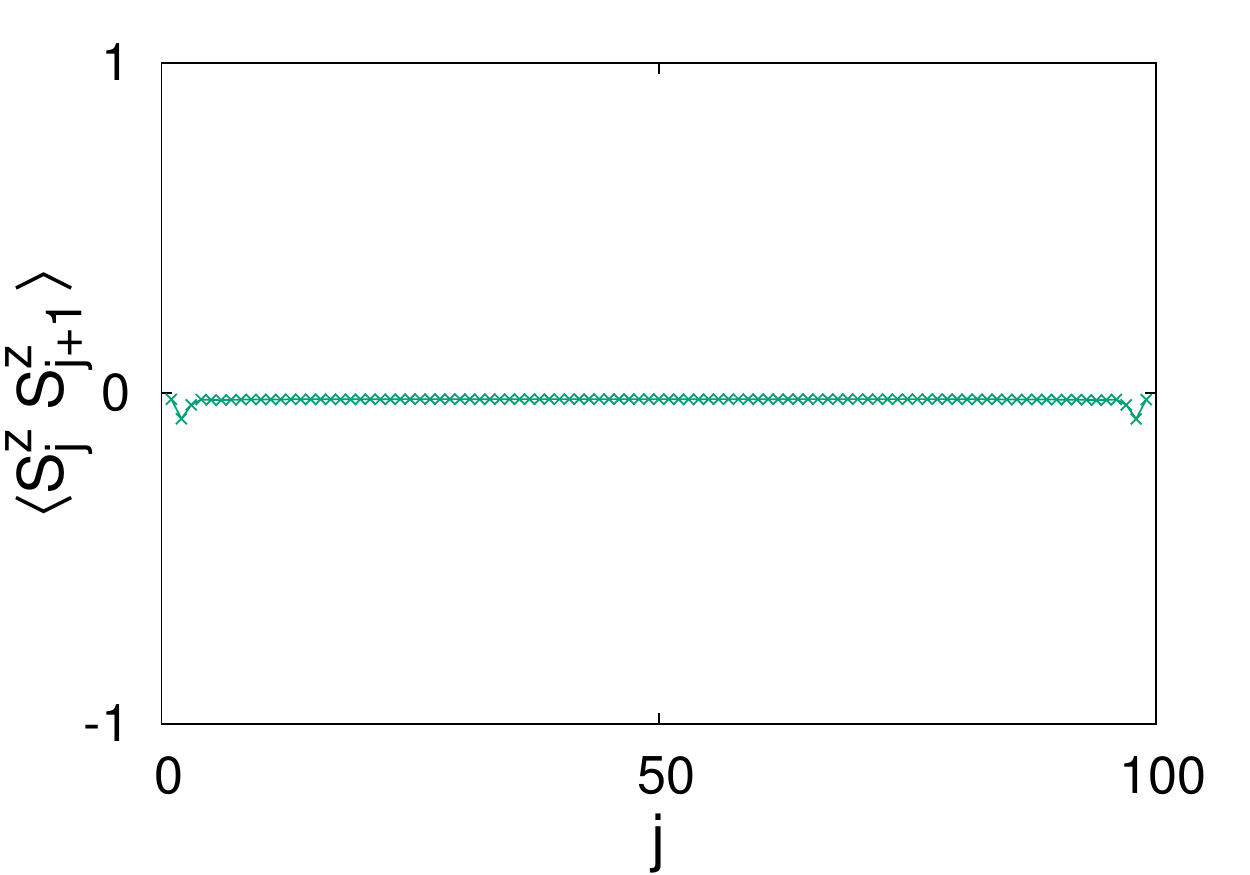}
\caption{}
%\label{label-random_v1_and_v2_gamma_60_theta_15}
\end{subfigure}\hfill
\begin{subfigure}[b]{0.24\textwidth}
\centering
    \includegraphics[trim={0cm 0cm 0.35cm 0cm},clip,width=\linewidth]{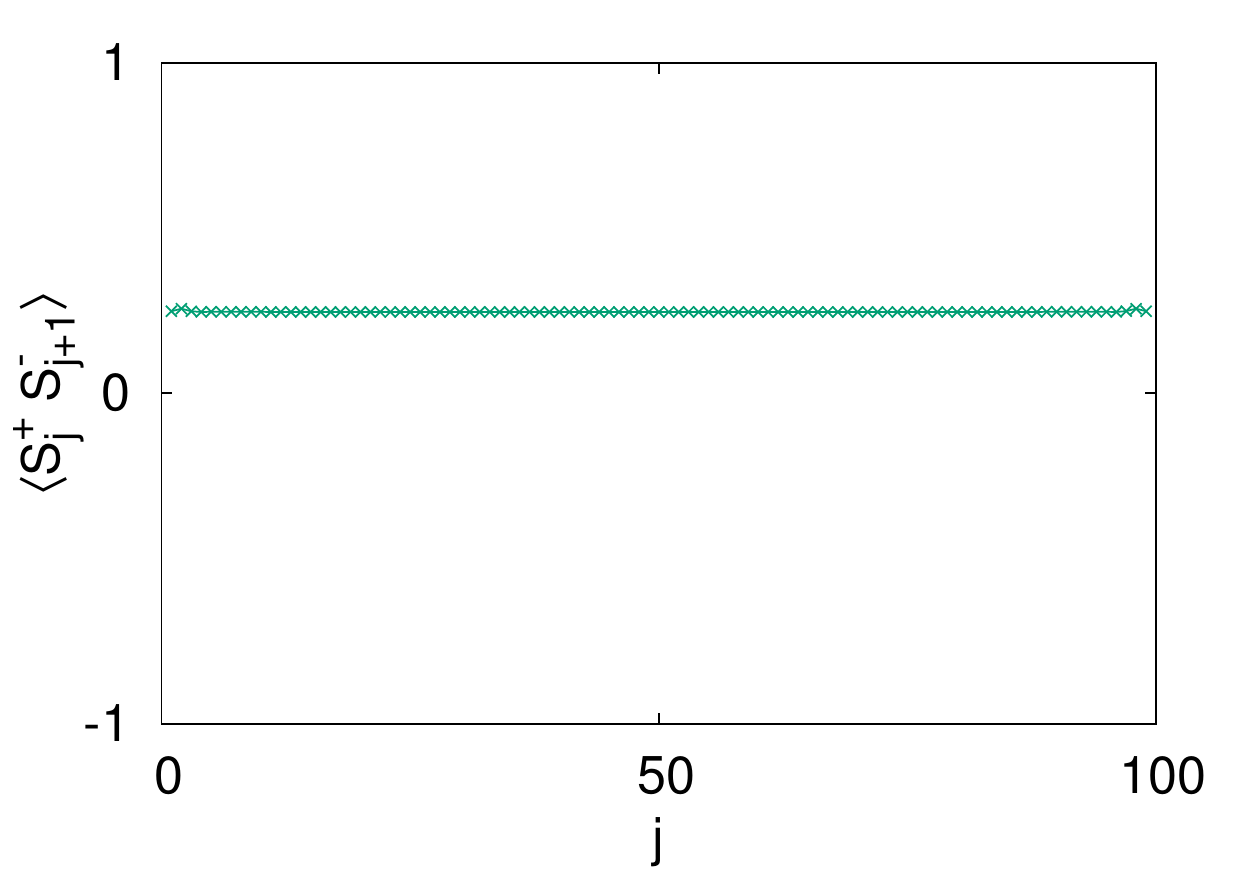}
\caption{} 
%\label{label-random_v1_and_v2_gamma_60_theta_15}
\end{subfigure}
\caption{Additional correlations for the SF phase. $(\gamma, \theta) = (\pi,0.2333 \pi)$. $(\alpha_o, \alpha_e, \alpha_2) = (-0.343, -0.343, -0.047)$. Region: FFA. $\ket{\text{init}} = \ket{\text{random}}$.}
\label{fig:correlations_gamma_180_theta_42_SF_phase_appendix}
\end{figure}

\fig{fig:correlations_gamma_180_theta_42_SF_phase_appendix} shows additional plots for the SF phase shown in \fig{label-random_v1_gamma_180_theta_42_loge_Sp_1_Sm_j_vs_loge_j}, which belongs to the frustrated region FFA.

\FloatBarrier 

\subsubsection{Ferromagnetic phase}

\begin{figure}[htb!]
\begin{subfigure}[b]{0.24\textwidth}
\centering
    \includegraphics[trim={0cm 0cm 0.35cm 0cm},clip,width=\linewidth]{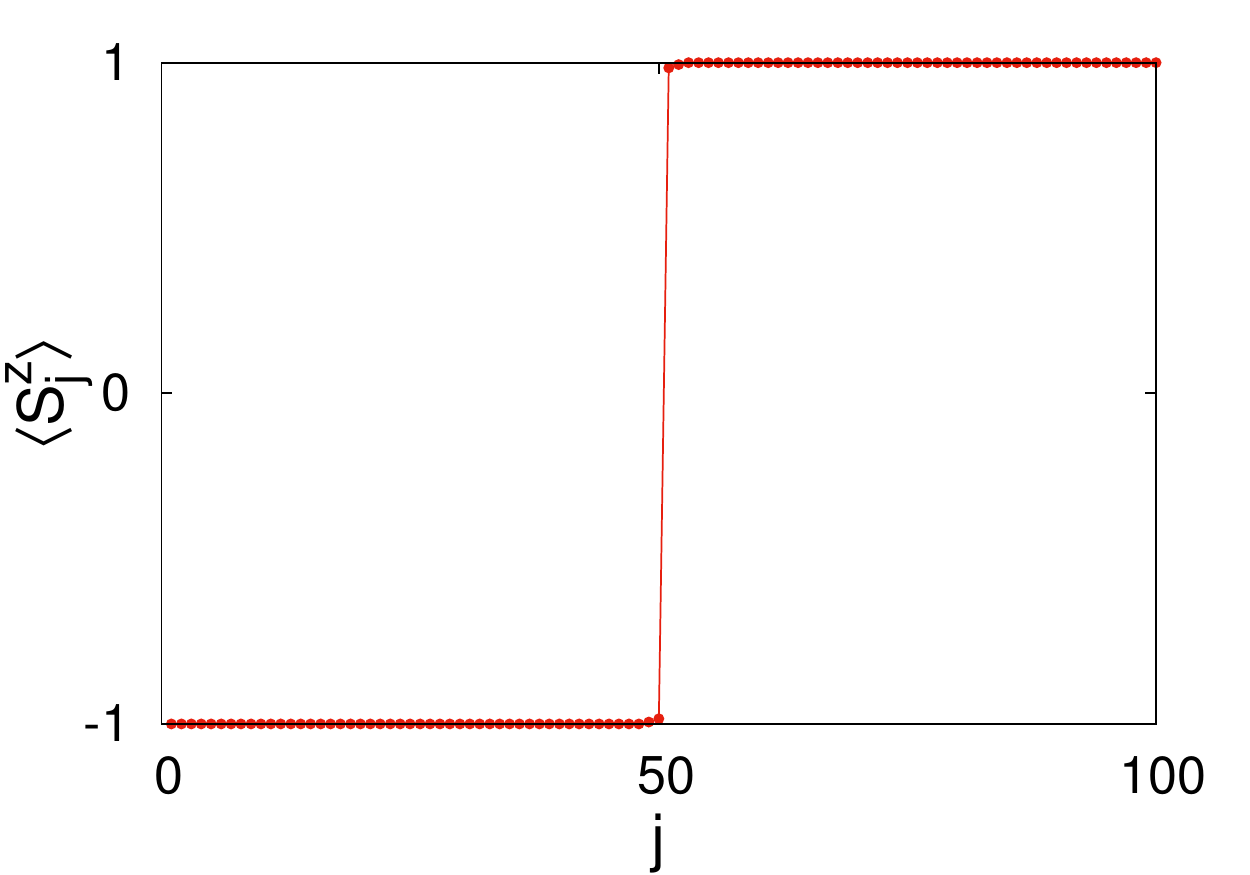}
\caption{}
%\label{label-random_v1_and_v2_gamma_60_theta_30}
\end{subfigure}\hfill
\begin{subfigure}[b]{0.24\textwidth}
\centering
    \includegraphics[trim={0cm 0cm 0.35cm 0cm},clip,width=\linewidth]{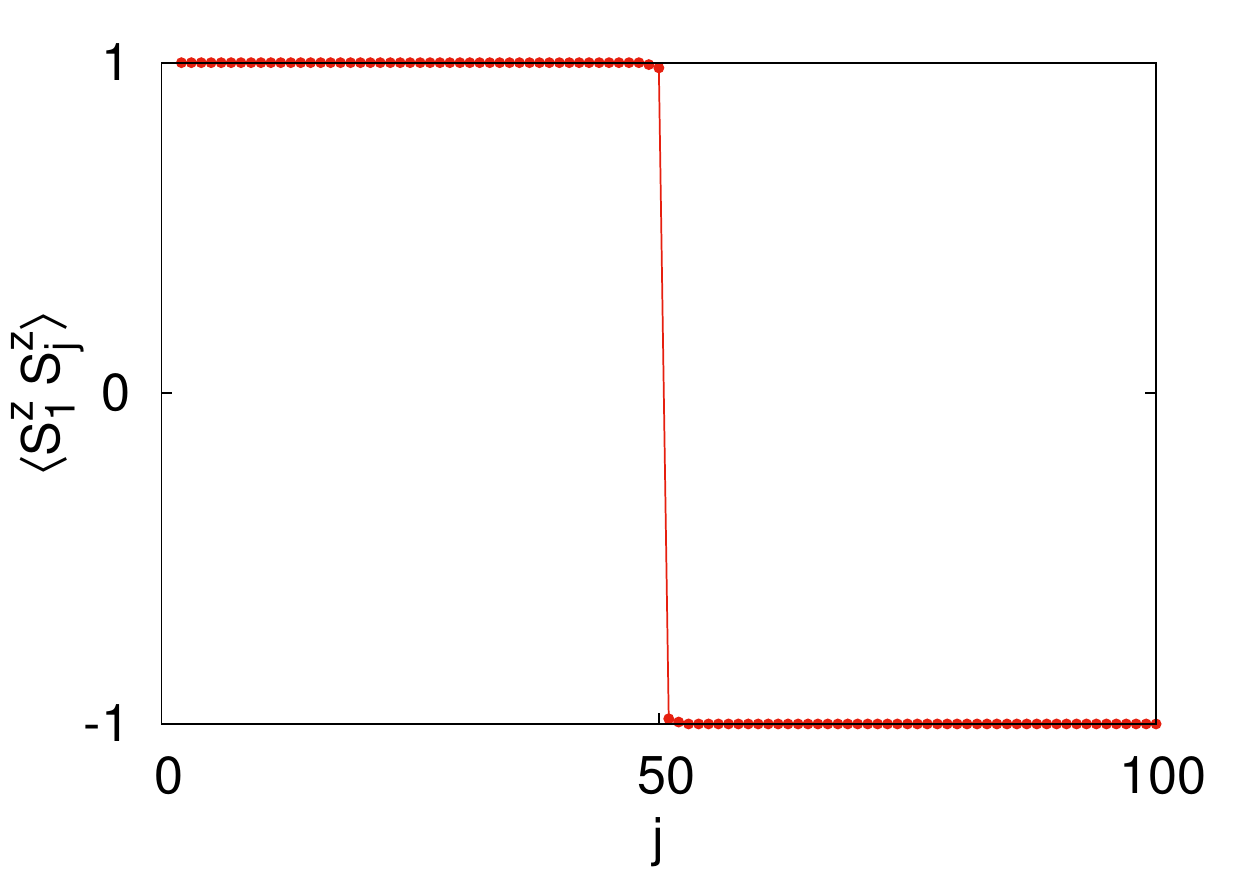}
\caption{}
%\label{label-random_v1_and_v2_gamma_60_theta_15}
\end{subfigure}\hfill
\begin{subfigure}[b]{0.24\textwidth}
\centering
    \includegraphics[trim={0cm 0cm 0.35cm 0cm},clip,width=\linewidth]{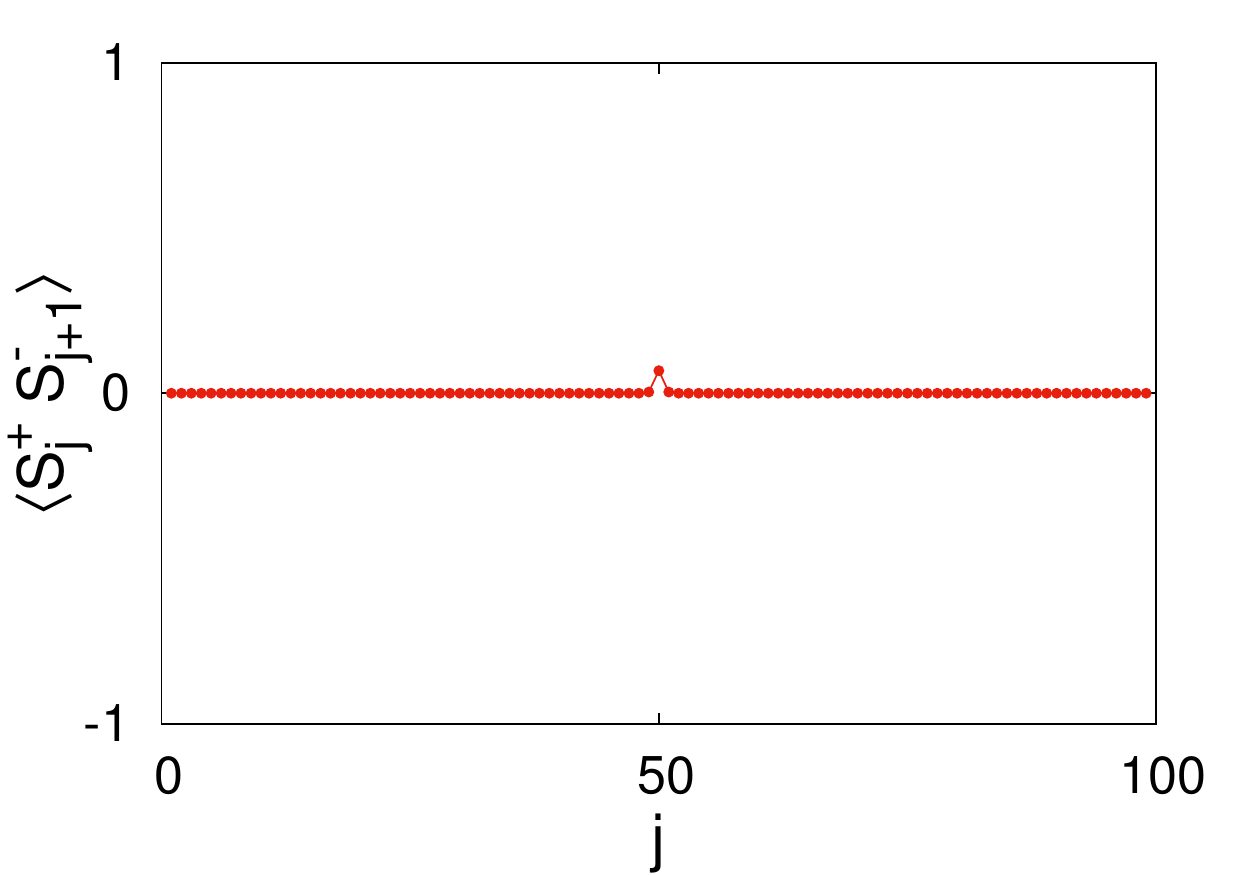}
\caption{} 
%\label{label-random_v1_and_v2_gamma_60_theta_15}
\end{subfigure}\hfill
\begin{subfigure}[b]{0.24\textwidth}
\centering
    \includegraphics[trim={0cm 0cm 0.35cm 0cm},clip,width=\linewidth]{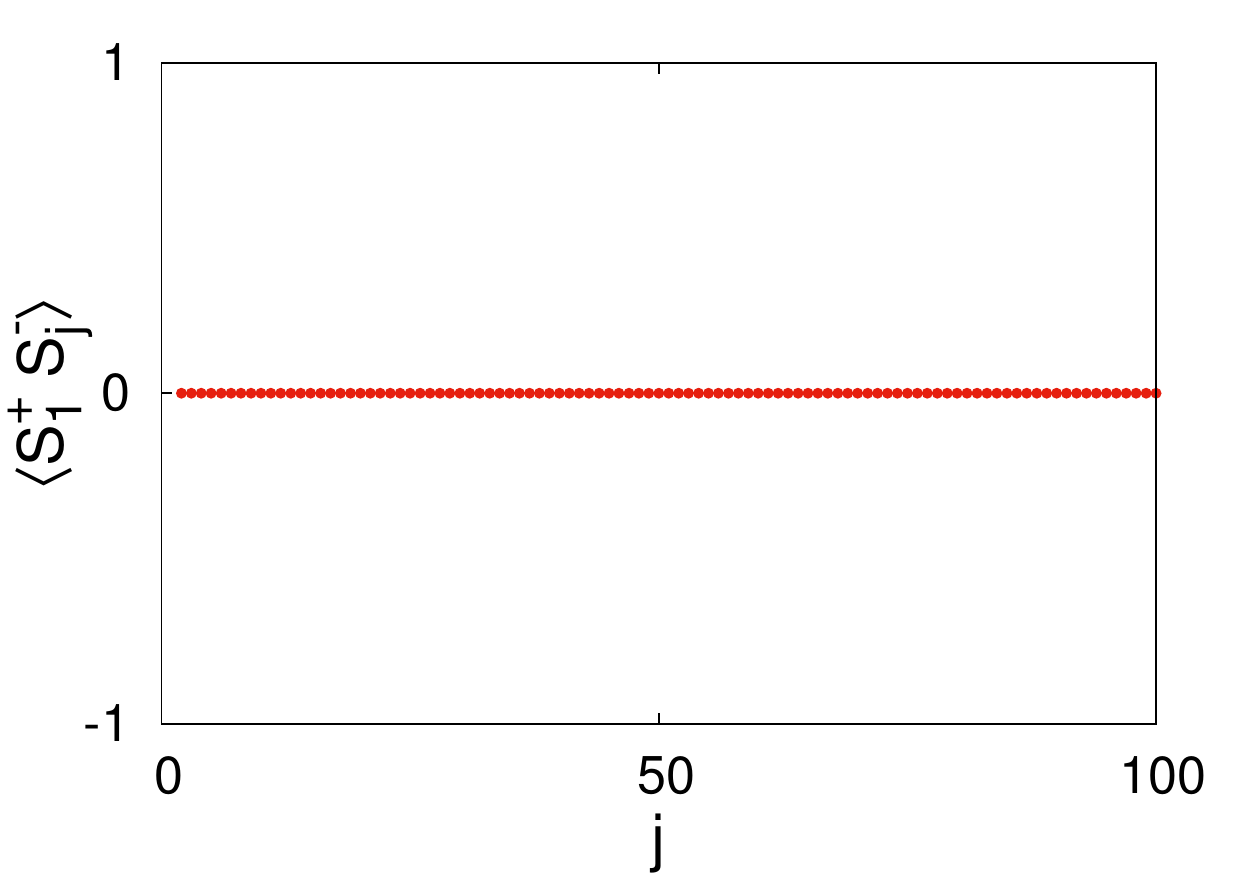}
\caption{} 
%\label{label-random_v1_and_v2_gamma_60_theta_15}
\end{subfigure}
\caption{Additional correlations for the FM phase. $(\gamma, \theta) = (\pi, \pi/2)$. $(\alpha_o, \alpha_e, \alpha_2) = (-2.000, -2.000, -0.276)$. Region: FFF. $\ket{\text{init}} = \ket{\ldots \downarrow\downarrow\downarrow\uparrow\uparrow\uparrow \ldots}$.}
\label{fig:correlations_gamma_180_theta_90_FM_phase_appendix}
\end{figure}

\fig{fig:correlations_gamma_180_theta_90_FM_phase_appendix} shows additional plots for the ferromagnetic phase shown in \fig{fig:gamma_180_theta_90_FM}, which belongs to the non-frustrated region FFF.

%\subsubsection{Antiferromagnetic phase}

\FloatBarrier 

\subsubsection{Antiferromagnetic phase: AFM1}

\begin{figure}[htb!]
\begin{subfigure}[b]{0.24\textwidth}
\centering
    \includegraphics[trim={0cm 0cm 0.35cm 0cm},clip,width=\linewidth]{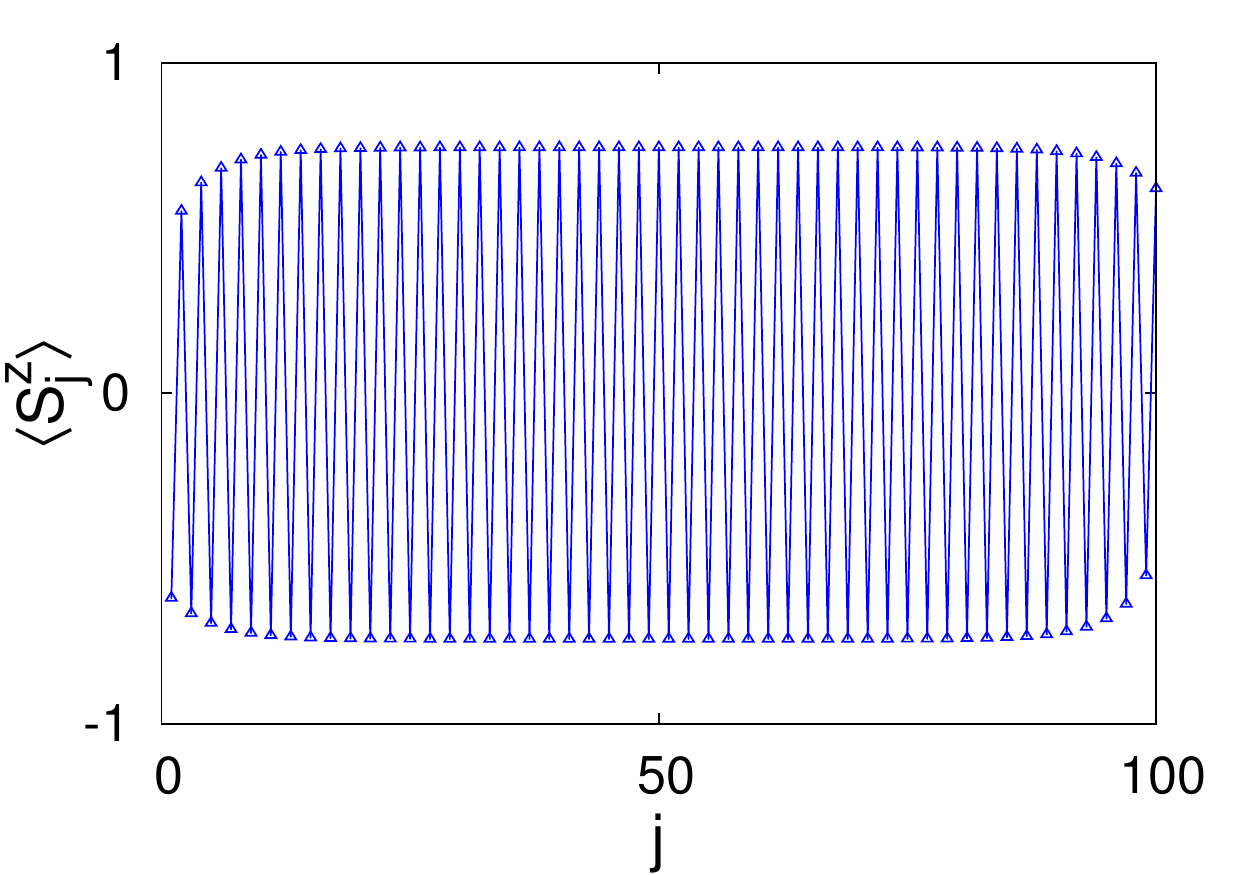}
\caption{}
%\label{label-random_v1_and_v2_gamma_60_theta_30}
\end{subfigure}\hfill
\begin{subfigure}[b]{0.24\textwidth}
\centering
    \includegraphics[trim={0cm 0cm 0.35cm 0cm},clip,width=\linewidth]{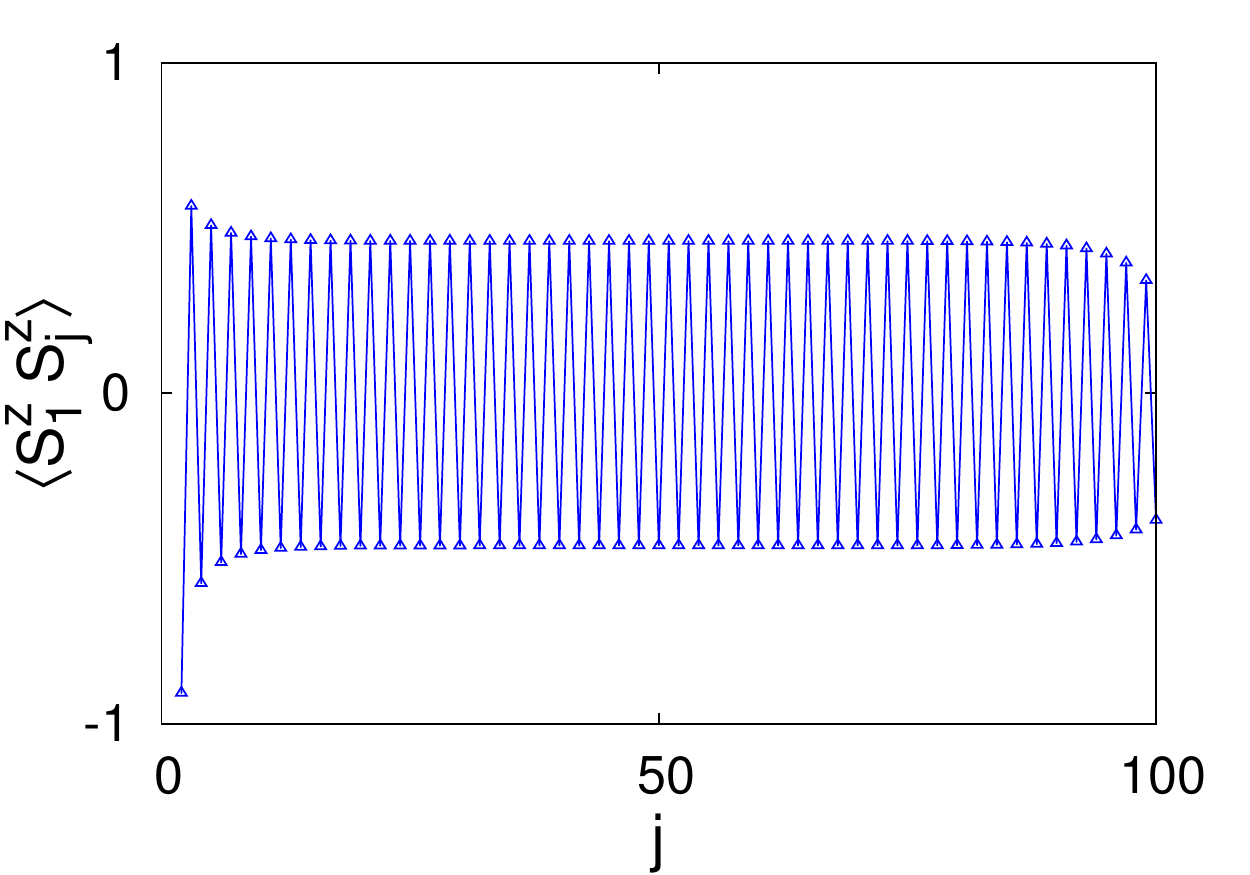}
\caption{}
%\label{label-random_v1_and_v2_gamma_60_theta_15}
\end{subfigure}\hfill
\begin{subfigure}[b]{0.24\textwidth}
\centering
    \includegraphics[trim={0cm 0cm 0.35cm 0cm},clip,width=\linewidth]{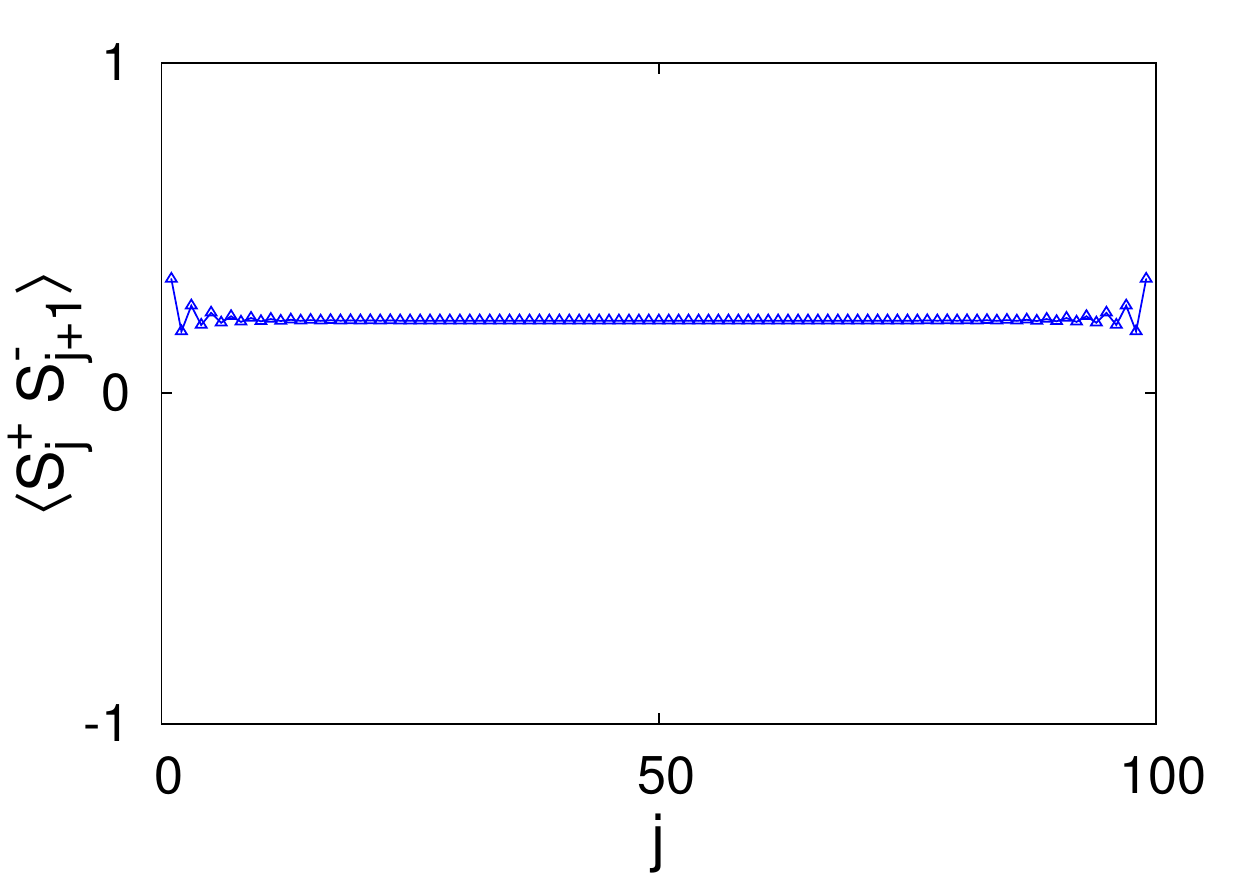}
\caption{} 
%\label{label-random_v1_and_v2_gamma_60_theta_15}
\end{subfigure}\hfill
\begin{subfigure}[b]{0.24\textwidth}
\centering
    \includegraphics[trim={0cm 0cm 0.35cm 0cm},clip,width=\linewidth]{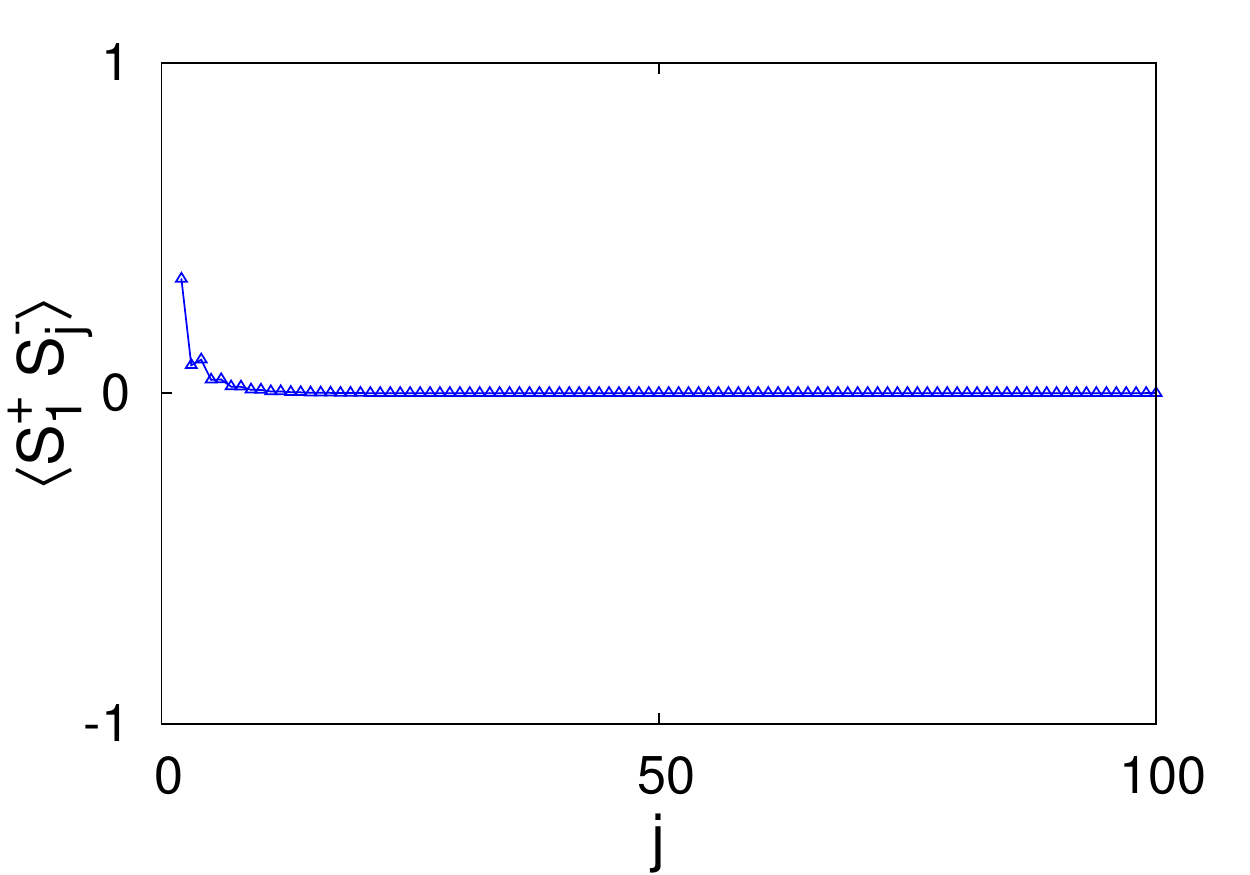}
\caption{} 
%\label{label-random_v1_and_v2_gamma_60_theta_15}
\end{subfigure}
\caption{Additional correlations for the AFM1 phase. $(\gamma, \theta) = (\pi,0)$. $(\alpha_o, \alpha_e, \alpha_2) = (1.000, 1.000, 0.138)$. Region: AAA. $\ket{\text{init}} = \ket{\downarrow\uparrow\downarrow\uparrow\downarrow\uparrow \ldots}$.}
\label{fig:correlations_gamma_180_theta_0_AFM1_phase_appendix}
\end{figure}

\fig{fig:correlations_gamma_180_theta_0_AFM1_phase_appendix} shows additional plots for the AFM1 phase shown in \fig{fig:gamma_180_theta_0_AFM}, which belongs to the frustrated region AAA.

\FloatBarrier 

\subsubsection{Antiferromagnetic phase: AFM2}

\begin{figure}[htb!]
\begin{subfigure}[b]{0.24\textwidth}
\centering
    \includegraphics[trim={0cm 0cm 0.35cm 0cm},clip,width=\linewidth]{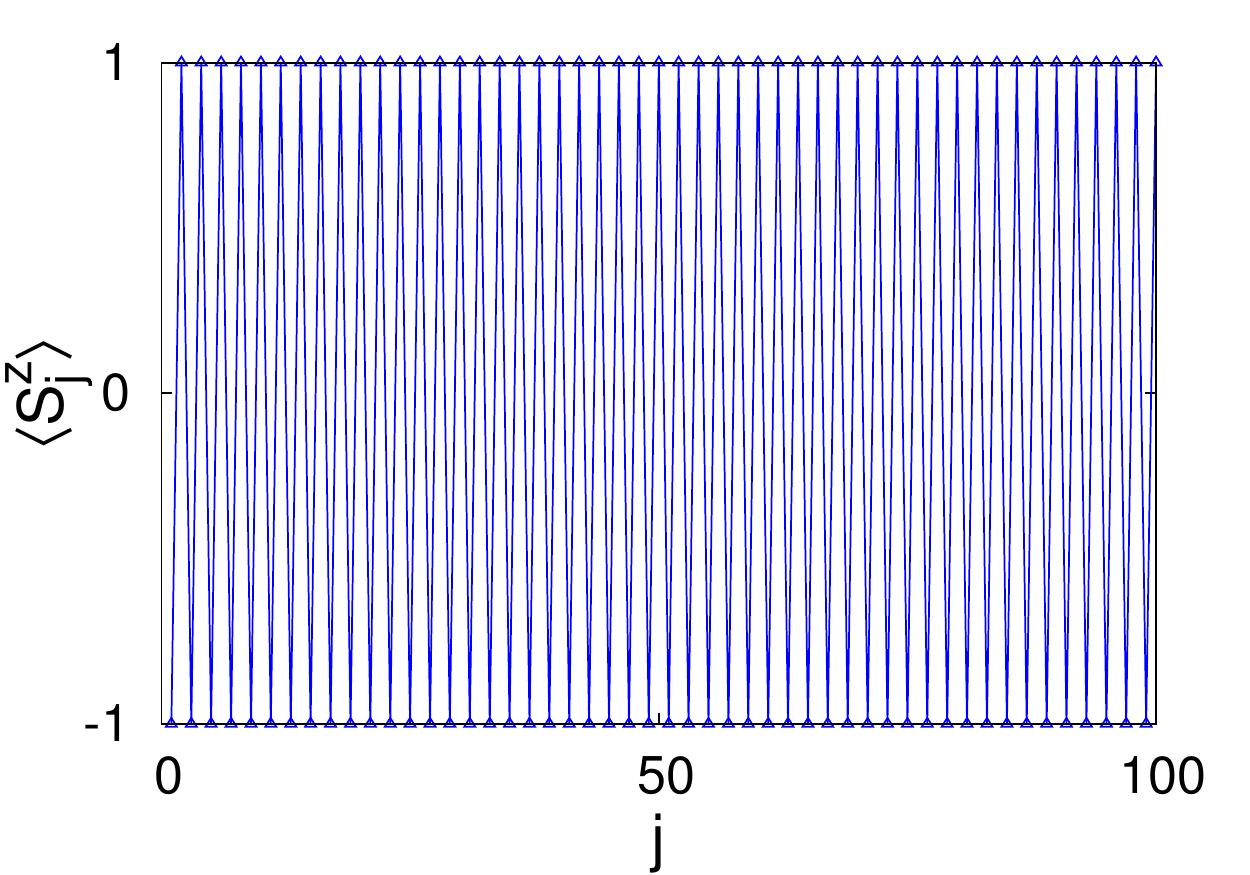}
\caption{}
%\label{label-random_v1_and_v2_gamma_60_theta_30}
\end{subfigure}\hfill
\begin{subfigure}[b]{0.24\textwidth}
\centering
    \includegraphics[trim={0cm 0cm 0.35cm 0cm},clip,width=\linewidth]{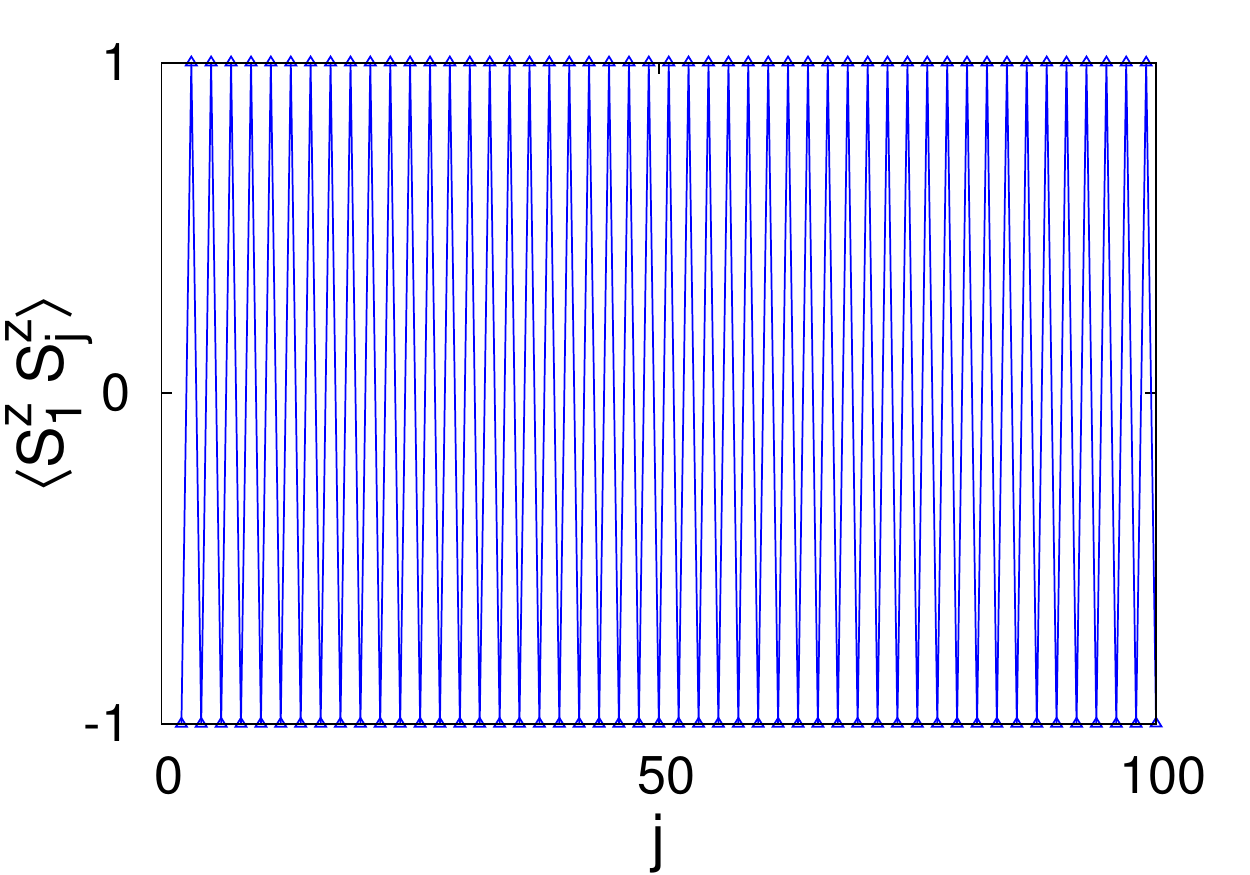}
\caption{}
%\label{label-random_v1_and_v2_gamma_60_theta_15}
\end{subfigure}\hfill
\begin{subfigure}[b]{0.24\textwidth}
\centering
    \includegraphics[trim={0cm 0cm 0.35cm 0cm},clip,width=\linewidth]{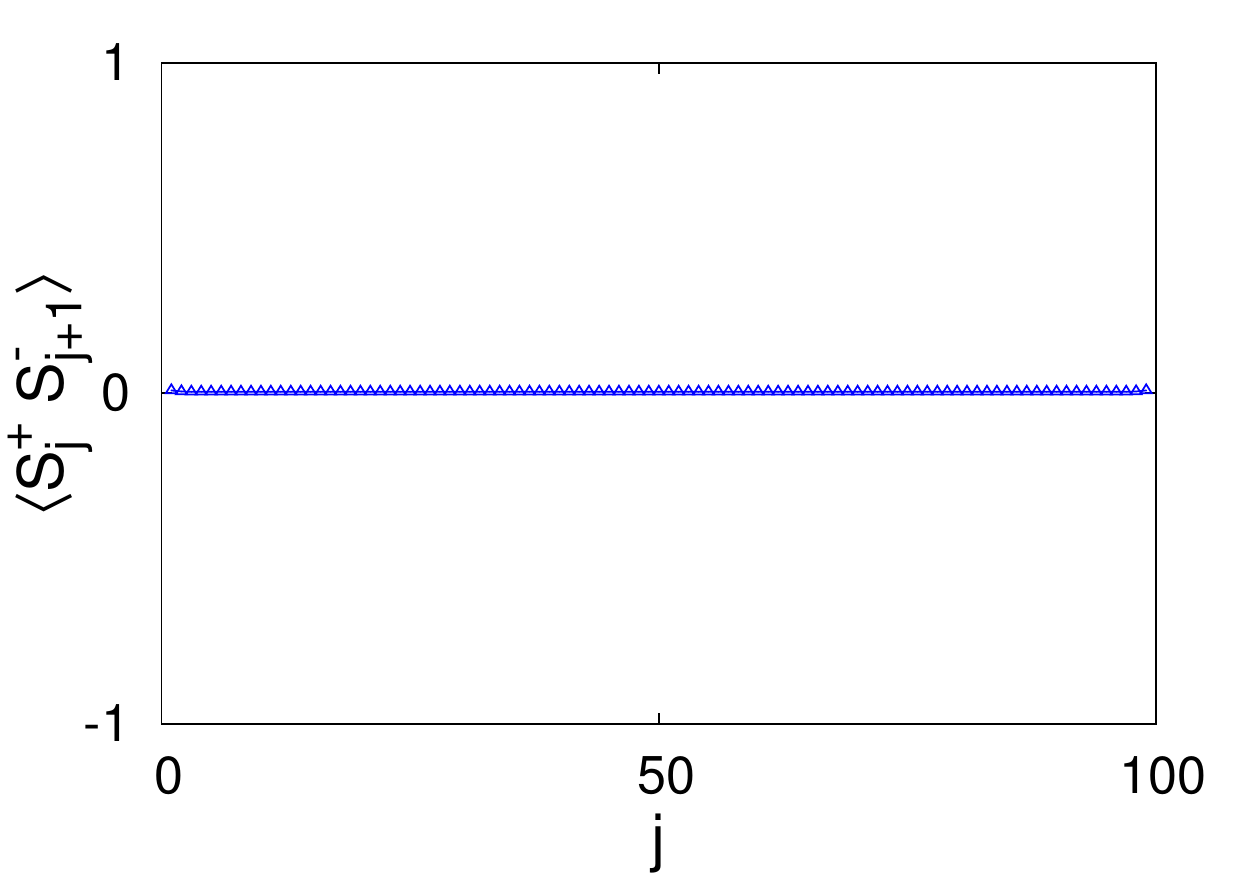}
\caption{} 
%\label{label-random_v1_and_v2_gamma_60_theta_15}
\end{subfigure}\hfill
\begin{subfigure}[b]{0.24\textwidth}
\centering
    \includegraphics[trim={0cm 0cm 0.35cm 0cm},clip,width=\linewidth]{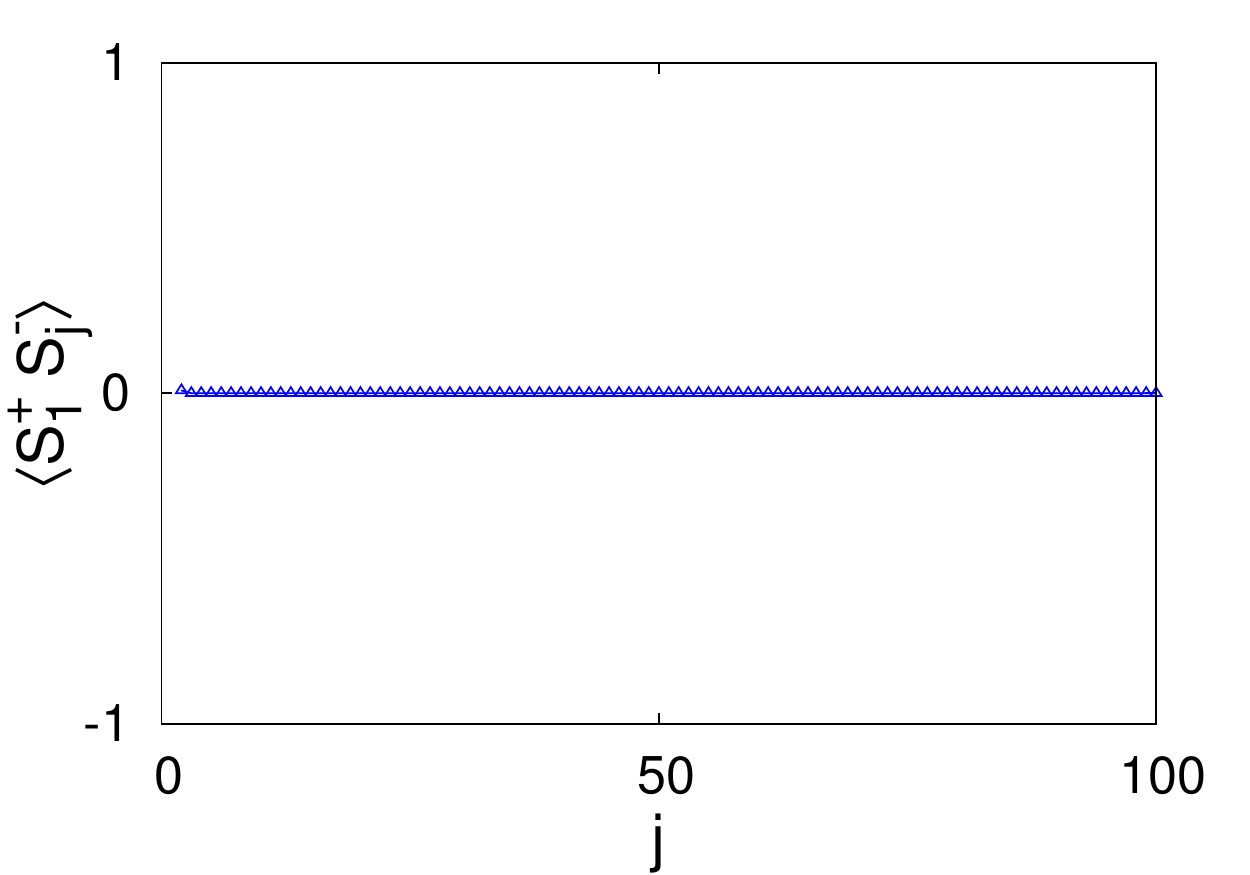}
\caption{} 
%\label{label-random_v1_and_v2_gamma_60_theta_15}
\end{subfigure}
\caption{Additional correlations for the AFM2 phase. $(\gamma, \theta) = (\pi/6,\pi/2)$. $(\alpha_o, \alpha_e, \alpha_2) = (0.799, 0.799, -13.740)$. Region: AAF. $\ket{\text{init}} = \ket{\downarrow\uparrow\downarrow\uparrow\downarrow\uparrow \ldots}$.}
\label{fig:correlations_gamma_30_theta_90_AFM2_phase_appendix}
\end{figure}

\fig{fig:correlations_gamma_30_theta_90_AFM2_phase_appendix} shows additional plots for the AFM2 phase shown in \fig{fig:gamma_30_theta_90_AFM}, which belongs to the non-frustrated region AAF.

\FloatBarrier 

\subsection{Transition between antiferromagnetic and superfluid phases}

In the phase diagram, it is hard to locate the exact boundary between AFM and SF phases for the finite system size $N = 100$. To understand the transition between these two phases, we neglect the hopping and interaction in the NNN direction (i.e., we set $\beta_2 = 0$ and $\alpha_2 = 0$.). We are interested in the situation where $\alpha_{o,e} > -1/4$, which means the pairwise interactions prefer antiparallel alignment of spins. As before, we set $\beta_1 = 1$ and for convenience, we consider $\alpha_o = \alpha_e$.

\begin{figure}[htb!]
\begin{subfigure}[b]{0.24\textwidth}
\centering
    \includegraphics[trim={0cm 0cm 0.35cm 0cm},clip,width=\linewidth]{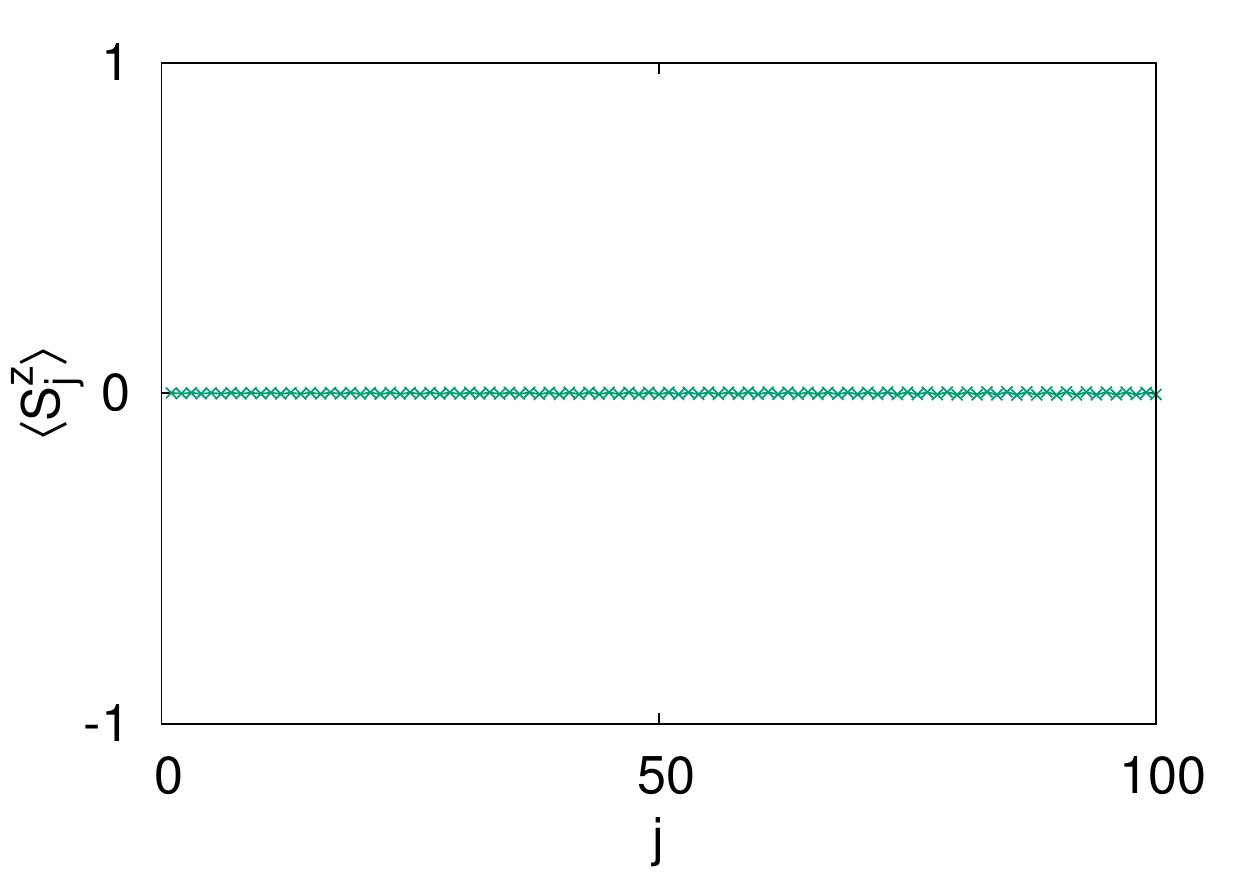}
\caption{}
%\label{label-random_v1_and_v2_gamma_60_theta_30}
\end{subfigure}\hfill
\begin{subfigure}[b]{0.24\textwidth}
\centering
    \includegraphics[trim={0cm 0cm 0.35cm 0cm},clip,width=\linewidth]{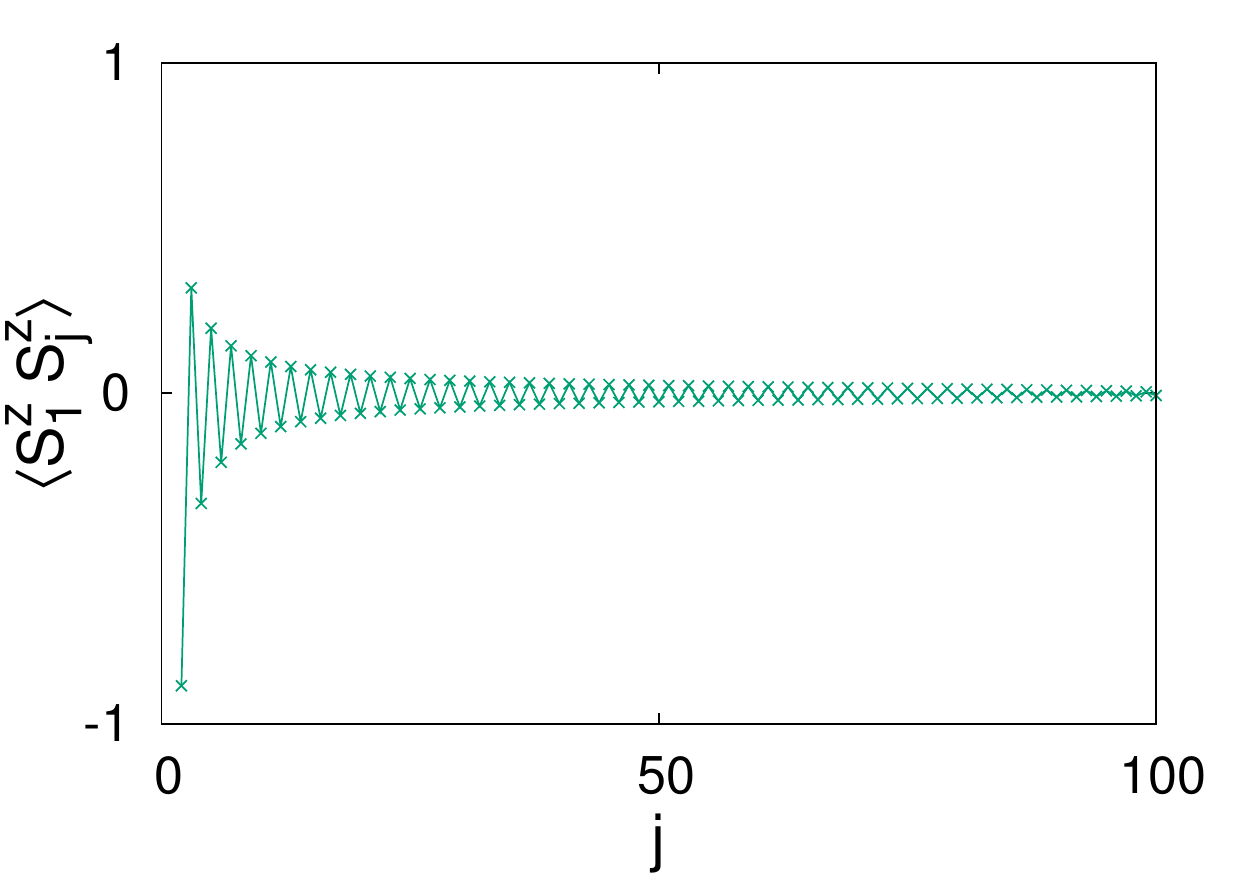}
\caption{}
%\label{label-random_v1_and_v2_gamma_60_theta_15}
\end{subfigure}\hfill
\begin{subfigure}[b]{0.24\textwidth}
\centering
    \includegraphics[trim={0cm 0cm 0.35cm 0cm},clip,width=\linewidth]{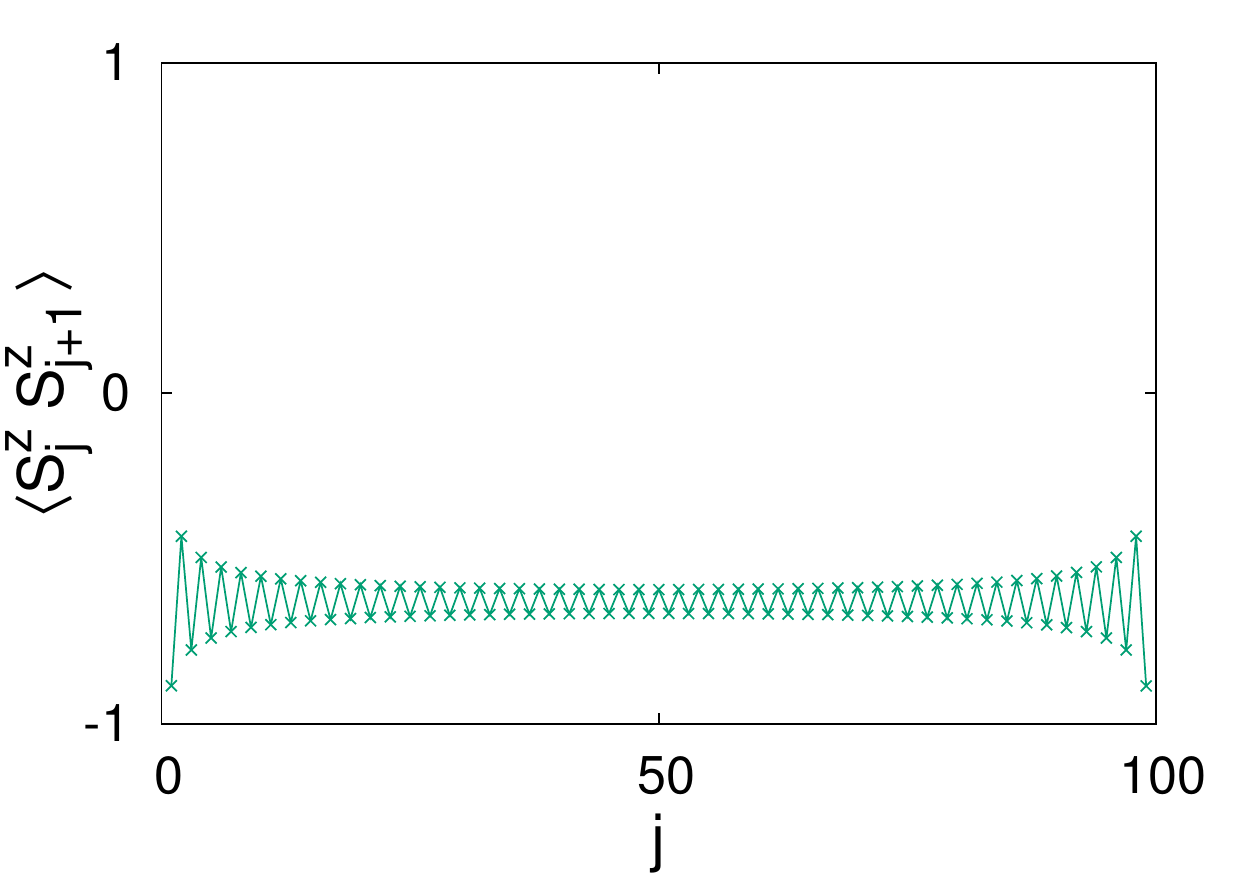}
\caption{} 
%\label{label-random_v1_and_v2_gamma_60_theta_15}
\end{subfigure}\hfill
\begin{subfigure}[b]{0.24\textwidth}
\centering
    \includegraphics[trim={0cm 0cm 0.35cm 0cm},clip,width=\linewidth]{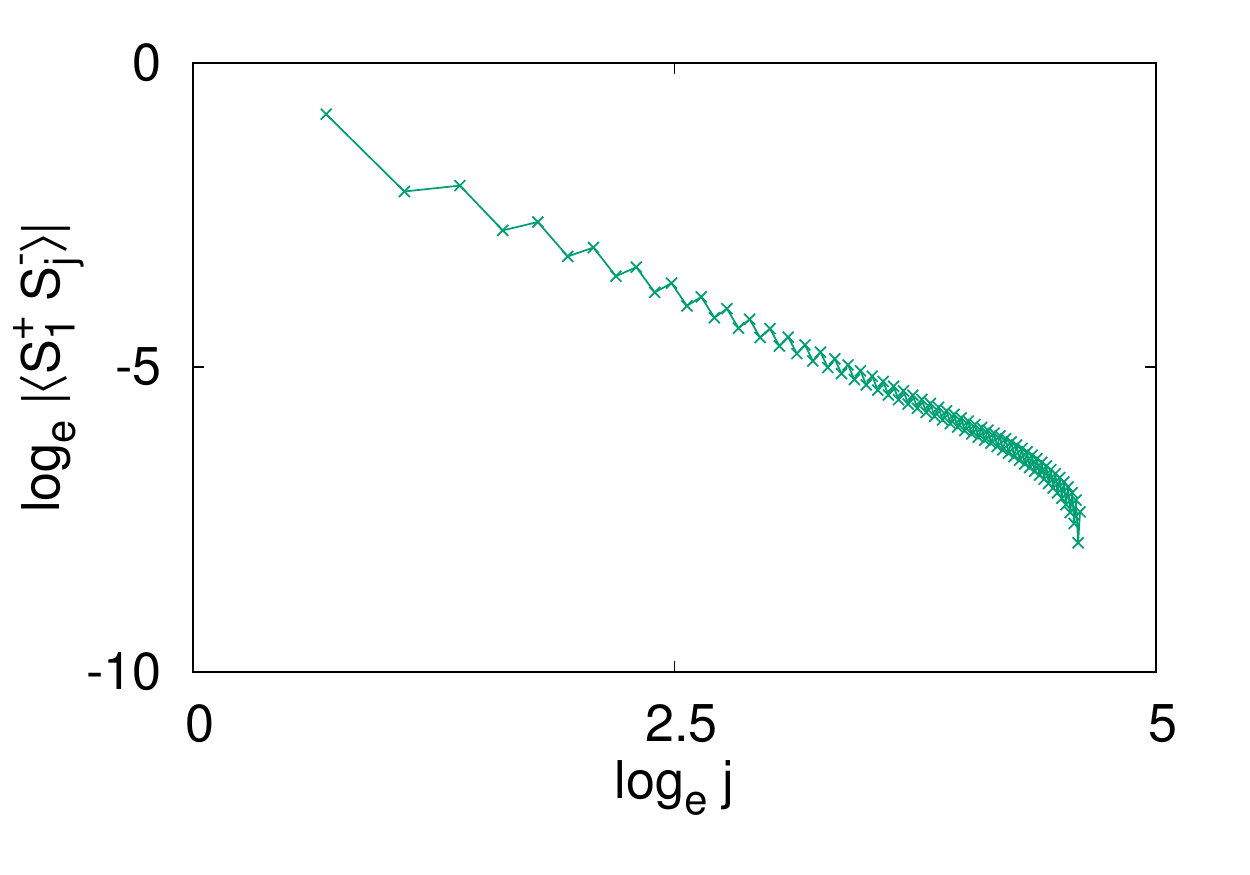}
\caption{} 
%\label{label-random_v1_and_v2_gamma_60_theta_15}
\end{subfigure}
\caption{Correlations for the SF phase with $\beta_1 = 1, \alpha_o = \alpha_e = 0.3$. $\ket{\text{init}} = \ket{\text{random}}$.}
\label{fig:correlations_B1_1_B2_0_ao_point3_ae_point3_SF_phase_appendix}
\end{figure}

\begin{figure}[htb!]
\begin{subfigure}[b]{0.24\textwidth}
\centering
    \includegraphics[trim={0cm 0cm 0.35cm 0cm},clip,width=\linewidth]{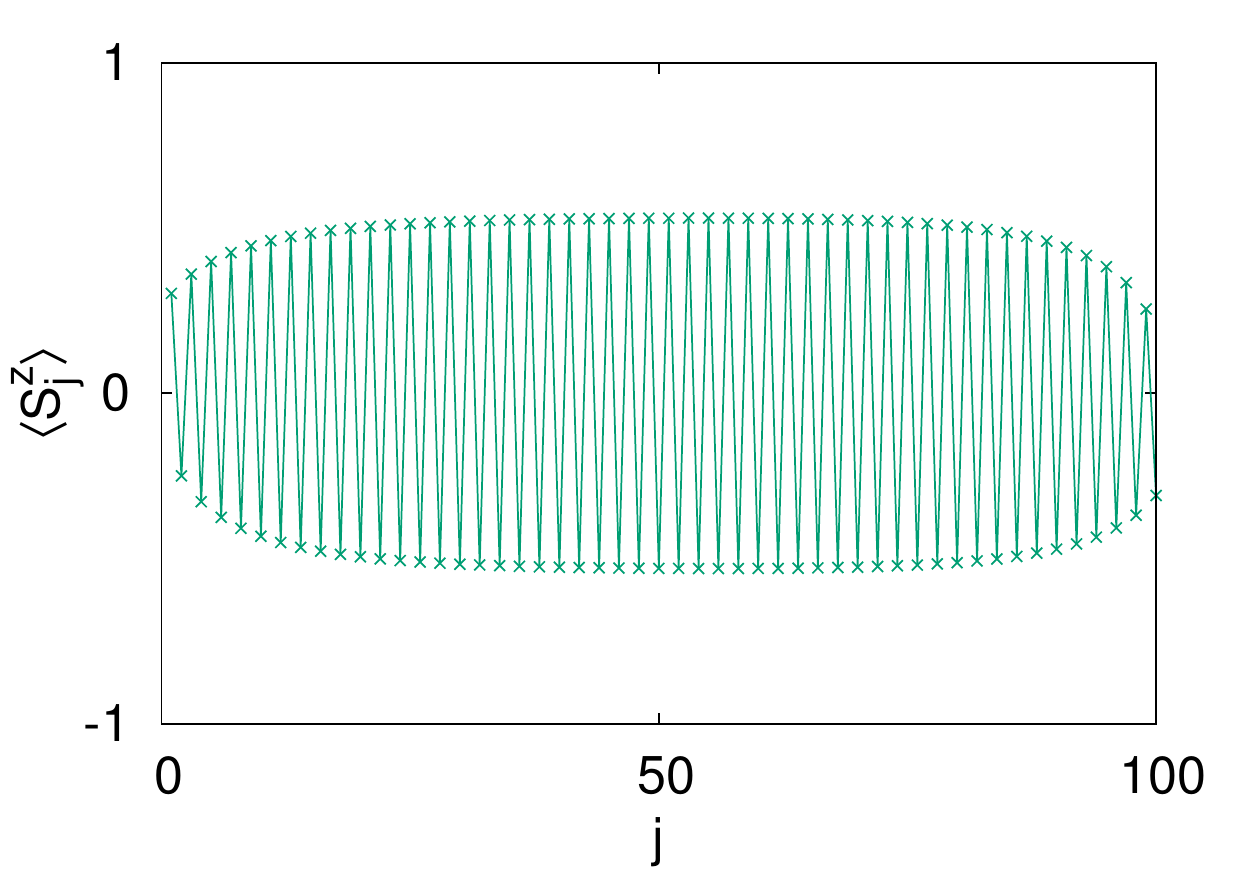}
\caption{}
%\label{label-random_v1_and_v2_gamma_60_theta_30}
\end{subfigure}\hfill
\begin{subfigure}[b]{0.24\textwidth}
\centering
    \includegraphics[trim={0cm 0cm 0.35cm 0cm},clip,width=\linewidth]{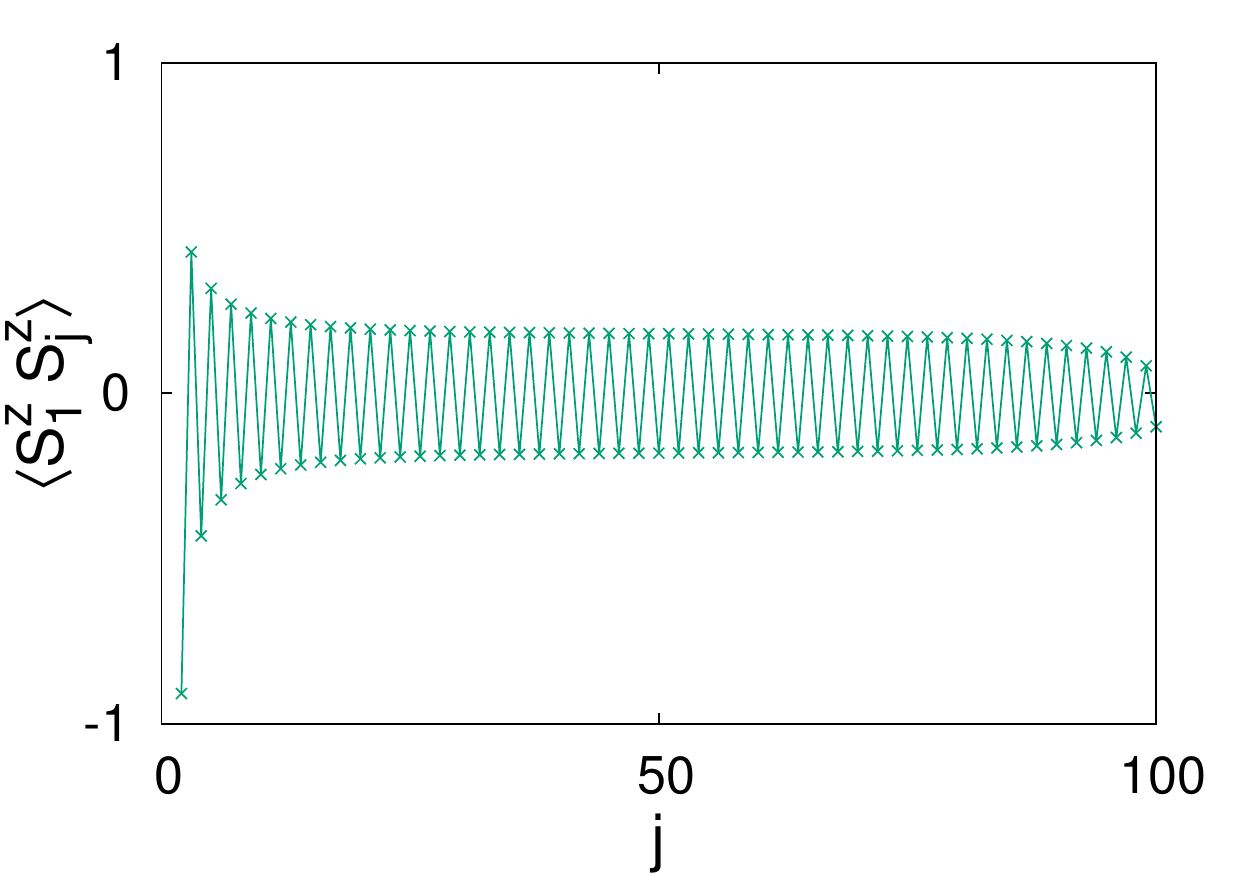}
\caption{}
%\label{label-random_v1_and_v2_gamma_60_theta_15}
\end{subfigure}\hfill
\begin{subfigure}[b]{0.24\textwidth}
\centering
    \includegraphics[trim={0cm 0cm 0.35cm 0cm},clip,width=\linewidth]{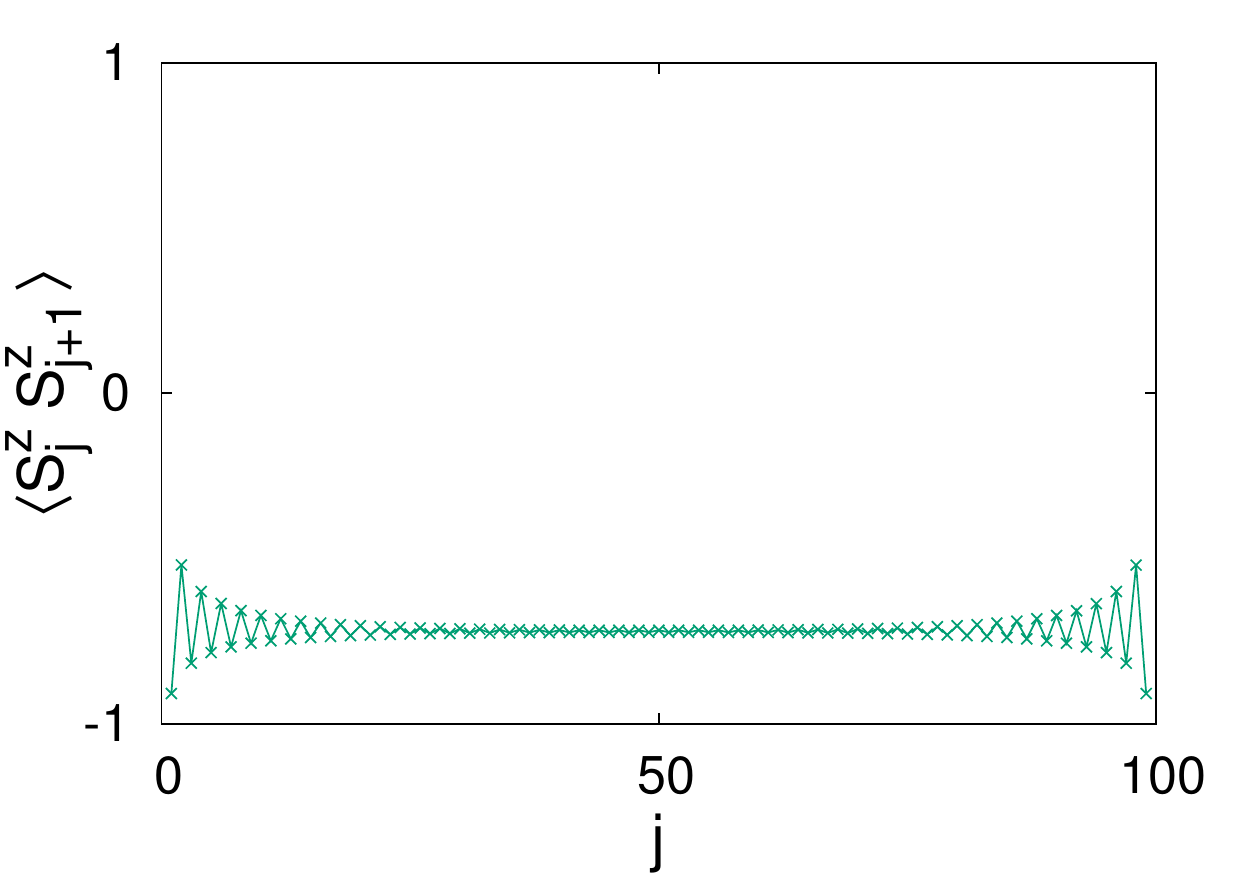}
\caption{} 
%\label{label-random_v1_and_v2_gamma_60_theta_15}
\end{subfigure}\hfill
\begin{subfigure}[b]{0.24\textwidth}
\centering
    \includegraphics[trim={0cm 0cm 0.35cm 0cm},clip,width=\linewidth]{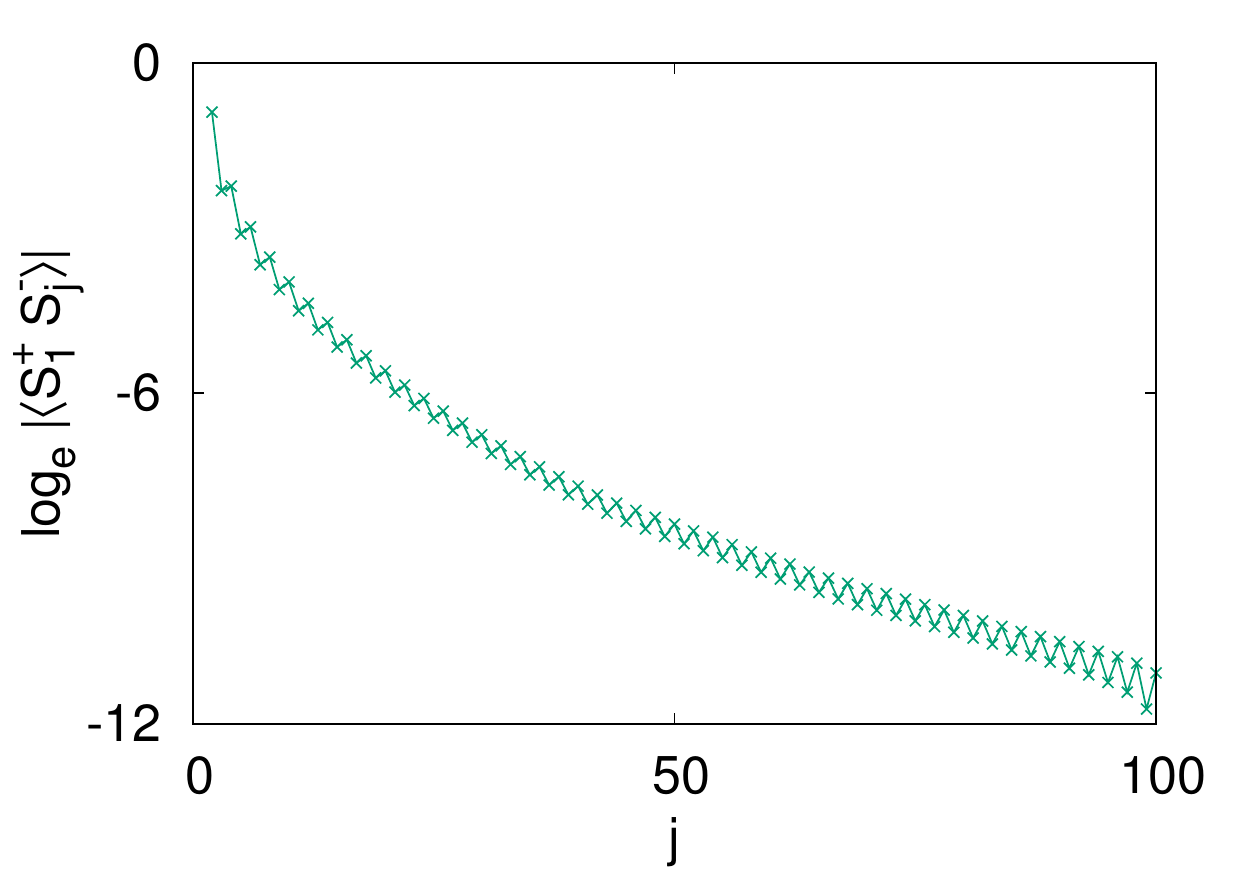}
\caption{} 
%\label{label-random_v1_and_v2_gamma_60_theta_15}
\end{subfigure}
\caption{Correlations for the AFM phase with $\beta_1 = 1, \alpha_o = \alpha_e = 0.4$. $\ket{\text{init}} = \ket{\text{random}}$.}
\label{fig:correlations_B1_1_B2_0_ao_point4_ae_point4_AFM_phase_appendix}
\end{figure}

\fig{fig:correlations_B1_1_B2_0_ao_point3_ae_point3_SF_phase_appendix} and \fig{fig:correlations_B1_1_B2_0_ao_point4_ae_point4_AFM_phase_appendix} show the various correlations for the cases $\alpha_{o,e} = 0.3$ and $\alpha_{o,e} = 0.4$. It is interesting to note that the nature of the correlations $\langle S^z_j S^z_{j+1} \rangle$ and $\langle S^z_1 S^z_j \rangle$ is not very different for the two cases; in fact, these correlations suggest the likelihood of an AFM phase. However, a SF phase in the former case is confirmed by the polynomial decay of the correlation $\langle S^+_1 S^-_j \rangle$ while an AFM phase in the latter is confirmed by the tendency of the spins to localize in lattice sites as indicated by the alternating sign for the values of the correlation $\langle S^z_j \rangle$ and the exponential decay of the correlation $\langle S^+_1 S^-_j \rangle$ which clearly indicates an insulating phase. 

Therefore, depending on the strength of $\beta_1$ (with hopping and interaction between nearest-neighbors only), the system can be in a SF or AFM phase when the pairwise interactions in the odd and even directions prefer antiparallel alignment. We also notice that there is a smooth crossover somewhere between  $\alpha_{o,e} = 0.3$ and $\alpha_{o,e} = 0.4$. Based on these results, it is safe to conclude that the nature of the transition between SF and AFM phases in the phase diagram (\fig{fig:phase_diagram}) is qualitatively the same.

%To crop a picture: trim={<left> <lower> <right> <upper>}

%To crop a picture: trim={<left> <lower> <right> <upper>}

\FloatBarrier

%\bibliographystyle{aipauth4-1} 
%Use of the above line changes the numbering of references? DON'T KNOW WHY.

%\clearpage

\bibliography{references_file}

\end{document}